\documentclass[fleqn,usenatbib]{mnras}

\usepackage{newtxtext,newtxmath}

\usepackage[T1]{fontenc}
\usepackage{comment}

\usepackage[flushleft]{threeparttable}
\usepackage{tablefootnote}
\DeclareRobustCommand{\VAN}[3]{#2}
\let\VANthebibliography\thebibliography
\def\thebibliography{\DeclareRobustCommand{\VAN}[3]{##3}\VANthebibliography}

\usepackage{graphicx}	
\usepackage{amsmath}	
\usepackage{multirow}
\usepackage{siunitx}
\usepackage{adjustbox}
\usepackage{orcidlink}

\title[Filament fragmentation in NGC 6334-43]{A nine-member protostellar system forming via filament fragmentation in the high mass protocluster NGC 6334-43}

\author[D. J. Taylor et al.]{D. J. Taylor,$^{1}$\thanks{E-mail: dt79@st-andrews.ac.uk}\orcidlink{0009-0009-2022-7484}
C. J. Cyganowski,$^{1}$\orcidlink{0000-0001-6725-1734}
C. L. Brogan$^{2}$\orcidlink{0000-0002-6558-7653}
T. R. Hunter $^{2}$\orcidlink{0000-0001-6492-0090}
B. A. McGuire $^{2, 3}$\orcidlink{0000-0003-1254-4817}
G. M. Williams $^{4}$\orcidlink{0000-0001-5933-2147}
\\
$^{1}$Scottish Universities Physics Alliance (SUPA), School of Physics and Astronomy, University of St Andrews, North Haugh, St Andrews KY16 9SS, UK\\
$^{2}$National Radio Astronomy Observatory, 520 Edgemont Road, Charlottesville VA 22903, USA\\
$^{3}$Department of Chemistry, Massachusetts Institute of Technology,
Cambridge, MA 02139, USA\\
$^{4}$Department of Physics, Aberystwyth University, Ceredigion, Cymru, SY23 3BZ, UK\\
}

\date{Accepted XXX. Received YYY; in original form ZZZ}

\pubyear{2026}

\begin{document}
\label{firstpage}
\pagerange{\pageref{firstpage}--\pageref{lastpage}}
\maketitle

\begin{abstract}
\noindent We present the serendipitous discovery of a nine-member system comprised of protostellar and candidate prestellar sources in $\sim$350 au-resolution images from Complex Chemistry in hot Cores with ALMA (CoCCoA). The system is bound in a stability analysis, has a mean separation between pairs of 7930 au, and appears to have formed via the fragmentation of a single large-scale filamentary structure traced by 1.20\,mm continuum and H$^{13}$CO$^+$ J = 3-2 emission. Two multiples within the nine-member system, a triple and a binary, have properties consistent with formation by core fragmentation on $\sim$1500-1700 au scales. The hot core NGC 6334-43 is resolved into two components (ALMA2a/ALMA2b) separated by 618 au and driving a bipolar outflow traced by $^{12}$CO J = 2-1 and SiO J = 5-4 in $\sim$1250 au-resolution archival Atacama Large Millimeter/submillimeter Array (ALMA) data. Only one other source in the nine-member system is clearly protostellar: ALMA6a, which drives an outflow traced by $^{12}$CO. The outflow properties of ALMA2a/ALMA2b and ALMA6a are consistent with high-mass and low-mass Class 0 sources respectively. By fitting the CH$_{3}$CN J = 13-12 emission towards ALMA2a, ALMA2b and ALMA6a, we derive M$_{\rm vir}$ = 4.5, 5.4 and 2.6 M$_{\odot}$ respectively. The other six sources in the nine-member multiple have M$_{\rm gas}$ = 0.50-1.87 M$_{\odot}$ and appear young, as indicated by their sparse mm-wavelength line emission and non-detection in published cm continuum observations. Our results highlight the potential of serendipitous discoveries in ALMA surveys to add to the small observational sample of young high-mass protomultiple systems.

\end{abstract}




\section{Introduction} \label{introduction}

Whilst on the main sequence, massive stars (M$_{\rm ZAMS}$ > 8 M$_{\odot}$, Zero-Age Main-Sequence) exhibit binarity or higher order multiplicity almost ubiquitously \citep[e.g][]{2014Sana, 2017MoeandDiStefano, Pauwels2023}. However, the relative prevalence of the distinct formation mechanisms that can lead to this multiplicity is unknown \citep[see recent review by][]{Offner2022}. Fragmentation of discs \citep[e.g.][]{Adams1989, Bonnell1994, Kratter2010, Oliva2020}, cores \citep[e.g.][]{Larson1972, Boss1979, Offner2010, Guszejnov2017} and filaments \citep[as observed by][]{Pineda2015} as well as gas mediated capture \citep[e.g.][]{Ostriker1994, Bate2003, Cournoyer2020} and N-body interactions \citep[e.g.][]{Bate2002,Moeckel2010, Meyer2014} have been suggested as mechanisms that may form high-mass multiple systems. Together, these mechanisms may act across a large range of scales and times.

Over the past 12 years, a wealth of surveys that successfully resolve the discs of individual protostars have been carried out in low-mass star forming clouds \citep[e.g.][]{Chen_2013, Tobin2016, Kounkel2016, Encalada_2021, Tobin2020, Tobin2022}. In such work, it is often suggested that both core and disc fragmentation contribute to the overall observed multiplicity of low-mass protostars. For example, the bimodal distribution of separations, with peaks at $\sim$100 au and $\sim$10$^{3}$ au, found in Orion by the VLA/ALMA Nascent Disk and Multiplicity (VANDAM) survey is primarily driven by the youngest, Class 0, sources \citep{Tobin2022}. As such, \cite{Tobin2022} suggest that both disc and core fragmentation, as well as inward migration acting on short timescales \citep[see e.g.][]{Bate_Bonnell1997,Stahler2010,Lee_2019}, are required to reproduce their observed separation distribution. However, understanding of how each formation mechanism acts and their relative importance is debated. For example, the separation distribution (with peaks at $\sim$75 au and $\sim$3000 au) of the VANDAM Perseus sample \citep{Tobin2016} is reproduced in simulations by \cite{Kuruwita2023} with only turbulent fragmentation and dynamical capture, i.e. without disc fragmentation.

Compared to their low-mass counterparts, high-mass protomultiple systems are much more challenging to identify and study. This is because larger distances to denser clouds that are highly extincted at infrared (IR) and optical wavelengths, as well as high-mass stars' rapid evolution whilst remaining deeply embedded in their natal environment, impede observational characterisation \citep[e.g.][]{Motte2018, Rosen2020_review}. As such, much understanding of high-mass multiple formation has to date been derived from simulations. Many works adopting initial conditions of isolated non-turbulent massive prestellar cores have found that multiples form primarily via disc fragmentation. For example, \cite{Krumholz2009} and \cite{Oliva2020} both found that fragmentation takes place within prominent spiral arm structures in the disc. Binaries with initial separations of 1000 - 2000 au are produced by both works, and \cite{Krumholz2009} also show that migration and gravitational interactions between fragments can close separations to <30 au. \citet{Mignon-Risse2023} have shown that magnetic fields have a strong influence on primordial system parameters; their non-ideal magneto-hydrodynamical (MHD) simulations result in initial binary separations 2$\times$ smaller than their non-magnetised runs. And whilst most studies produce high-mass ratio multiples, \cite{Mignon-Risse2021} produce binaries with a more even mass ratio (q = 1.1$-$1.6) via disc fragmentation in their non-ideal MHD super-Alfv\'enic cases. There, fragmentation of spiral arms results in binaries with separations of 300$-$700 au, which in turn host their own 100$-$200 au radii discs.

Factors including magnetic fields may also resist disc fragmentation. For instance, in the ideal MHD regime, \cite{Commercon2022} find that the inclusion of magnetic fields prevents the production of multiples entirely. The sub-Alfv\'enic turbulence adopted by \cite{Mignon-Risse2021} in their non-ideal MHD simulations also results in single stellar systems. Additionally, \cite{Meyer2024} show that `stellar wobbling' (star-disc interactions) acts against the development of gravitational instabilities in the disc.

Introducing additional physics into simulations may promote the production of multiples via core fragmentation. The work of \cite{Rosen2020}, which assumes an isolated prestellar core and includes outflows, radiation feedback and turbulence, finds that binaries are produced via core fragmentation at early times and via disc fragmentation at late times. 

Certainly, however, the commonly assumed initial conditions of isolated massive prestellar cores are observationally very rare \citep[see][and references therein, for the complete sample]{Valeillemanet2025}. A notable exception in this regard is \cite{He2023}, who conduct zoom-in simulations of a giant molecular cloud (GMC), and find that multiples residing in shared circumstellar discs (radii of 200$-$6000 au) were formed via `quasi-spherical' or filament fragmentation at early times, before their accretion onto a dominant circumstellar disc. Notably, this result suggests that the observation of a multiple within a shared disc is not in itself sufficient to conclude that the system has formed via disc fragmentation.

Close comparison between the multiples produced in simulations and system parameters obtained from observations may be highly complex as observed system properties are highly dependent on the evolutionary stage of the protostar(s). Parameters such as the mass ratio, separation and disc properties of protomultiples of all masses may dramatically change on short timescales due to a variety of physical processes, including gas-dynamical friction \citep{Bate_Bonnell1997}, disc migration \citep[e.g.][]{Tokovinin2019, Munoz2019, Siwek2023}, mass accretion \citep[e.g.][]{Lund2018}, magnetic braking \citep[e.g.][]{Harada2021} and interaction with the disc or nearby protostars \citep[e.g.][]{Ramirez2021}. Simulations of whole clouds populated by low-mass multiple systems find that the transition between a widely and closely separated system can occur on timescales of just 10$^{4}$$-$10$^{5}$ years due to mass accretion and gas-dynamical friction \citep{Offner2010, Offner_2016, Lee_2019}. Moreover, \cite{Cournoeyer-Cloutier2024} simulate three massive clusters (initial M$_{\rm gas}$ of 2 $\times$ 10$^{4}$, 8 $\times$ 10$^{4}$ and 3.2 $\times$ 10$^{5}$ M$_{\odot}$) and find that the timescales such processes act on are environment-dependent: in their most massive cloud, the separations are reduced on the shortest timescale. Thus, using observed system properties to distinguish between formation mechanisms \citep[e.g.][]{Offner2022} requires identifying and studying the youngest multiple systems. 

To date, only a few very young high-mass multiple systems have been identified \citep{Cyganowski2022,Kong2023, Barnes2023}, with most previous studies characterising systems that are already IR-bright, host UC/HC H{\footnotesize II} regions or display rich hot core line emission \citep[e.g.][]{Kraus2017, Beuthar2017, Ilee_2018, Pomohaci2019, Koumpia2021, Ahmadi2023}. Recently, \cite{Li2024, Li2025} showed that bound high-order protomultiple systems can form via distinct formation mechanisms. \cite{Li2024} find that a triple, quadruple and quintuple protostellar system have each formed via core fragmentation on scales of $\sim$731 au in G333.23$-$0.06. \cite{Li2025} show that a septuple system with a mean deprojected separation of 332 au has formed via disc fragmentation in NGC 6334I(N). 

The limited sample of young protobinaries and protomultiples in high-mass star forming environments known in the literature motivates searching for additional examples. Notably, one of the youngest massive protobinaries discovered to date, the G11.92$-$0.61 MM2 system, was identified serendipitously in Atacama Large Millimeter/submillimeter Array (ALMA) observations targeting the nearby line-rich disc G11.92$-$0.61 MM1 \citep{Cyganowski2022}. In this work, we use the Complex Chemistry in hot Cores with ALMA (CoCCoA) survey \citep{Chen2023, Chen2025} observations of the field containing the NGC 6334-43 hot core to characterise the protostellar multiplicity and report that a bound nine-member system appears to have formed via the fragmentation of a single large-scale filamentary structure.

\begin{figure}
    \centering
	\includegraphics[trim=0 0 0 0, clip, width=\columnwidth]{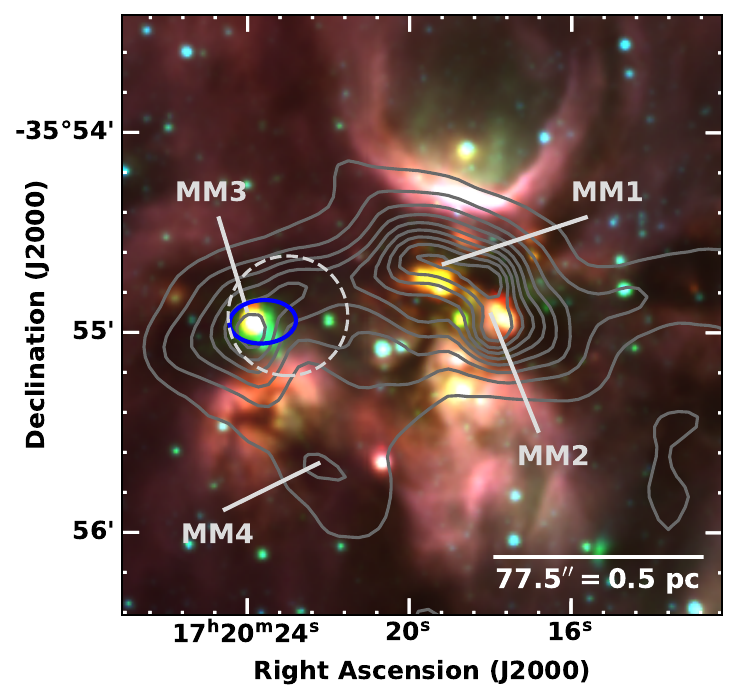}
       \caption{Three colour \emph{Spitzer} Galactic Legacy Infrared Midplane Survey Extraordinaire (GLIMPSE) \citep{Benjamin2003,Churchwell2009} image (red = 8.0\,$\mu$m, green = 4.5\,$\mu$m, blue = 3.6\,$\mu$m),  
       overlaid with 9$^{\prime\prime}$-resolution 450 $\mu$m SCUBA \citep{DiFrancesco2008} contours (silver, levels: [0.1, 0.2, 0.3, 0.4, 0.5, 0.6, 0.7, 0.8, 0.9] $\times$ I$_{\rm peak} =$ 187 Jy beam$^{-1}$. Submm sources identified by \citet{Sandell1999} are
       labelled in white and the FWHM extent of AGAL351.251+00.652 \citep{Contreras2013} is marked with a blue ellipse. The 15\% response level of the CoCCoA ALMA primary beam is shown as a light grey dashed circle.}
    \label{fig:rgb}
\end{figure}

\subsection{NGC 6334-43} \label{NGC6334-43}

The CoCCoA survey was designed to study complex organic molecules (COMs) towards hot cores in 25 high-mass star forming regions. As shown in Figure~\ref{fig:rgb}, the field of view (FOV) of the CoCCoA NGC 6334-43 ALMA observations lies towards MM3, one of four sub-mm continuum peaks identified to the east of NGC 6334 IV = A by \cite{Sandell1999} in James Clerk Maxwell Telescope (JCMT) observations (see also \citealt{Persi2008book}). Subsequent single-dish mm-wavelength spectral line observations showed that MM3 hosts a hot core, NGC 6334-43 \citep{McGuire2014, Weaver2017}, which was included in the CoCCoA sample. MM3 also corresponds to the ATLASGAL source AGAL351.251+00.652 \citep[][see Figure~\ref{fig:rgb}]{Contreras2013}, which has v$_{\rm LSR}$ measurements of $-$0.42 km s$^{-1}$ \citep{Rathbone2016}, $-$0.82 km s$^{-1}$ \citep{Urquhart_2019} and $-$1.14 km s$^{-1}$ \citep{Wienen2018}. Two maser parallaxes have been reported for the NGC 6334 high-mass star-forming region: 0.789 $\pm$ 0.161 mas \citep{Chibeuze2014} and 0.752 $\pm$ 0.069 mas \citep{Wu2014}. As in \citet{Reid2019} we adopt a weighted mean of these measurements, giving d = 1.33$^{+0.11}_{-0.13}$ kpc. 

To date, the NGC 6334-43 region has been studied at sub-arcsecond-resolution only at near-mid infrared (IR) \citep{Persi2009} and cm wavelengths \citep{Yanza2025}. Within the CoCCoA FOV, \citet{Yanza2025} report three VLA cm continuum detections and \citet{Persi2009}, using the Perssons Auxiliary Nasmyth Infrared Camera (PANIC), report seven IR sources, which are detected only in the K$_{s}$ band, where emission may arise from reprocessed or scattered light from a protostar or relate to outflows. \cite{Persi2009} also detect three H$_{2}$ emission knots, which are known to trace gas shocked by outflows \citep{Neufeld1998,Maret2008}. In addition, two masers with accurate Australia Telescope Compact Array (ATCA) positions have been reported near NGC 6334-43: a 1.665 GHz OH maser \citep{Brooks_Whiteoak2001} and a 6.668 GHz Class II CH$_{3}$OH maser \citep{Caswell2010}. Class II CH$_{3}$OH masers are radiatively excited \citep{Cragg1992, Sobolev1994, Sobolev1997} and exclusively associated with high-mass star formation \cite[e.g.][]{Minier2003, Xu2008} and OH masers are often found in their vicinity \cite[e.g.][]{Caswell1997, Caswell1998}.  

In this paper, we present the first high resolution mm-wavelength study of the NGC 6334-43 field, complementing our $\sim$350 au (0\farcs264)-resolution CoCCoA data with previously-unpublished archival ALMA $\sim$1250 au (0\farcs938)-resolution observations of the outflow tracers $^{12}$CO and SiO. We detail observational information relevant to this work in Section~\ref{sec:Observations}. We present our results in Section~\ref{sec:Results}, where we discuss the relation of the newly-resolved mm sources to detected outflows and known IR, cm continuum, and maser emission. In Section~\ref{sec:analysis} we present analysis of a large-scale H$^{13}$CO$^{+}$ structure, modelling of the CH$_{3}$CN emission seen towards the identified mm sources and derived properties of the mm sources and their outflows. Section~\ref{sec:Discussion} discusses the nature of the outflow-driving sources and the stability and likely formation mechanism(s) of the nine-member system and its sub-systems. We summarise our main conclusions in Section~\ref{conclusions}. 
    
\section{Observations} \label{sec:Observations}

\subsection{CoCCoA survey}\label{sec:Coccoa_obs}

In this paper, we use data from the CoCCoA survey (PI: B. McGuire) taken as part of ALMA project 2019.1.00246.S 
and described in \citet{Chen2023}. The CoCCoA survey data include two tunings of the Band 6 receiver \citep{Kerr2014}, each with four correlator spectral windows (spws). These cover 238.007$-$241.733 GHz for the lower tuning and 258.009$-$261.735 GHz for the upper tuning, with small overlaps between spws. 

Each spw was imaged with 0.3\,MHz channels, corresponding to velocity channel widths of 0.373$-$0.377 km s$^{-1}$ for the lower tuning and 0.344$-$0.348 km s$^{-1}$ for the upper tuning. We also jointly image two of the lower-tuning spws (from 238.400 GHz to 239.336 GHz), also with 0.3\,MHz channels, to produce an image cube covering the full CH$_{3}$CN (13-12) ladder for the fitting analysis presented in Section~\ref{CH3CN fitting}.  As noted in \citet{Chen2023}, the spectral resolution of the data are $\sim$0.6 km s$^{-1}$, twice the channel width due to the Hanning smoothing applied online in the ALMA correlator \citep{Escoffier2007}.

The phase centre of the NGC 6334-43 ALMA pointing is 17$^{\rm h}$20$^{\rm m}$23.000$^{\rm s}$ $-$35$^{\circ}$54$^{\prime}$55.00$^{\prime\prime}$ and the full width at half maximum (FWHM) of the primary beam at 1.20\,mm is $\sim$23$^{\prime \prime}$ ($\sim$0.15 pc). 
As described in \citet{Chen2023}, the continuum images and line cubes for the 14 CoCCoA fields presented in that work were convolved to a common angular resolution of 0\farcs330 $\times$ 0\farcs330 (P.A. = $-$0.0$^{\circ}$). In this work, we make use of both the convolved images and image cubes described in \citet{Chen2023} and the unconvolved, native-resolution images and cubes for the NGC 6334-43 field. 

The unconvolved lower tuning, upper tuning and combined continuum images have synthesised beam sizes of 0\farcs276 $\times$ 0\farcs257 (P.A. = 78.4$^{\circ}$), 0\farcs305 $\times$ 0\farcs236 (P.A. = $-$87.3$^{\circ}$) and 0\farcs280 $\times$ 0\farcs248 (P.A. = 89.1$^{\circ}$) respectively.
The rms noise ($\sigma$) near the centre of the CoCCoA FOV in the unconvolved primary beam corrected continuum images is $\sigma_{\rm centre}$ = 0.102 mJy beam$^{-1}$, 0.166 mJy beam$^{-1}$ and 0.088 mJy beam$^{-1}$ for the lower tuning, upper tuning and combined continuum respectively. In the unconvolved images prior to correction for the primary beam response $\sigma_{\rm centre,nonpb}$ = 0.101 mJy beam$^{-1}$, 0.163 mJy beam$^{-1}$ and 0.086 mJy beam$^{-1}$ for the lower tuning, upper tuning and combined continuum images respectively. 

We note that the rms varies significantly across the CoCCoA FOV, and discuss individual sources with reference to the local noise in Section~\ref{individual_sources}. 

The unconvolved spectral line cubes from the lower tuning dataset have synthesised beam sizes ranging from 0\farcs286$-$0\farcs298 $\times$ 0\farcs267$-$0\farcs276 (P.A. between 75.8 and 78.6$^{\circ}$). The unconvolved spectral line cubes from the upper tuning dataset have synthesised beam sizes ranging from 0\farcs317$-$0\farcs321 $\times$ 0\farcs249$-$0\farcs252 (P.A. between $-$87.6$^{\circ}$ and $-$85.8$^{\circ}$). 

Local rms noise values for line detection towards each source are listed in Table~\ref{tab:fitted_line_emission_properties}. The rms noise of the CH$_{3}$CN cube used in Section~\ref{CH3CN fitting}, which is smoothed to  0\farcs330 $\times$ 0\farcs330, is 4.05 mJy beam$^{-1}$ in the vicinity of ALMA2a/b and 2.57 mJy beam$^{-1}$ near ALMA6a (which is significantly nearer the pointing centre). 

For our continuum analysis, we use the unconvolved image as we seek to resolve compact components of candidate protomultiple systems. For our line analysis, we primarily use the convolved image cubes, because most lines of interest are weak (see Section~\ref{CoCCoA_line_emission}) and the convolved cubes provide better signal-to-noise for weak, extended emission. The exception is our search for disc signatures (see Appendix~\ref{Appendix-discs}), for which we used both the convolved and unconvolved line cubes. 

Unless otherwise stated, all measurements are taken from images that have been corrected for the primary beam response. However, as the noise increases sharply towards the edge of the primary beam, some figures show line data prior to correction for the primary beam response in order to better visualise the line morphology across the full CoCCoA FOV.

\subsection{Archival ALMA observations}\label{sec:archive_obs}

To complement the CoCCoA data, we use lower-resolution observations of $^{12}$CO (2-1) and SiO (5-4) from the ALMA archive.  We downloaded the raw 12m and 7m array Band 6 mosaic data from project 2017.1.00180.S (PI: Louvet) for field 6334\_-\_MDC\_9, which covers the CoCCoA NGC 6334-43 FOV. 
The 12m data were observed as a single execution with 43 antennas on Apr 8 2018, while seven executions were observed with the 7m array from Oct 26 2017 to Apr 9 2018, with 9-12 antennas.  We reprocessed the raw data using version 2024.1.0.8 of the ALMA science pipeline \citep{Hunter2023_pipeline} in CASA \citep{casa2022} version 6.6.1.17, which includes correction for renormalization issues $>$2\% and improved continuum-finding performance.
Phase-only iterative self-calibration was performed manually on the resulting 7m and 12m psuedocontinuum datasets, and the solutions applied to the respective continuum-subtracted line data.

In this paper, we consider only the $^{12}$CO (2-1) and SiO (5-4) lines from these observations. Combined 7m and 12m line cubes for each line were made using multiscale clean, automasking \citep{Kepley20} and Briggs weighting with robust=0.5. For $^{12}$CO (2-1), additional manual masking was done to optimise the cleaning of the complex emission. 

The synthesised beamsize and channel width of the resulting cubes are 1\farcs010 $\times$ 0\farcs827 (P.A. = $-$61.6$^{\circ}$) and 0.64 km s$^{-1}$ for $^{12}$CO and 1\farcs090 $\times$ 0\farcs880 (P.A. = $-$64.2$^{\circ}$) and 0.4 km s$^{-1}$ for SiO respectively. The rms noise of the primary-beam corrected SiO cube is $\sim$ 21 mJy beam$^{-1}$, while the rms of the primary-beam corrected $^{12}$CO cube varies significantly channel-to-channel, ranging from $\sigma$ $\sim$ 16 mJy beam$^{-1}$ for channels without strong emission to  $\sigma$ $\sim$ 34 mJy beam$^{-1}$ for channels with strong emission. 

\begin{figure}
    \centering
	\includegraphics[trim=0 0 0 0, clip, width=\columnwidth]{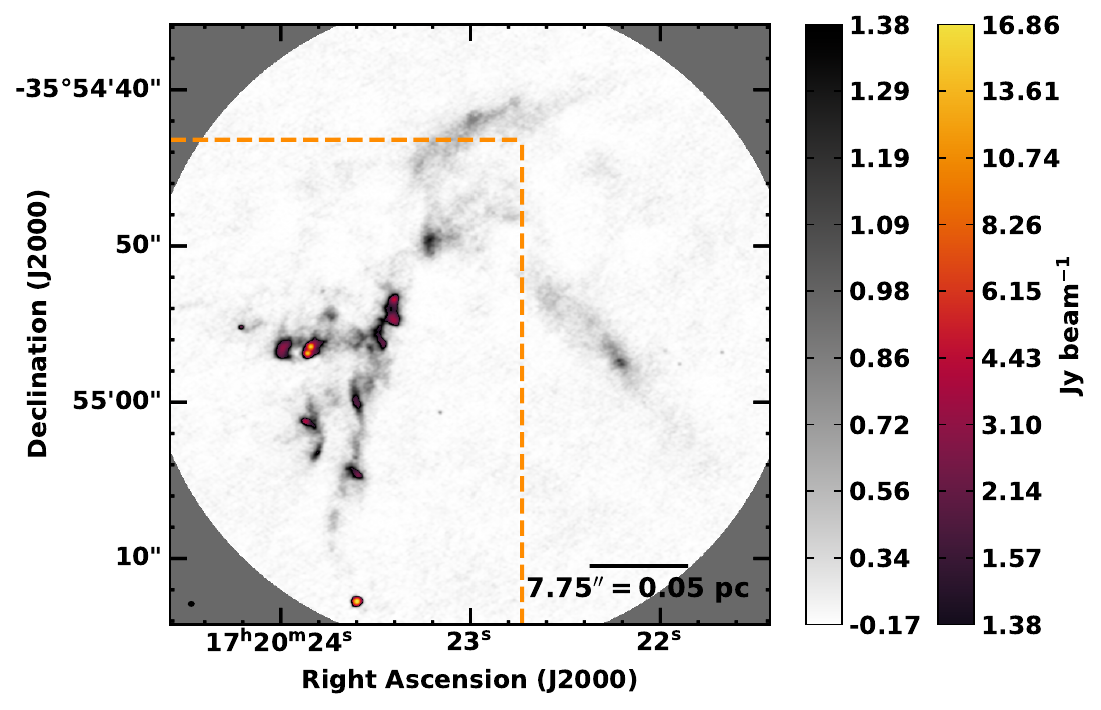}
       \caption{CoCCoA ALMA 1.20\,mm continuum image of the NGC 6334-43 field (colourscale, using a double colourbar). The image shown has not been corrected for the primary beam response and the greyscale shows emission from $-$2$\sigma_{\rm centre,nonpb}$, where $\sigma_{\rm centre,nonpb}$ = 0.086 mJy beam$^{-1}$ (see Section~\ref{sec:Coccoa_obs}), to 16$\sigma_{\rm centre,nonpb}$ on a power law stretch with exponent = 2; the colourscale shows emission from 16$\sigma_{\rm centre,nonpb}$ to 196$\sigma_{\rm centre,nonpb}$ on a power law stretch with exponent = 0.5.   The ALMA synthesised beam is shown at bottom left and the dashed orange lines mark the FOV of Figure~\ref{fig:continuum_plot}a.}
    \label{fig:full_FOV_continuum}
\end{figure}

\section{Results} \label{sec:Results}

\subsection{ALMA 1.20 mm Continuum Emission} \label{continuum emission}

\begin{figure*}
    \centering
	\includegraphics[trim=0 0 0 0, clip, width = \textwidth]{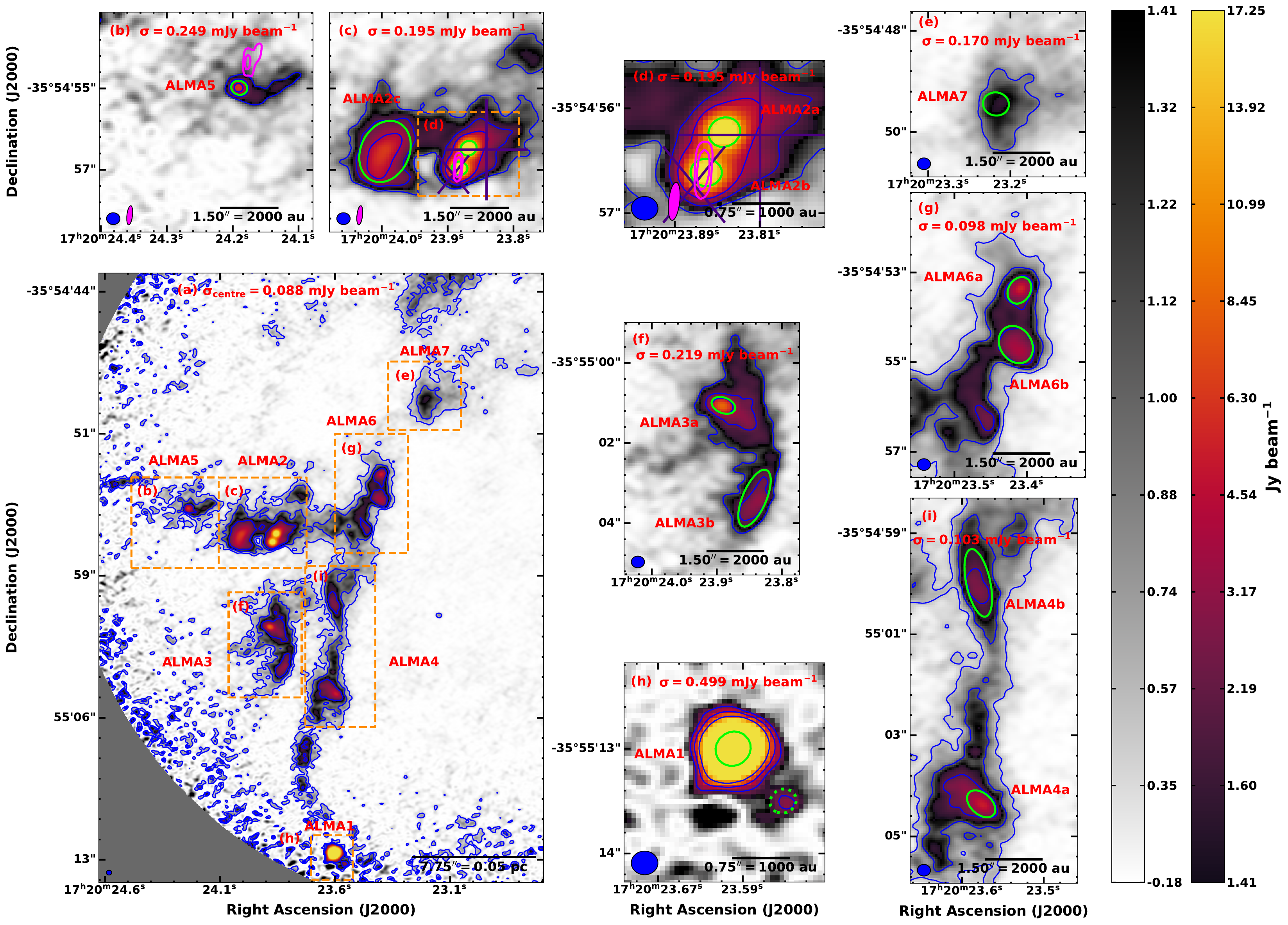}
    \caption{(a) Primary beam corrected CoCCoA ALMA 1.20\,mm continuum image (colourscale and blue contours, levels: [5, 10, 20]$\times \sigma_{\rm centre}$, printed in panel), with candidate protomultiple systems and ALMA1, ALMA5 and ALMA7 labelled (see Section~\ref{continuum emission}). The colourscale uses a double colourbar as in Figure~\ref{fig:full_FOV_continuum} but using $\sigma_{\rm centre}$ measured in the primary beam corrected image ($\sigma_{\rm centre}$ = 0.088 mJy beam$^{-1}$, see Section~\ref{sec:Coccoa_obs}). Orange dashed lines mark the fields of view of the zoomed panels (b), (c), and (e-i). (b-i) Zoom views of the CoCCoA ALMA 1.20\,mm continuum image (colourscale and blue contours, levels [5, 10, 20]$\times \sigma$, where $\sigma$ is the local noise measured near each system, printed in each panel). Panel (d) is a zoom view of the region shown with dashed orange lines in (c). Green ellipses show the FWHM extents of the Gaussian fits described in Section~\ref{continuum emission} (see Table~\ref{tab:fitted_source_properties}), with the dotted ellipse in (h) indicating a tentative detection (see Section~\ref{alma1}). Primary beam corrected VLA 3.00\,cm continuum contours from \citet{Yanza2025} are overlaid in panels (b-d) (pink, levels: [5, 15]$\sigma$, where $\sigma$ = 0.83 $\times$ 10$^{-5}$ Jy beam$^{-1}$ and is measured near ALMA2a/b). Positions of OH \citep{Brooks_Whiteoak2001} and Class-II CH$_{3}$OH \citep{Caswell2010} masers are plotted as an indigo plus and cross respectively in panels (c-d); the extent of these symbols reflects the reported positional uncertainties. The ALMA synthesised beam (blue) is shown at lower left in each panel, and the VLA  3.00\,cm synthesised beam (magenta) is shown in panels (b-d).}
    \label{fig:continuum_plot}
\end{figure*}

The 1.20\,mm CoCCoA image of the NGC 6334-43 field is presented in Figure~\ref{fig:full_FOV_continuum}. As shown in Figure~\ref{fig:full_FOV_continuum}, the vast majority of the 1.20\,mm continuum flux density ($\sim$90\%) is concentrated in the south-east portion of the observed FOV (indicated by orange dashed lines in Figure~\ref{fig:full_FOV_continuum}, which outline the FOV of Figure~\ref{fig:continuum_plot}a). The most striking feature in Figure~\ref{fig:continuum_plot}a is a large-scale filamentary structure, which is a single structure at 5$\sigma_{\rm centre}$. The structure has a main component along the north-south axis and two branches to the east. In this paper we focus on the identification and characterization of candidate protomultiple systems associated with this large-scale filamentary structure. We identify candidate protomultiple systems from the 1.20\,mm continuum image as compact sources $\geq$ 8$\sigma$ that are closely spaced on the plane of the sky and share some common structure $\geq$ 5$\sigma$, where $\sigma$ is the local noise near each candidate system (see Section~\ref{individual_sources} and Figure~\ref{fig:continuum_plot}). At a resolution of 0\farcs280 $\times$ 0\farcs248 (373 $\times$ 331 au) we find four such systems. For completeness, we also characterise three additional $\geq$ 8$\sigma$ apparently isolated sources within the south-east portion of the FOV. We show a zoom view of each system and the three additional single sources in Figure~\ref{fig:continuum_plot}. Systems and sources are named in order of decreasing peak intensity and are labelled in Figure~\ref{fig:continuum_plot}.

The north-south component of the large-scale filamentary structure extends $\sim$0.12 pc, from $\sim$0.04 pc south of ALMA4a through ALMA4b and ending with ALMA6 at its tip. The first eastern branch extends east $\sim$0.04 pc through the ALMA2 system and the second eastern branch extends $\sim$0.03 pc east and $\sim$0.03 pc south from ALMA4b through the ALMA3 system. We estimate the width at the 5$\sigma_{\rm centre}$ level of the large-scale filamentary continuum structure as 1500$-$3000 au, where the range reflects measurements at multiple locations along the N-S filament and the two branches. We estimate the flux density within the 5$\sigma_{\rm centre}$ contour of the filamentary structure in the residual image produced after the fitting of compact sources (see Section~\ref{individual_sources}) using CASA's \textsc{imstat} routine \citep{casa2022} as $\sim$ 0.543 $\pm$ 0.006 Jy. The uncertainty is estimated as $\sigma$ $\times$ $\sqrt{N_{\rm beams}}$, where N$_{\rm beams}$ is the number of pixels in the region divided by the number of pixels in the beam \citep{Cantwell2016, Thwala2019}.

We fit each mm continuum source with a single two-dimensional Gaussian using CASA's \textsc{imfit} task \citep{casa2022}; for most sources, we also include a zero-level offset in the fit to account for background emission from the larger-scale structure (see Table~\ref{tab:fitted_source_properties}). Additional details on the fitting of individual sources are given in Section~\ref{individual_sources}. The fitted source properties are presented in Table~\ref{tab:fitted_source_properties} and the FWHM extent of each source is shown in Figure~\ref{fig:continuum_plot}.

In general, a single Gaussian component is sufficient to recover compact sources from their surrounding structure and leave noise-like residuals. The exceptions are ALMA4a, where we use a two-component fit in order to retrieve a compact component from an extended continuum structure that resides within the large-scale filamentary structure, and ALMA7 (see Section~\ref{alma7}).

\begin{table*}
	\centering
	\caption{Properties of 1.20\,mm continuum sources}
	\label{tab:fitted_source_properties}
        \setlength{\tabcolsep}{4pt}
    \begin{tabular*}{\linewidth}{@{\extracolsep{\fill}}cccccccccc} 
        \hline
        & & & \multicolumn{6}{c}{Fitted properties} & \\\cline{4-9}\rule{0pt}{2.5ex}
        Source& \ \ Local $\sigma$ $^{a}$& S/N$^{a}$ & Offset$^{b}$ & \multicolumn{2}{c}{\ \ Position (J2000)$^{b}$} & \ \ I$_{\nu}$$^{b}$ & \ \ S$_{\nu}$$^{b}$& \ \ Size$^{b}$\\
        & (mJy & & (mJy & $\alpha (\textsuperscript{h m s})$ & $\delta ( ^{\circ \ \prime \ \prime\prime}) $ & (mJy & (mJy) & (mas $\times$ mas $\text{[P.A.} (^{\circ})]$)\\
        & beam$^{-1}$) & & beam$^{-1}$) & & & beam$^{-1}$) & &\\
        \hline        
        ALMA1&0.499& 281 &-&17:20:23.60174&$-$35:55:12.6999&139 (1)&266 (2)& 272 $\times$ 228 [150] (3 $\times$ 3 [3]) \\
        \hline
        \ \ ALMA2a$^{c}$&0.195&142&2.17 (0.04)&17:20:23.8439&$-$35:54:56.449&24.2 (0.2)&35.9 (0.5)&211 $\times$ 152 [139] (7 $\times$ 8 [6])\\
        
        \ \ ALMA2b$^{c}$&0.195&125&2.17 (0.04)&17:20:23.8596&$-$35:54:56.873&22.4 (0.2)&29.9 (0.4)&169 $\times$ 139 [90] (7 $\times$ 6 [10])\\
        
        ALMA2c$^{d}$&0.195&33&-&17:20:23.9878&$-$35:54:56.52& - & - &1310 $\times$ 980 [158] (30 $\times$ 20 [3])\\

        \hline
        ALMA3a$^{e}$&0.219&40&1.25 (0.02)&\ \ 17:20:23.867$^{e}$&$-$35:55:01.225$^{e}$&6.3 (0.3) & 16 (1)&450 $\times$ 210 [67] (30 $\times$ 20 [4])\\
    
        ALMA3b&0.219&15&0.58 (0.02)&\ \ 17:20:23.814&$-$35:55:03.17&2.4 (0.2)&22 (2)&1300 $\times$ 420 [157] (100 $\times$ 50 [3])\\
        \hline
        
        \ \ ALMA4a$^{f}$&0.103&51&0.59 (0.01)&17:20:23.5964&$-$35:55:04.607&3.8 (0.1)&17 (1)&650 $\times$ 360 [47] (20 $\times$ 10 [2])\\
        
        ALMA4b&0.103&26&0.61 (0.02)&17:20:23.6014&$-$35:54:59.90&2.0 (0.1)&21 (1)&1500 $\times$ 430 [12] (100 $\times$ 20 [1])\\
        \hline
        ALMA5&0.249&26&0.73 (0.01)&17:20:24.2088&$-$35:54:55.186&5.4 (0.3)&8 (1)&200 $\times$ 140 [60] (40 $\times$ 50 [40])\\
        \hline
        \ \ ALMA6a$^{c}$&0.098&56&0.61 (0.01)&17:20:23.4066 &$-$35:54:53.476 &4.0 (0.1)&14.4 (0.4)&510 $\times$ 350 [152] (20 $\times$ 10 [4])\\
        
        \ \ ALMA6b$^{c}$&0.098&41&0.61 (0.01)&17:20:23.4129 &$-$35:54:54.626&3.0 (1.0)&24 (1)&790 $\times$ 610 [31] (20 $\times$ 40 [5])\\
        \hline
        ALMA7$^{d,f}$ & 0.170 & 9 & - & 17:20:23.225 & $-$35:54:49.7 & - &
        - & 500 $\times$ 400 [80] (300 $\times$ 400 [200])\\
        \hline
        \end{tabular*}
   \begin{tablenotes}
    \small
     \item \bf{Notes} 
     \item{ \textnormal{$^{a}$ $\sigma$ measured locally in an emission-free region at a similar distance from the centre of the CoCCoA FOV to the source. S/N is the peak pixel value divided by the local $\sigma$.}}
     \item{ \textnormal{$^{b}$ From two-dimensional Gaussian fits using CASA's \textsc{imfit} routine. Uncertainties are indicated by the number of significant figures and/or given in parentheses. `Offset' is the zero-level offset and `-' indicates fits in which a zero-level offset was not needed. `Size' is the deconvolved FWHM size.}}
     \item{ \textnormal{$^{c}$ Sources in the same multiple system that were fit simultaneously using a two-component Gaussian fit.}}
     \item{ \textnormal{$^{d}$ Gaussian fitting provides a reasonable representation of source position and size, but not of the flux (see Sections~\ref{alma2} and ~\ref{alma7}) so we do not report I$_{\nu}$ or S$_{\nu}$.}}
     \item{ \textnormal{$^{e}$ Fit using a fixed position, set to the location of the continuum peak, to fit the compact emission.}}
     \item{ \textnormal{$^{f}$ Parameters of the more compact of two components returned by a simultaneous two-component Gaussian fit (see Sections~\ref{alma4} and ~\ref{alma7}).}}
    \end{tablenotes}
\end{table*}

\subsection{Line emission} \label{Line_emission_all}

\subsubsection{CoCCoA line emission} \label{CoCCoA_line_emission}

In this paper, we focus on the molecular line emission associated with the newly-identified compact mm sources. Hot core line emission in the CoCCoA sample has been studied by \cite{Chen2023}, who include ALMA2b in their sample; we do not consider the hot core line emission in detail here. To identify line emission associated with the mm continuum sources (excluding the ALMA2a/b hot cores) we extracted a primary-beam corrected spectrum from the pixel corresponding to each fitted centroid position (see Table~\ref{tab:fitted_source_properties}). For this analysis, we used the convolved (0\farcs330 $\times$ 0\farcs330) CoCCoA line cubes for better sensitivity to weak emission, and we define a detection as two or more adjacent channels with $\geq$4$\sigma$ emission. In evaluating detections, $\sigma$ is measured for each source and for each channel as the rms of a local line-free region. An average $\sigma$ for each line and source is given in Table~\ref{tab:fitted_line_emission_properties}. 

We detect sixteen unique molecular line transitions across ALMA1, 2c, 3a/b, 4a/b, 5, 6a/b and 7 (see Table~\ref{tab:line_emission_properties}). ALMA6a is the most line rich of the sources other than ALMA2a/b, and we do not detect line emission towards ALMA1 (see detections listed by source in Table~\ref{tab:fitted_line_emission_properties}). Many of the detected molecules (HC$^{15}$N, H$^{13}$CO$^+$, C$^{34}$S, H$^{13}$CN and HN$^{13}$C) are common tracers of dense gas \citep{Shirley_2015, Akel2022}. Additionally, HCN can be present in outflows \citep{Walker-Smith2014}, which CH$_{3}$OH, C$^{34}$S and H$_{2}$CS are also known to trace \citep[][respectively]{ZHOU2021, Ortega_2023, Minh_2011}. SO often traces shocks in star forming regions \citep[e.g.][]{Sakai2014, Palau2017,Vangelder2021, Garufi2022}, while CH$_{3}$CN traces hot cores \citep[e.g.][]{Purcell2006, Brouillet2022}, hot corinos \citep[e.g.][]{Cazaux2003, Peaches_Yang_2021}, and outbursting protostars \citep[e.g. V883 Ori,][]{Jeong_2025} and can trace protoplanetary discs \citep[e.g.][]{O_berg_2015, Bergner2018}. Lastly, CH$_{3}$CCH has been observed towards warm cores \cite[e.g.][]{Miettinen2006,Calcutt2019} while CH$_{3}$OCHO is typically abundant towards hotter, more evolved, cores \citep[e.g.][]{Qin2022,Bonfand2024,Mininni2025}.

\begin{table}
	\centering
	\caption{Line emission discussed in this paper}
	\label{tab:line_emission_properties}
    \begin{tabular*}{\columnwidth}{@{\extracolsep{\fill}}ccc} 
        \hline
        Species and line transition & $\nu_{0}$$^{a}$ & E$_{\rm upper}$$^{a}$ \\
        & (GHz) & (K) \\
        \hline        
        \multicolumn{3}{c}{Lower tuning} \\
        \hline
        CH$_{3}$CN J = 13(7)-12(7) & 238.9127154 & 291.0  \\
        CH$_{3}$CN J = 13(4)-12(4) & 239.0642988 & 194.6 \\
        CH$_{3}$CN J = 13(3)-12(3) &  239.0964966 & 144.6  \\
        CH$_{3}$CN J = 13(2)-12(2) & 239.1195044 & 108.9\\
        CH$_{3}$CN J = 13(1)-12(1) & 239.1333129 & 87.5 \\
        CH$_{3}$CN J = 13(0)-12(0) & 239.1379164 & 80.3 \\
        CH$_{3}$CCH J = 14(1)-13(1) & 239.2477272 & 93.3\\
        CH$_{3}$CCH J = 14(0)-13(0) & 239.2522968 & 86.1 \\
        CH$_{3}$OH J = 3-2 5(1)$^{+}$-4(1)$^{+}$ A $v_{t}$=0 & 239.746219 & 49.1 \\
        H$_{2}$CS J = 7(0,7)-6(0,6) & 240.26632 & 46.1\\
        C$^{34}$S J = 5-4& 241.016194 & 34.7\\
        CH$_{3}$OH J = 3-2 5(0)-4(0) E $v_{t}$=0 & 241.700159 & 47.9 \\
        \hline
        \multicolumn{3}{c}{Upper tuning}\\
        \hline
        HC$^{15}$N J = 3-2 & 258.157100 & 24.8 \\
        SO J = 6(6)-5(5) & 258.255813 &  56.5 \\
        H$^{13}$CN J = 3-2 & 259.0118211 & 24.9\\
        CH$_{3}$OCHO J = 24-23 & \ \ 259.343$^{b}$ & 158.2 \\
        H$^{13}$CO$^+$ J = 3-2 & 260.255339 & 25.0\\
        HN$^{13}$C J = 3-2 & 261.2633101& 25.1 \\
        \hline
    \end{tabular*}
    \begin{tablenotes}
    \small
     \item \bf{Notes} 
     \item{ \textnormal{$^{a}$ From the Jet Propulsion Laboratory \citep[JPL:][]{JPL_1998} entries in the Splatalogue spectral line database (\hyperlink{https://splatalogue.online/}{https://splatalogue.online/})}.}
     \item{\textnormal{$^{b}$ Eight unresolved J$_{K_{a}, K_{c}}$ = 24$_{*,24}$-23$_{*,23}$ transitions of methyl formate near this frequency.}}  
    \end{tablenotes}
\end{table}

\begin{table*}
	\centering
	\caption{Parameters of Gaussian fits to CoCCoA line emission}
	\label{tab:fitted_line_emission_properties}
    \begin{tabular}{ccccccc}
        \hline
        & & &  \multicolumn{3}{c}{Gaussian fit parameters} & \\\cline{4-6}
        Source& Line & \ \ Local $\sigma$ $^{a}$ & Amplitude & Centre & Width ($\sigma$) \\
        & & (mJy beam$^{-1}$) & (mJy beam$^{-1}$) & (km s$^{-1}$) & (km s$^{-1}$) \\
        \hline
        \multicolumn{6}{c}{Lower tuning} \\
        \hline
        ALMA2c & CH$_{3}$OH (239.7 GHz) & 4.25 & 43 (6) & 0.5 (0.1) & 0.74 (0.09)\\
        ALMA2c & CH$_{3}$OH (241.7 GHz) & 3.56 & 52 (5) & 0.76 (0.06) & 0.58 (0.05)\\
        ALMA2c$^{b}$ & C$^{34}$S & 4.96 & - & - & - \\
        \hline
        ALMA3a$^{c}$ & CH$_{3}$OH (239.7 GHz) & 4.25 & 33 (4) & 0.1 (0.1) & 1.5 (0.1) \\
        ALMA3a & CH$_{3}$OH (241.7 GHz) & 3.56 & 37 (3) & 0.3 (0.1) & 1.13 (0.09)\\
        ALMA3a & CH$_{3}$CCH ($k$ = 0) & 3.39 & 22 (3) & 0.4 (0.1) & 0.48 (0.05) \\
        ALMA3a$^{b}$ & C$^{34}$S & 4.96 & - & - & -\\
        \hline
        ALMA5$^{d}$ & CH$_{3}$OH (239.7 GHz) & 7.20 & - & - & - \\
        ALMA5$^{d}$ & CH$_{3}$OH (241.7 GHz) & 6.03 & - & - & - \\
        \hline
        ALMA6a & CH$_{3}$OH (239.7 GHz) & 2.44 & 40 (1) & $-$2.11 (0.06) & 1.11 (0.05)\\
        ALMA6a & CH$_{3}$OH (241.7 GHz) & 2.08 & 36 (1) & $-$2.15 (0.04) & 0.90 (0.04)\\
        ALMA6a & CH$_{3}$CCH ($k$ = 0) & 2.03 & 12 (1) & $-$2.3 (0.1) & 0.7 (0.1)\\
        ALMA6a & CH$_{3}$CCH ($k$ = 1) & 2.19 & 10 (1) & $-$2.5 (0.1) & 0.8 (0.1)\\
        ALMA6a$^{b}$ & C$^{34}$S & 3.14 & - & - & -\\
        ALMA6a & H$_{2}$CS & 2.96 & 21 (2) & $-$2.66 (0.08) & 0.75 (0.08)\\
        \hline
        ALMA6b & CH$_{3}$OH (239.7 GHz) & 2.44 & 26 (2) & $-$1.94 (0.05)& 0.60 (0.05)\\
        ALMA6b & CH$_{3}$OH (241.7 GHz) & 2.08 & 17 (2) & $-$1.83 (0.07) & 0.59 (0.07) \\
        \hline
        \multicolumn{6}{c}{Upper tuning} \\
        \hline
        ALMA2c & HC$^{15}$N & 6.83& 34 (5) & 0.95 (0.09) & 0.56 (0.09)\\
        ALMA2c & SO & 6.85& 52 (7) & 0.6 (0.1) & 0.8 (0.1) \\
        \ \ ALMA2c$^{c}$& H$^{13}$CN & 6.22 & 42 (9) & 1.3 (0.1) & 0.38 (0.09) \\
        ALMA2c & H$^{13}$CO$^+$ & 6.83 & 40 (1) & 1.2 (0.1) & 0.4 (0.1) \\
        \hline
        ALMA3a & H$^{13}$CO$^+$ & 6.83 & 59 (4)& $-$0.4 (0.1) & 1.2 (0.1)\\
        \hline
        ALMA3b & H$^{13}$CO$^+$ & 6.83 & 60 (5) & $-$1.37 (0.05) & 0.48 (0.05)\\
        \hline
        \ \ ALMA4b$^{c}$ & H$^{13}$CN& 4.05 & 33 (3) & $-$1.1 (0.1) & 1.1 (0.1)\\
        ALMA4b & H$^{13}$CO$^+$ & 4.46 & 82 (3) & $-$1.16 (0.02) & 0.46 (0.02)\\
        ALMA4b & HN$^{13}$C & 4.46 & 66 (4) & $-$1.37 (0.02) & 0.40 (0.02)\\
        \hline
        \ \ ALMA5$^{c}$ & H$^{13}$CN & 10.3 & 42 (4)& 3.6 (0.2) & 1.5 (0.3)\\
        \hline
        ALMA6a & HC$^{15}$N & 3.56 & 23 (2) & $-$1.90 (0.08) & 0.84 (0.08)\\
        ALMA6a & SO & 3.44 & 18 (2) & $-$2.1 (0.1) & 0.8 (0.1)\\
        ALMA6a & H$^{13}$CN & 3.23 & 50 (2) & $-$2.47 (0.03) & 0.86 (0.04)\\
        ALMA6a & H$^{13}$CO$^+$ & 3.69 & 42 (2) & $-$2.26 (0.03) & 0.61 (0.04)\\
        ALMA6a & HN$^{13}$C & 3.39 &  35 (2) & $-$2.43 (0.05) & 0.65 (0.05)\\
        \hline
        ALMA6b & HN$^{13}$C & 3.39 & 19 (3) & $-$2.58 (0.07) & 0.42 (0.07)\\
        \hline
    \end{tabular}
   \begin{tablenotes}
    \small
     \item \bf{Notes} 
     \item{$^{a}$ \textnormal{Mean rms in a line-free region over 100 channels centred on the rest frequency of the line measured in the primary beam corrected cubes.}}
     \item{$^{b}$ \textnormal{We do not report fitting parameters as the line is significantly affected by missing short-spacing data (see Section~\ref{CoCCoA_line_emission}).}}
     \item{$^{c}$ \textnormal{A restricted channel range was required for \textsc{pyspeckit} to best fit the line feature. The velocity ranges around each rest frequency are as follows; ALMA2c H$^{13}$CN from $-$0.70 km s$^{-1}$ to 3.48 km s$^{-1}$, ALMA4b H$^{13}$CN from $-$3.13 km s$^{-1}$ to 0.70 km s$^{-1}$, ALMA5 H$^{13}$CN from 0.70 km s$^{-1}$ to 6.26 km s$^{-1}$ and ALMA3a CH$_{3}$OH from $-$ 3.39 km s$^{-1}$ to 3.39 km s$^{-1}$.}}
     \item{$^{d}$ \textnormal{Line not fit as it displays a non-Gaussian profile, likely related to an outflow (see Section~\ref{alma5}).}}
    \end{tablenotes}
\end{table*}

\begin{figure*}
	\includegraphics[trim=0 0 0 490, clip, width = \textwidth]{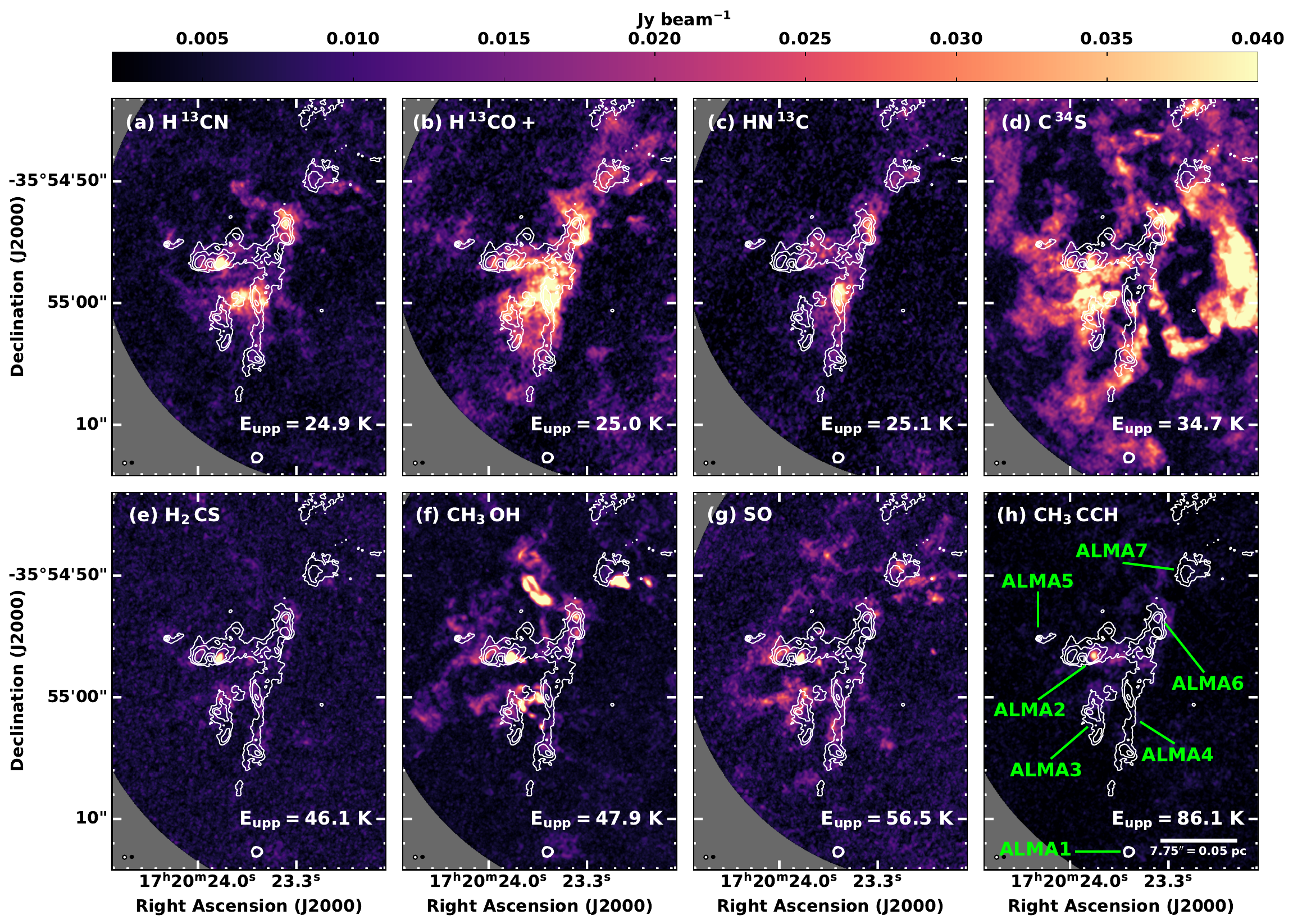}
    \caption{Peak intensity (moment 8) maps for selected lines in Table~\ref{tab:line_emission_properties} that exhibit widespread emission, overlaid with CoCCoA ALMA 1.20\,mm continuum contours (in white, uncorrected for the primary beam response, levels: [5, 10, 20]$\sigma_{\rm centre,nonpb}$, where $\sigma_{\rm centre,nonpb}$ = 0.086 mJy beam$^{-1}$). The FOV matches Figure~\ref{fig:continuum_plot}a. The molecule name is given at upper left in each panel and the panels are ordered by increasing E$_{\rm upper}$, displayed at lower right in each panel. The candidate systems discussed in Section~\ref{individual_sources} are labelled in green on panel (h). The ALMA line and continuum synthesised beams are at bottom left in each panel, in white and black respectively.}
    \label{fig:mom8_FOV_widespread}
\end{figure*}

\begin{figure*}
	\includegraphics[trim=0 0 0 220, clip, width = \textwidth]{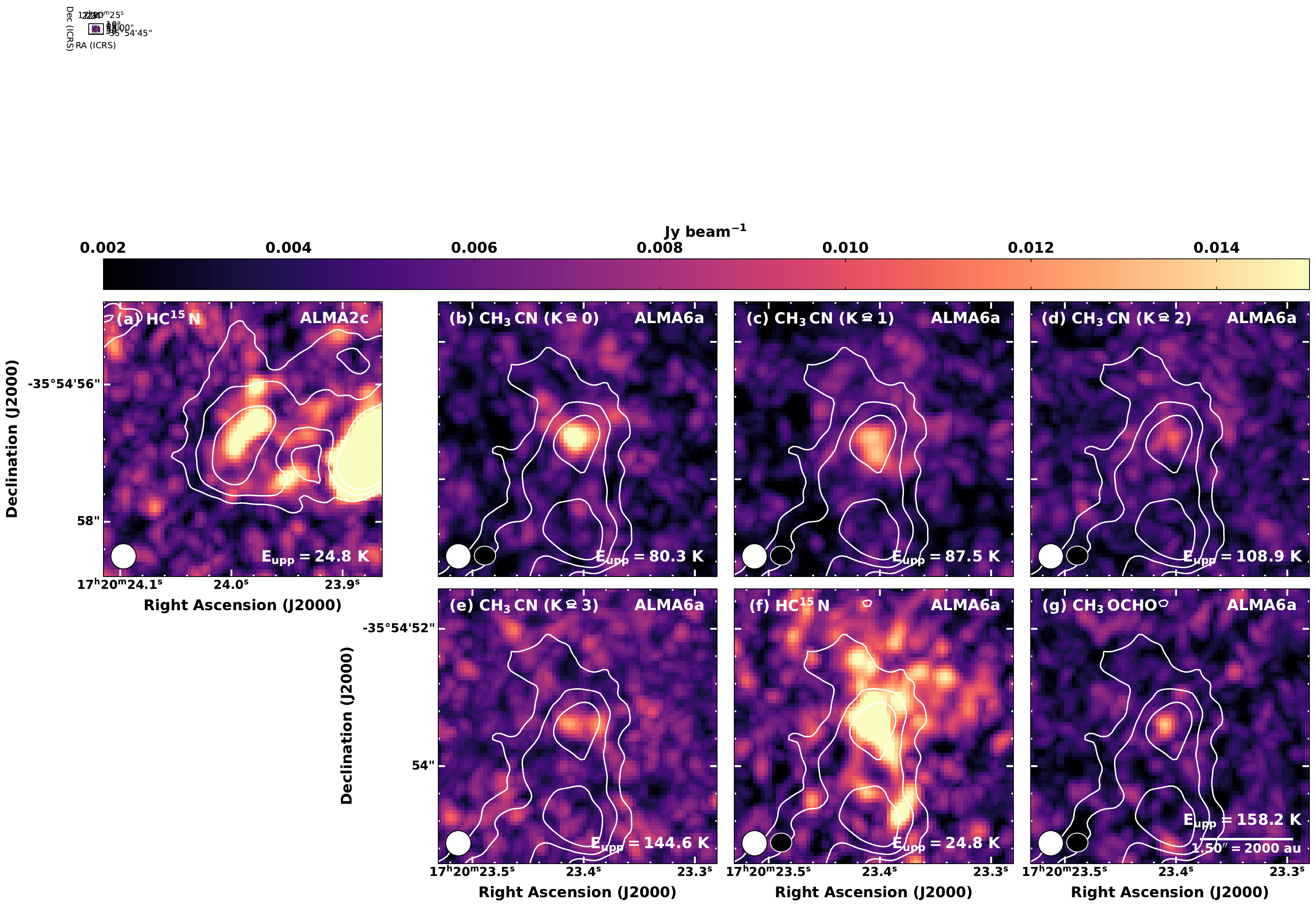}
    \caption{Zoom view of peak intensity (moment 8) maps for all compact lines in Table~\ref{tab:line_emission_properties}, overlaid with CoCCoA ALMA 1.20\,mm continuum contours (in white, uncorrected for the primary beam response, levels: [5, 10, 20]$\sigma_{\rm centre,nonpb}$, where $\sigma_{\rm centre,nonpb}$ = 0.086 mJy beam$^{-1}$). The molecule name and E$_{\rm upper}$ are given at upper left and lower right respectively in each panel. The source name is indicated at upper right. The ALMA line and continuum synthesised beams are at bottom left in each panel, in white and black respectively.}
    \label{fig:mom8_FOV_compact}
\end{figure*}

\begin{figure*}
	\includegraphics[trim =0 0 0 490, clip, width = \textwidth]{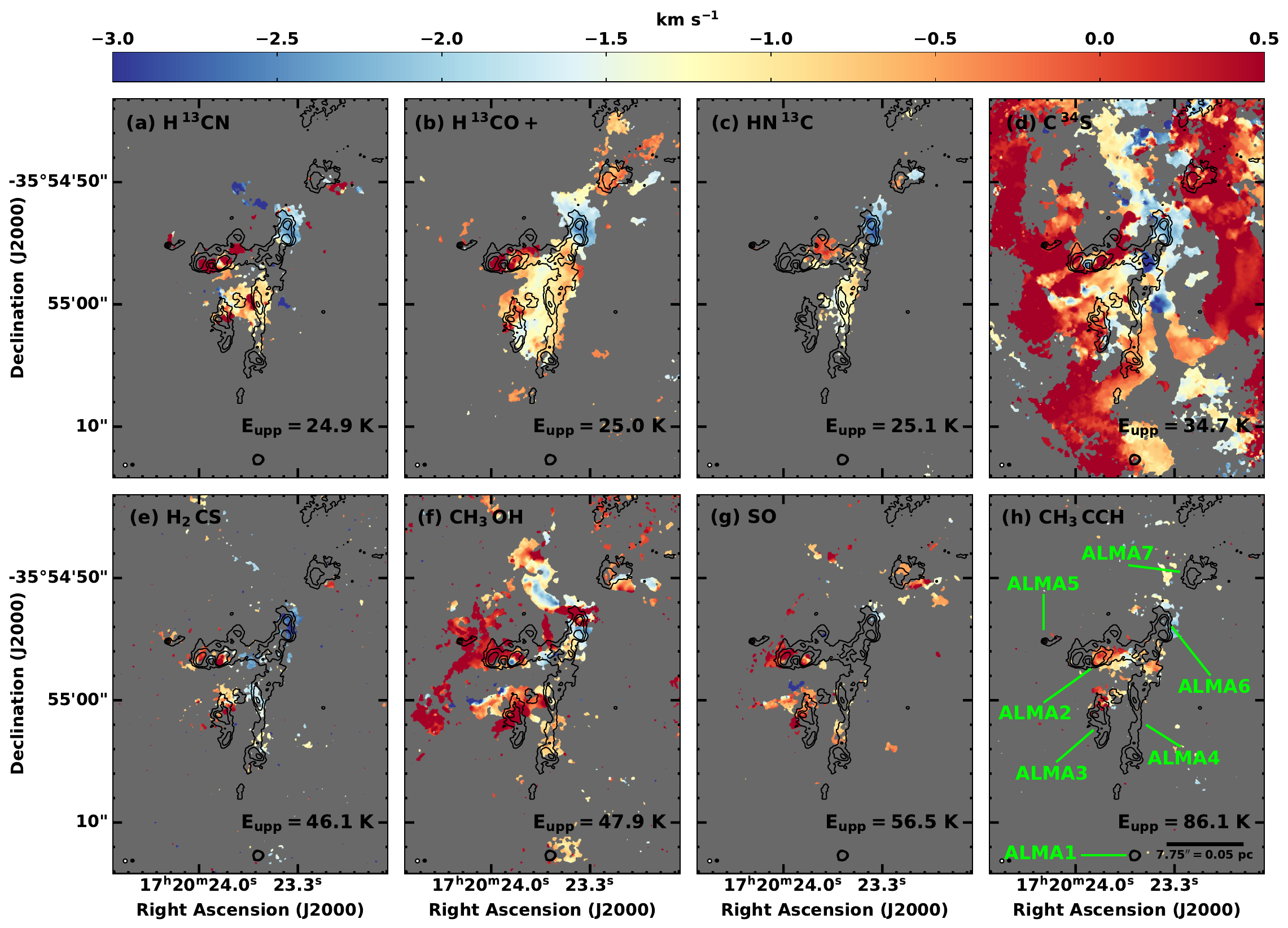}
    \caption{Intensity weighted velocity (moment 1) maps for lines shown in Figure~\ref{fig:mom8_FOV_widespread}, overlaid with CoCCoA ALMA 1.20\,mm continuum contours (in black, uncorrected for the primary beam response, levels: [5, 10, 20]$\sigma_{\rm centre,nonpb}$, where $\sigma_{\rm centre,nonpb}$ = 0.086 mJy beam$^{-1}$). The FOV matches Figure~\ref{fig:continuum_plot}a. Panels(e,f,h) are masked at 4$\sigma$ and all other panels are masked at 3$\sigma$. We use the average $\sigma$ measured towards the ALMA2, ALMA3, ALMA4 and ALMA6 systems (see Table~\ref{tab:fitted_line_emission_properties}); 4.78, 5.34, 5.29, 2.83, 2.70, 2.19, 5.39 and 1.99 mJy beam$^{-1}$ for panels(a-g) respectively. The colourscale is centred on a v$_{\rm LSR}$ = $-$1.25 km s$^{-1}$. The molecule name is given at upper left in each panel and the panels are ordered by increasing E$_{\rm upper}$, displayed at lower right in each panel. The candidate systems discussed in Section~\ref{individual_sources} are labelled in green on panel (h). The ALMA line and continuum synthesised beams are at bottom left in each panel, in white and black respectively.}
    \label{fig:mom1_FOV}
\end{figure*}

Peak intensity (moment 8) maps illustrating the range of morphologies exhibited by these sixteen lines are shown in Figures~\ref{fig:mom8_FOV_widespread} and ~\ref{fig:mom8_FOV_compact}. Lines that show widespread, extended emission are presented in Figure~\ref{fig:mom8_FOV_widespread}, and lines that show more compact emission are shown in Figure~\ref{fig:mom8_FOV_compact}. Figure~\ref{fig:mom1_FOV} presents intensity weighted velocity (moment 1) maps of the lines shown in Figure~\ref{fig:mom8_FOV_widespread}.

In Figure~\ref{fig:mom8_FOV_widespread}, H$^{13}$CO$^+$ emission and HN$^{13}$C emission (although more weakly) both trace the large-scale filamentary continuum structure. Their structures both extend northwards past ALMA6. Figure~\ref{fig:mom1_FOV}b shows that H$^{13}$CO$^+$ traces large-scale velocity gradients along the filamentary continuum structure (see Section~\ref{Large-scale H13CO+}). Interestingly, the spatial distribution of H$^{13}$CN, H$_{2}$CS (albeit much more faintly), CH$_{3}$OH and SO are similar and appear morphologically distinct from the continuum. In particular, we note that an extended structure to the north-east of ALMA6, most strongly seen in CH$_{3}$OH and associated with an outflow (see Section~\ref{CO_results}), is a common feature. A lack of short-spacing data appears to severely limit the apparently extremely widespread C$^{34}$S emission. As such, we do not consider its morphology in our following analysis. Lastly, the CH$_{3}$CCH emission in the field is weak and not closely associated with the continuum. In Section~\ref{individual_sources} we discuss particular line emission features relevant to each source. 

Using \textsc{pyspeckit} \citep{Ginsburg2011, Ginsburg2022} we fit Gaussian profiles to the line detections (apart from the CH$_{3}$CN observed towards ALMA6a, the fitting of which we detail in Section~\ref{CH3CN fitting}) and present the fit results in Table~\ref{tab:fitted_line_emission_properties}. We do not consider hyperfine structure in any cases; the nine H$^{13}$CO$^+$ J = 3-2 hyperfine components span less than half a channel width and three of the six hyperfine components of H$^{13}$CN (F = 4-3, F = 3-2 and F = 2-1) dominate the relative intensity (>96\%) and span only a single channel. We note that the observed CH$_{3}$OCHO towards ALMA6a features eight unresolved J$_{K_{a}, K_{c}}$ = 24$_{*,24}$-23$_{*,23}$ transitions across $\sim$four channels. As such, we do not report Gaussian fit parameters for this feature. Other than the detections of CH$_{3}$OH towards ALMA5, all line detections are well represented by single Gaussians, although in some cases we restrict the fitted velocity range in order to best represent the feature of interest (see notes in Table~\ref{tab:fitted_line_emission_properties}).

We note that in some cases emission is visible towards continuum sources in Figure~\ref{fig:mom8_FOV_widespread} where corresponding line fits are not included in Table~\ref{tab:fitted_line_emission_properties}. This is because inclusion in Table~\ref{tab:fitted_line_emission_properties} is based on the line strength at the fitted continuum position (see Table~\ref{tab:fitted_source_properties}), which does not always coincide with the strongest line emission in the vicinity of a continuum source. 

For sources with detected CH$_{3}$CN emission (ALMA2a,b and ALMA6a) we obtain v$_{\rm LSR}$ measurements from our LTE modelling (see Section~\ref{CH3CN fitting}). For other sources, we use the Gaussian fits to the lines detected towards the mm continuum peak (see Table~\ref{tab:fitted_line_emission_properties}). Where possible, we derive a v$_{\rm LSR}$ from compact line detections (HC$^{15}$N towards ALMA2c, see Figure~\ref{fig:mom8_FOV_compact}a). For sources without compact line emission (ALMA3a, ALMA3b, ALMA4b and ALMA6b) we report an average of their fitted lines centres from Table~\ref{tab:fitted_line_emission_properties}. As these particular values are derived from lines exhibiting widespread emission, we caveat the v$_{\rm LSR}$ as potentially related to the filament rather than the compact sources. 

We do not detect any features $\geq$4$\sigma$ in two adjacent channels towards ALMA4a or ALMA7. For ALMA4a, we obtain an estimate of its v$_{\rm LSR}$ by fitting the marginally detected H$^{13}$CO$^+$ profile (>3$\sigma$ and >4$\sigma$ in adjacent channels) towards the mm continuum peak. The fitting parameters returned are an amplitude of 42 $\pm$ 5 mJy, a centre of $-$1.35 $\pm$ 0.06 km s$^{-1}$ and a width of 0.42 $\pm$ 0.06 km s$^{-1}$. For ALMA7, we fit the marginally detected H$^{13}$CN profile (>3$\sigma$ and >4$\sigma$ in adjacent channels) towards the mm continuum peak. The fitting parameters returned are an amplitude of 18 $\pm$ 2 mJy, a centre of $-$0.8 $\pm$ 0.1 km s$^{-1}$ and a width of 0.60 $\pm$ 0.09 km s$^{-1}$. 

We note that we do not report a v$_{\rm LSR}$ for ALMA5 because we find that the line detections (see Table~\ref{tab:fitted_line_emission_properties}) are most likely related to its outflow (see Section~\ref{alma5}) rather than being associated with the compact continuum source.

\subsubsection{Archival ALMA $^{12}$CO and SiO emission} \label{CO_results}

\begin{figure*}
	\includegraphics[trim=0 0 0 0, clip, scale=0.46]{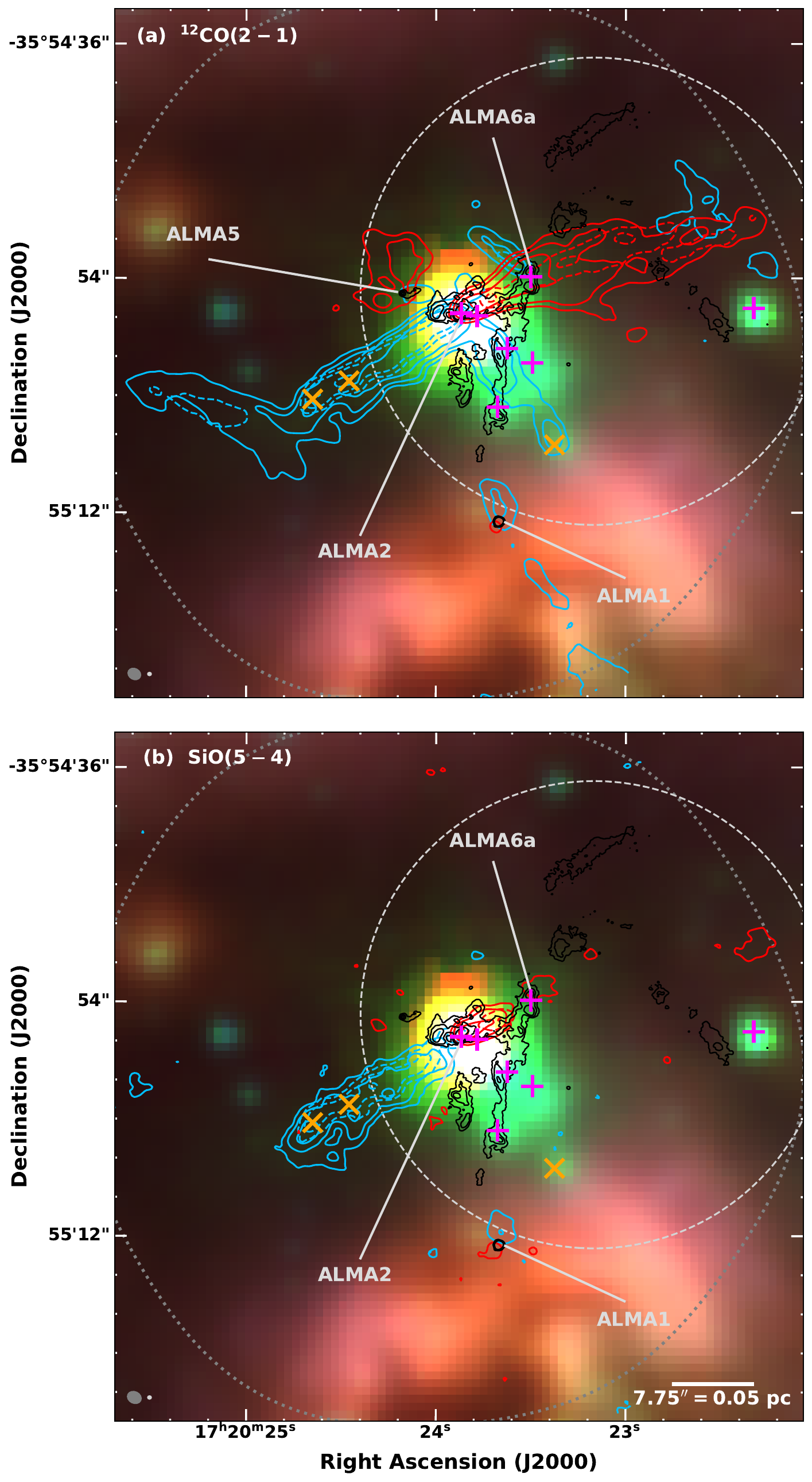}
    \caption{Both panels: Three colour \emph{Spitzer} GLIMPSE \citep{Benjamin2003,Churchwell2009} image (red = 8.0\,$\mu$m, green = 4.5\,$\mu$m and blue = 3.6\,$\mu$m), overlaid with non-primary beam corrected ALMA 1.20\,mm continuum contours (black, levels: [5, 10, 20]$\sigma_{\rm centre,nonpb}$, where $\sigma_{\rm centre,nonpb}$ = 0.086 mJy beam$^{-1}$). We show low and high velocity emission as solid and dashed lines respectively. The 15\% response level of the CoCCoA ALMA primary beam is shown as a light grey dashed circle and the 25\% response level of the archival ALMA mosaic is shown in dotted dark grey. Positions of H$_2$ emission knots and IR sources from \citet{Persi2009} are shown as orange crosses and magenta plusses respectively. Panel (a): Contours of non-primary beam corrected ALMA $^{12}$CO emission with levels of 0.35 Jy beam$^{-1}$ km s$^{-1}$ $\times$ [4, 12, 36] (cyan, red). High velocity blueshifted emission from $-$109.72 km s$^{-1}$ to $-$50.84 km s$^{-1}$, low velocity blueshifted emission from $-$50.20 km s$^{-1}$ to $-$6.04 km s$^{-1}$, low velocity redshifted emission from 6.76 km s$^{-1}$ to 50.28 km s$^{-1}$, and high velocity redshifted emission from 50.92 km s$^{-1}$ to 139.24 km s$^{-1}$. The archival ALMA $^{12}$CO beam (dark grey) is shown alongside the CoCCoA continuum beam (light grey). Panel (b): Contours of non-primary beam corrected ALMA SiO emission with levels of 0.09 Jy beam$^{-1}$ km s$^{-1}$ $\times$ [4, 12, 36] (cyan, red). High velocity blueshifted emission from $-$101.8 km s$^{-1}$ (cut off by the width of the line cube) to $-$49.8 km s$^{-1}$, low velocity blueshifted emission from $-$49.4 km s$^{-1}$ to $-$3.0 km s$^{-1}$, low velocity redshifted emission from 4.2 km s$^{-1}$ to 50.2 km s$^{-1}$, and high velocity redshifted emission from 50.6 km s$^{-1}$ to 70.6 km s$^{-1}$. The archival ALMA SiO beam (dark grey) is shown alongside the CoCCoA continuum beam (light grey).}
    \label{fig:Louvet_outflow_plot}
\end{figure*}

We display integrated intensity maps of the red- and blue-shifted $^{12}$CO and SiO emission from the $\sim$1250 au-resolution archival ALMA observations (see Section~\ref{sec:archive_obs}) in Figure~\ref{fig:Louvet_outflow_plot}. The velocity ranges shown are unique to each tracer as $^{12}$CO traces the large-scale gas structures over a larger velocity interval than SiO. The emission from both tracers is dominated by a large-scale east-west outflow that presents at high velocities (up to $\sim$139 km s$^{-1}$ relative to ALMA2a/b's v$_{\rm lsr}$) and appears to be driven by ALMA2a/b. ALMA1 drives a relatively small bipolar outflow seen in both tracers. Additionally, ALMA5 drives a compact redshifted lobe and ALMA6a drives a single blueshifted outflow lobe, both seen in $^{12}$CO emission. We discuss the association of the $^{12}$CO and SiO outflows shown in Figure~\ref{fig:Louvet_outflow_plot} with mm continuum sources and multiple systems in Section~\ref{individual_sources}. 

Notably, five IR sources from \cite{Persi2009} appear to be associated with extended 4.5\,$\mu$m emission (green in Figure~\ref{fig:Louvet_outflow_plot}). The remaining two IR sources in the field appear more closely associated with ALMA2a/b, where multiband emission is seen. Emission specifically in the 4.5\,$\mu$m band used in the GLIMPSE \citep{Benjamin2003,Churchwell2009} survey distinctly traces outflow features \citep[see][]{Cyganowski_2008} because the band contains H$_{2}$ and CO lines \citep[see Figure 1 in][]{Reach2006} and has a relatively low contribution from polycyclic aromatic hydrocarbon (PAH) emission \citep{Smith2006}. Further, both an H$_{2}$ and Br$\gamma$ line can contribute significantly to the K$_{s}$ band flux \citep[e.g.][]{Varricatt2010, Lee_2012}. Therefore, the morphology seen in Figure~\ref{fig:Louvet_outflow_plot} suggests that line emission associated with outflows may contribute significantly to the K$_{s}$ band emission observed by \cite{Persi2009}. As such, we do not include comment on the relative positions of the mm continuum sources and the IR sources. Figure~\ref{fig:Louvet_outflow_plot} also shows that the H$_{2}$ emission knots from \cite{Persi2009} are coincident with strong blueshifted $^{12}$CO outflow emission. 

\subsection{Notes on individual systems and sources} \label{individual_sources}

\begin{table*}
	\centering
	\caption{Derived source properties}
	\label{tab:derived_source_properties}
        \begin{adjustbox}{width=1\textwidth}
    \begin{tabular}{cccccccccc}
        \hline
        \multirow{2}{*}{Source} & \ \ Projected separation$^{a}$ & Size$^{b}$ &\ \ T$_{\rm b}$$^{c}$& $\alpha_{\rm 1.25-1.15mm}$$^{b}$ &Assumed T$_{\rm dust}$$^{b}$&$\tau_{dust}$$^{b}$& $\text{M}_{\text{gas}}$$^{b}$ & M$_{\rm vir}$$^{b}$&v$_{\rm LSR}$$^{b}$\\
        &(au)&(au $\times$ au) & (K)& &(K)& & $(\textup{M}_\odot)$ & $(\textup{M}_\odot)$& (km s$^{-1}$)\\
        \hline
        ALMA1& - &360 $\times$ 300&84 (3)& 1.5 (0.9) &100$-$200 & 1.84$-$0.55 &2.61$-$0.75&-& - \\
        \hline
        ALMA2a&\multirow{2}{*}{\centering 618 (4)}&280 $\times$ 200&22 (2)& 2.1 (0.7)& 215 & 0.10& 0.08 & 4.5 (0.7) &\ \ 0.75 \\
        
        ALMA2b&&230 $\times$ 180&25 (3)&2.4 (0.6)&244 & 0.11&0.06&5.4 (0.3)& \ \ 0.30\\
        
        \ \ \ ALMA2c$^{d}$&\ \ 2330 (20)$^{e}$&1700 $\times$ 1300&-&4 (1)&20& - &1.87 & -&0.95\\

        \hline
        ALMA3a&\multirow{2}{*}{\centering 2800 (100)}&600 $\times$ 280&3.0 (0.8)&3 (1)&20&0.12&0.50& - &0.1\\
        
        ALMA3b&&1700 $\times$ 600&0.8 (0.3)&4 (2)&20&0.04&0.66&-&$-$1.37\\
        \hline
        ALMA4a&\multirow{2}{*}{\centering 6300 (100)}&870 $\times$ 480&1.5 (0.1)&4 (1)&20&0.08&0.52&-&$-$1.35$^{f}$\\
        
        ALMA4b&&1900 $\times$ 600&0.7 (0.3)&6 (2)&20&0.03&0.63&-&$-$1.2\\
        \hline
        ALMA5& - &260 $\times$ 180&\ \ 5 (6) &2 (1)&20&0.32&0.26 &-& -\\
        \hline
        ALMA6a&\multirow{2}{*}{\centering 1530 (30)}&680 $\times$ 460&1.6 (0.2)&2.6 (1.5)&110&0.01&0.06& 2.6 (1.5) & $-$1.65\\
        
        ALMA6b&&1000 $\times$ 800&0.96 (0.09)&3 (1)&20&0.05&0.71 & -&$-$2.1\\
        \hline
        ALMA7$^{d}$& - &600 $\times$ 600&- & 6 (3) &20&-&0.58 & -&$-$0.8$^{f}$\\
        \hline
    \end{tabular}
        \end{adjustbox}
   \begin{tablenotes}
    \small
     \item \bf{Notes} 
     \item{ \textnormal{$^{a}$ Calculated at d = 1.33 kpc (see Section~\ref{NGC6334-43}) using the centroid positions in Table~\ref{tab:fitted_source_properties} and Astropy's \textsc{SkyCoord} task \citep{Astropy2022}. The quoted uncertainty is the difference between the maximum and minimum projected separations that can be derived within the uncertainties of the fitted centroid positions.}}
     \item{ \textnormal{$^{b}$ See Section~\ref{continuum emission} for detail on Size, Section~\ref{Deriving further source properties} for detail on $\alpha_{\rm 1.25-1.15mm}$, Assumed T$_{\rm dust}$, $\tau_{dust}$, $\text{M}_{\text{gas}}$ and M$_{\rm vir}$ and see Section~\ref{CoCCoA_line_emission} for detail on v$_{\rm LSR}$. The number of significant figures reported for Size reflects the uncertainties in Table~\ref{tab:fitted_source_properties}.}}
     \item{ \textnormal{$^{c}$ Calculated assuming the Rayleigh-Jeans approximation from the integrated flux densities and the FWHM of the sources in Table~\ref{tab:fitted_source_properties}.}}
     \item{ \textnormal{$^{d}$ Gaussian fit parameters from Table~\ref{tab:fitted_source_properties} are used to derive all relevant quantities except M$_{\rm gas}$, where we use a different method to estimate the integrated flux density (see Section~\ref{Deriving further source properties}).}}
     \item{ \textnormal{$^{e}$ Separation relative to ALMA2a.}}
     \item{ \textnormal{$^{f}$ The v$_{\rm LSR}$ of ALMA4a and ALMA7 are derived from marginal detections (see Section~\ref{CoCCoA_line_emission}).}}
    \end{tablenotes}
\end{table*}

In the following subsections, we describe the morphology of each system with reference to the local rms noise of the continuum measured in its vicinity (see Table~\ref{tab:fitted_source_properties}). We also comment upon the CoCCoA line emission morphology, outflows associated with each system and, where relevant, recently-published high-resolution 3.00\,cm (see Figure~\ref{fig:continuum_plot}b-d), 1.36\,cm and 0.91\,cm data presented in \cite{Yanza2025}\footnote{We apply a positional offset correction to the \citet{Yanza2025} images as the position of the calibrator used in their observations, J1744-3116, is reported to a higher precision in the ALMA catalogue than in the VLA catalogue. The resultant correction is $\sim$0\farcs31 $\sim$190$^{\circ}$ east of north.}. The only CoCCoA mm continuum source with a cm-$\lambda$ counterpart is ALMA2b. We report its flux density from \cite{Yanza2025} along with 5$\sigma$ upper limits for other CoCCoA sources, calculated from local measurements of the cm rms, in Table~\ref{tab:cm data table}. Lastly, for sources that display compact line emission we search for indications of Keplerian rotation that could indicate the presence of disc(s). As this search was limited by the resolution of the CoCCoA observations, we present it in Appendix~\ref{Appendix-discs}; at the resolution of our observations, we find no clear evidence for disc-like signatures.

\begin{table}
	\centering
	\caption{Cm continuum properties of CoCCoA mm sources$^{a}$}
	\label{tab:cm data table}
        \setlength{\tabcolsep}{2.8pt}
    \begin{tabular}{c|cc|cc|cc} 
        \hline
        \multirow{2}{*}{Source} & \multicolumn{2}{c|}{3.00\,cm (0\farcs217)} & \multicolumn{2}{c|}{1.36\,cm (0\farcs124)} & \multicolumn{2}{c}{0.91\,cm (0\farcs083)} \\
         & (\textmu Jy)$^{b}$ & $\alpha_{\rm cm-mm}$$^{c}$ & (\textmu Jy)$^{b}$ & $\alpha_{\rm cm-mm}$$^{c}$ & (\textmu Jy)$^{b}$ & $\alpha_{\rm cm-mm}$$^{c}$ \\
        \hline
        ALMA1 & <60 & >2.4 & <236 & >2.0 & ... & ...\\
        ALMA2a & <42 & >2.0 & <176 & >2.0 & ... & ... \\
        ALMA2b & 250 (20) & 1.4 & 230 (40) & 1.9 & ... & ...\\
        ALMA2c & <63 & >1.5 & <198 & >1.5 & ... & ... \\
        ALMA3a & <47 & >1.5 & <161 & >1.5 & ... & ...\\
        ALMA3b & <55 & >1.2 & <184 & >1.1 & ... & ...\\
        ALMA4a & <63 & >1.3 & <161 & >1.3 & ... & ...\\
        ALMA4b & <63 & >1.1 & <167 & >1.0 & ... & ... \\
        ALMA5 & <40 & >1.5 & <163 & >1.4 & ... & ...\\
        ALMA6a & <41 & >1.4 & <166 & >1.3 & <169 & >1.6\\
        ALMA6b & <55 & >1.3 & <180 & >1.2 & <189 & >1.4\\
        ALMA7 & <58 & >1.0 & <154 & >1.0 & <196 & >1.0 \\
        \hline
    \end{tabular}
    \begin{tablenotes}
    \small
     \item \bf{Notes} 
     \item{ \textnormal{$^{a}$ Data from \cite{Yanza2025}, with angular resolutions given in the headings. All of the CoCCoA mm continuum sources lie within the full width at half power (FWHP) primary beam response of the 3.00\,cm image. Whilst the FOV (defined within the 20\% attenuation limit of the primary beam) of the 1.36\,cm image covers all sources, only ALMA6 and ALMA7 lie within the FWHP response of the primary beam. Only ALMA6 and ALMA7 lie in the FOV (but outside the FWHP) of the 0.91\,cm image.}}
     \item{ \textnormal{$^{b}$ We report the integrated flux density and uncertainty (in parentheses) from entry 22 of table A1 of \citet{Yanza2025} for ALMA2b and our measured 5$\sigma$ upper limits for all other sources, calculated from local measurements of the cm rms.}}
    \item{ \textnormal{$^{c}$ Spectral index $\alpha_{\rm cm-mm}$ calculated using the listed cm-$\lambda$ flux density or limit and the 1.20\,mm I$_{\nu}$ from Table~\ref{tab:fitted_source_properties}. For ALMA2c and ALMA7, we use the 1.20\,mm peak intensity, I$_{\nu}$, measured using CASA's \textsc{imstat} task: 6.50 and 1.53 mJy beam$^{-1}$ respectively. For cm nondetections this calculation gives a lower limit to the spectral index.}}
     \end{tablenotes}
\end{table}

\subsubsection{ALMA1} \label{alma1}

Figure~\ref{fig:continuum_plot}h shows ALMA1, a bright and compact source located $\sim$8.1$^{\prime \prime}$ (11,000 au) south of ALMA4a. To our knowledge, ALMA1 has no prior mention in the literature. The residual of the 2D Gaussian fit includes a $\sim$9$\sigma$ residual $\sim$0\farcs3 (400 au) from the fitted centroid of ALMA1, indicating the possible presence of further substructure unresolved by our observations. We note the presence of a faint ($\sim$6$\sigma$) source $\sim$0\farcs8 (1100 au) south-west of ALMA1 (marked as a dotted ellipse in Figure~\ref{fig:continuum_plot}h). However, this is a tentative detection in a high-noise region at the 6\% level of the CoCCoA primary-beam and we do not consider it further in our analysis. No line emission is detected towards ALMA1, which is unsurprising given its position.

As shown in Figure~\ref{fig:Louvet_outflow_plot}, ALMA1 drives a single, relatively compact, bipolar outflow seen in both $^{12}$CO and SiO emission. The directions of the red and blueshifted lobes are consistent between $^{12}$CO and SiO, suggesting the presence of a single driving protostar. SiO emission indicates an active outflow, as gas-phase SiO only persists for $\sim$10$^{4}$ years post-shock, due to either re-accretion onto dust grains or oxidation into SiO$_{2}$, which significantly decreases its abundance \citep{desForets197}. Therefore, ALMA1 is likely actively accreting \citep[see reviews of][]{Frank2014,Pascucci2023}. 

\subsubsection{ALMA2} \label{alma2}

The hot core targeted by the CoCCoA survey, NGC 6334-43, is resolved by ALMA into two components separated by 0\farcs465 (618 au in projected separation), which we designate ALMA2a and ALMA2b (see Figure~\ref{fig:continuum_plot}d). In the 1.20\,mm continuum image ALMA2a and ALMA2b share a significant common structure up to $\sim$58$\sigma$. While ALMA2b is well fit by a single Gaussian component (peak residual <2$\sigma$), a compact $\sim$7$\sigma$ residual at the fitted centroid position of ALMA2a may suggest the presence of additional substructure unresolved by our observations. ALMA2b displays stronger and richer line emission than ALMA2a (see a sample of each spectrum in Figure~\ref{fig:ALMA2ab_WEEDS}).  

A third extended component, ALMA2c (see Figure~\ref{fig:continuum_plot}c), lies 1\farcs76 (2330 au in projected separation) from ALMA2a and shares a weaker common structure with the binary up to $\sim$8$\sigma$. We note that while a Gaussian fit provides a reasonable representation of the shape of ALMA2c (see Figure~\ref{fig:continuum_plot}c), it does not represent the flux well as both an extended positive and negative residual structure remain, peaking at $\sim$6$\sigma$. We thus instead use CASA's \textsc{imstat} task to estimate the flux density of ALMA2c within its 9$\sigma$ continuum contour (where it is distinct from ALMA2a/b) as 64 $\pm$ 4 mJy. ALMA2c is detected in four lines. H$^{13}$CN and H$^{13}$CO$^+$ outline a common structure with ALMA2a/b, whereas ALMA2c appears distinct from ALMA2a/b in SO (Figure~\ref{fig:mom8_FOV_widespread}). Compact HC$^{15}$N emission is also detected towards ALMA2c (Figure~\ref{fig:mom8_FOV_compact}a). 

ALMA2a/b are associated with the only two reported masers in the CoCCoA FOV: an OH maser \citep{Brooks_Whiteoak2001} offset $\sim$0\farcs4 (500 au) to the west of the fitted continuum peak of ALMA2a and a Class II CH$_{3}$OH maser \citep{Caswell2010} $\sim$0\farcs2 (200 au) south-east of ALMA2b. As noted above, ALMA2b is the only CoCCoA mm continuum source with a cm-$\lambda$ counterpart in the \citet{Yanza2025} continuum images (see Figure~\ref{fig:continuum_plot}c,d). \cite{Yanza2025} report $\alpha$ = $-$0.1 $\pm$ 0.6 for this cm source, consistent with optically thin free-free emission \citep{Ignace2004}. 

As noted above (Section~\ref{CO_results}), a large-scale bipolar outflow driven by ALMA2a/b and seen in both $^{12}$CO and SiO is the dominant feature in Figure~\ref{fig:Louvet_outflow_plot}. At the resolution of these archival data, it is unclear whether this outflow is driven by ALMA2a, ALMA2b, or the combined system. In $^{12}$CO the low and high velocity (solid and dashed lines in Figure~\ref{fig:Louvet_outflow_plot}) redshifted emission trace a similar structure, although the high velocity component is more collimated. The blueshifted emission is more complex. 

The high velocity blueshifted $^{12}$CO emission is clearly separated into two distinct structures (Figure~\ref{fig:Louvet_outflow_plot}): a `collimated' component along the main outflow axis and an off-axis component to the far east (hence the `SE' component). Low velocity blueshifted emission follows a similar off-axis direction towards the far south-east, however, as the low velocity blueshifted emission does not split into two separate structures (see Figure~\ref{fig:Louvet_outflow_plot}), we label it as either solely `collimated' or together as `collimated+SE'. A third low velocity blueshifted component extends to the south-west of ALMA2a/b, in the same direction as the 4.5\,$\mu$m emission (green in Figure ~\ref{fig:Louvet_outflow_plot}) and with a peak coincident with an H$_{2}$ emission knot \citep{Persi2009}. This `SW' feature could potentially be a blueshifted counterpart to the redshifted $^{12}$CO emission to the north/north-east of ALMA5 (see Section~\ref{alma5}), and part of a second bipolar outflow driven by the ALMA2a/b system. 

In SiO, the hot core binary's bipolar outflow is well-collimated. The low velocity blueshifted SiO has a peak of emission towards the far south-east end of the outflow axis that is coincident with an H$_{2}$ emission knot \citep{Persi2009} (see far south-east orange cross on Figure~\ref{fig:Louvet_outflow_plot}). The high velocity blueshifted SiO emission closely follows the low velocity component, and also displays strong emission at the location of the H$_{2}$ emission knots. The low velocity redshifted component does not trace the full extent of the outflow, although it is coincident with the far west end of the redshifted $^{12}$CO emission. The high velocity redshifted component only extends $\sim$4\farcs2 (5600 au) west of ALMA2a/b. The blueshifted SW feature is not seen in SiO emission. The detection of the outflow in SiO indicates that ALMA2a/b is likely actively accreting (see discussion in Section~\ref{alma1}). 

The IR source IR-MM3 (IRS 8E) \citep{Persi2009} is coincident with ALMA2a/b. \cite{Persi2009} derive an SED for IR-MM3 (IRS 8E), finding $\alpha_{IR}$ = 3.5 and a bolometric luminosity of 985 $L_{\odot}$ (which scales to 672 $L_{\odot}$ for our adopted maser-derived distance). The scaled luminosity does not change their interpretation of the source as a B3 to B5 ZAMS star. This interpretation would be consistent with free-free emission for the cm source coincident with ALMA2b. However, both our detection of two hot core components and the relation of the IR sources from \cite{Persi2009} to outflows in the field (see Section~\ref{CO_results}) may affect their SED modelling. 

\subsubsection{ALMA3} \label{alma3}

Figure~\ref{fig:continuum_plot}f shows ALMA3a and ALMA3b, which lie 2\farcs08 apart in the north-south direction (2800 au in projected separation) and are linked by continuum emission at the $\sim$5$\sigma$ level. We fix the fitted centre of ALMA3a in order to retain the compact component. However, a condensed $\sim$6$\sigma$ residual remains less than 0\farcs1 (100 au) from the fixed fitted centroid. ALMA3b is well fit by a single Gaussian but we note that the fit is highly elliptical (1\farcs3 $\times$ 0\farcs42 = 1700 $\times$ 560 au). We do not detect any compact molecular line emission associated with the spatial FWHM of either ALMA3a/b; line detections listed in Table~\ref{tab:fitted_line_emission_properties} are in common with nearby widespread emission (see Figure~\ref{fig:mom8_FOV_widespread}).   

\subsubsection{ALMA4} \label{alma4}

Figure~\ref{fig:continuum_plot}i shows ALMA4a and ALMA4b, which lie 4\farcs7 apart (6300 au in projected separation) and share a common $\sim$7$\sigma$ continuum structure. We fit ALMA4a using a zero-level offset and two Gaussian components as the compact source resides within both large-scale filamentary structure and a more local extended structure. The larger of the two Gaussian components is extended along the filament direction and removes emission primarily to the north-east of ALMA4a. The fit for the more compact Gaussian component is reported in Table~\ref{tab:fitted_source_properties} and shown in Figure~\ref{fig:continuum_plot}i. The residual of the two-component fit has a $\sim$7$\sigma$ peak <0\farcs1 (100 au) from the fitted ALMA4a centroid position. ALMA4b is well fit by a single Gaussian but, as with ALMA3b, the fit is highly elliptical (1\farcs4 $\times$ 0\farcs43 = 1900 $\times$ 600 au). We do not detect any compact molecular line emission associated with the FWHM of either ALMA4a/b; line detections listed in Table~\ref{tab:fitted_line_emission_properties} are in common with nearby widespread emission (see Figure~\ref{fig:mom8_FOV_widespread}. The ALMA4 system is coincident with the SW blueshifted component of the ALMA2a/b outflow (see Figure~\ref{fig:Louvet_outflow_plot}) and hence line emission coincident with these continuum sources could potentially be related to the outflow. We note that both ALMA3 and ALMA4 appear as large separation (>2000 au) systems, each with one compact and one extended component. 

\subsubsection{ALMA5} \label{alma5}

ALMA5 (Figure~\ref{fig:continuum_plot}b) is a compact source that does not exhibit obvious multiplicity at the resolution of our observations. It lies $\sim$3\farcs0 (4000 au) from ALMA2c and $\sim$4\farcs6 (6100 au) from ALMA2a. It is well fit by a single Gaussian. ALMA5 drives a compact outflow lobe that is seen in redshifted $^{12}$CO emission but is undetected in SiO (see Figure~\ref{fig:Louvet_outflow_plot}). Inspection of the $^{12}$CO line cube shows that the redshifted emission is stronger to the north than to the east, suggesting the outflow has a north-south orientation. We detect H$^{13}$CN and CH$_{3}$OH towards ALMA5, although the emission in each case appears to be related to the outflow (see also Section~\ref{alma6}); in particular see that H$^{13}$CN and CH$_{3}$OH extend north from ALMA5 in Figure~\ref{fig:mom8_FOV_widespread}, are redshifted in Figure~\ref{fig:mom1_FOV} and that the H$^{13}$CN profile fit towards ALMA5 has a relatively large fitted width (see Table~\ref{tab:fitted_line_emission_properties}). A possible alternative interpretation of the $^{12}$CO morphology (Figure~\ref{fig:Louvet_outflow_plot}a) is that ALMA5 drives an east-west outflow comprised of the eastern redshifted emission and weak, relatively low velocity ($>$-10 km s$^{-1}$), blueshifted emission seen to the west of ALMA5. In this scenario, (at least some) of the northern redshifted emission would be the redshifted counterpart to the blueshifted SW lobe associated with ALMA2a/b (see Section~\ref{alma2}), comprising a second bipolar outflow driven by ALMA2a/b.

While the mm continuum source ALMA5 is undetected in the \citet{Yanza2025} cm images, entry 23 in their table A1 (henceforth A23) is a 3.00\,cm detection that is coincident with ALMA5's $^{12}$CO outflow (lying $\sim$0\farcs8, 1100 au, north of ALMA5). A23 is not detected at 1.36\,cm, which implies $\alpha$$<$$-$0.8, a value consistent with synchrotron emission. Theoretical work \citep[e.g.][]{Padovani_2015, Padovani_2016, Cecere2016} has found that such non-thermal emission could arise at the locations of shocks, where jets have impacted very dense parts of the surrounding medium and accelerated particles \citep{Anglada2018}, a picture supported by the observations presented in \cite{Cerrasco2010}. As examples of synchrotron jet candidates are still not numerous \citep[e.g.][]{Osorio_2017, Hunter2018, Tychoniec2018, Sanna2019}, ALMA5's outflow could be an interesting candidate for further study, most especially in the 3$-$30\,cm range, where synchrotron emission would be stronger. 

\subsubsection{ALMA6} \label{alma6}

Figure~\ref{fig:continuum_plot}g shows ALMA6a and ALMA6b, a 1\farcs15-separation binary (1530 au in projected separation). The two compact components reside within a common continuum structure at the northern tip of the larger-scale north-south filament. ALMA6a and ALMA6b are only reasonably well described by their joint Gaussian fit (see Table~\ref{tab:fitted_source_properties}), which is complicated due to a connective bridge of emission. The residual of the joint fit has a compact $\sim$10$\sigma$ peak $<$0\farcs1 (100 au) from the fitted centroid of ALMA6a and a compact $\sim$8$\sigma$ peak $\sim$0\farcs2 (300 au) from the fitted centroid of ALMA6b (towards the south-west extreme of its FWHM extent). These residuals may indicate the presence of unresolved substructure(s) in both sources. 

In the CoCCoA data, ALMA6a is significantly more line-rich than ALMA6b (see Table~\ref{tab:fitted_line_emission_properties}). H$^{13}$CN and HN$^{13}$C trace a shared structure between ALMA6a/b (see Figure~\ref{fig:mom8_FOV_widespread}), whilst HC$^{15}$N, CH$_{3}$OCHO and CH$_{3}$CN are spatially restricted to ALMA6a (see Figure~\ref{fig:mom8_FOV_compact}). Notably, HN$^{13}$C, H$^{13}$CO$^+$, SO and CH$_{3}$CCH all peak to the west side of ALMA6 and not within the FWHM of either source. Whilst in Figure~\ref{fig:mom8_FOV_widespread} H$^{13}$CN, H$_{2}$CS and CH$_{3}$OH do peak towards ALMA6a, only HC$^{15}$N, CH$_{3}$OCHO and CH$_{3}$CN are both spatially restricted to the compact component of ALMA6a and clearly peak towards its fitted continuum centroid position (see Figure~\ref{fig:mom8_FOV_compact}). 

Interestingly, the ALMA6a/b system is associated with two narrow features in the CoCCoA continuum image: a bent arm extending $\sim$1\farcs5 (2000 au) east and $\sim$2\farcs0 (2700 au) south from the fitted centroid of ALMA6b, and a weaker counterpart extending $\sim$1\farcs2 (1600 au) east and $\sim$1\farcs3 (1700 au) north from the fitted centroid of ALMA6a (see Figure~\ref{fig:continuum_plot}g). The southern extension is distinct from the large-scale filamentary emission at the $\sim$12$\sigma$ level. The northern extension is weaker, peaking at $\sim$8$\sigma$. We estimate the integrated flux densities (with the same method used for ALMA2c in Section~\ref{alma2}) of these features as 24 $\pm$ 1 mJy (within the 12$\sigma$ contour of the southern extension) and 6.3 $\pm$ 0.3 mJy (within the 5$\sigma$ contour of the northern extension). The opposition of the structures and the bend in the southern extension are reminiscent of spiral arms \citep[see e.g.][]{Sanna2021, Cacciapuoti2023}. Notably, H$^{13}$CO$^+$ emission is present along the spiral-arm-like structure to the south of ALMA6 and shows a clear velocity gradient along its extent, from roughly $-$2.0 to $-$0.5 km s$^{-1}$ (Figure~\ref{fig:mom1_FOV}b). Velocity gradients along elongated structures have been interpreted as streamers \citep{pineda2023,Tobin2024}, recognised as important channels that connect cloud to disc scales and feed the growth of YSOs. 

ALMA6a drives a blue-shifted outflow lobe, traced by $^{12}$CO, that extends $\sim$5\farcs5 (7300 au) northeast from the continuum source (Figure~\ref{fig:Louvet_outflow_plot}, Section~\ref{CO_results}). As shown in Figure~\ref{fig:mom8_FOV_widespread}, CH$_{3}$OH \citep[which commonly traces outflows, see][]{Arce2007} and H$^{13}$CN also trace this outflow, exhibiting blueshifted emission northeast of ALMA6a (Figure~\ref{fig:mom1_FOV}). Interestingly, HCN is known to trace the most energetic and youngest (Class 0) outflows \citep[e.g.][]{Jorgensen_2004, Tafalla2013, Walker-Smith2014}. 

A notable feature of Figures~\ref{fig:mom8_FOV_widespread} and ~\ref{fig:Louvet_outflow_plot} is a `bead' of emission at the far north-east extent of the outflow lobe, seen in H$^{13}$CN, SO and SiO. Strong molecular emission towards the end of an outflow lobe has been previously observed \citep[e.g.][]{Arce2008_ADD,Tafalla2013, Gomez-Ruiz_2013, Tychoniec2019}, and can be interpreted as due to shocks produced where the outflow interacts with ambient material. This scenario is supported by the broader fitted linewidth of H$^{13}$CN at the location of the `bead'; $\sim$12 km s$^{-1}$ compared to only 0.86 km s$^{-1}$ toward ALMA6a (Table~\ref{tab:fitted_line_emission_properties}). Also of interest in Figure~\ref{fig:Louvet_outflow_plot} is compact $^{12}$CO emission $\sim$2$^{\prime \prime}$ north of the `bead's' location. Whilst we do not consider this feature as part of the ALMA6a outflow, we note that the position is coincident with C$^{34}$S, SO and CH$_{3}$OH emission (see Figure~\ref{fig:mom8_FOV_widespread}), and may indicate the location of a second shock. 

\subsubsection{ALMA7} \label{alma7}

Figure~\ref{fig:continuum_plot}e shows ALMA7, a comparatively weak source located $\sim$4.5" (6000 au) north-west from ALMA6a. ALMA7 is not part of the large-scale filamentary continuum structure, which breaks between ALMA6 and ALMA7 (see Figure~\ref{fig:continuum_plot}a), but does appear to reside within the same large-scale H$^{13}$CO$^{+}$ structure as the ALMA2, ALMA3, ALMA4, and ALMA6 systems.

As for ALMA4a (Section~\ref{alma4}), we fit ALMA7 with two Gaussian components to represent compact emission within a local extended structure.  The smaller fitted component, shown in Figure~\ref{fig:continuum_plot}e, provides a reasonable representation of the position and size of the compact emission from ALMA7, but does not represent its flux well as the emission from the local extended structure is overrepresented by its Gaussian fit. Instead, we estimate the integrated flux density of ALMA7  within its 5$\sigma$ continuum contour, using the same method as for ALMA2c in Section~\ref{alma2}, as 20 $\pm$ 2 mJy.

Weak H$^{13}$CN, HN$^{13}$C, and SO emission are present towards the extended continuum structure associated with ALMA7 (Figure~\ref{fig:mom8_FOV_widespread}).  However, this continuum structure is partially coincident with the redshifted lobe of the ALMA2 outflow, and extended line emission in the vicinity of ALMA7 could be related to the outflow.  No compact line emission is associated with ALMA7.

\section{Analysis}
\label{sec:analysis}

\subsection{Large-scale H$^{13}$CO$^{+}$ emission} \label{Large-scale H13CO+}

As noted in Section~\ref{CoCCoA_line_emission}, widespread H$^{13}$CO$^+$ emission traces the large-scale filamentary structure seen in the mm continuum emission. 
H$^{13}$CO$^+$ is widely used to investigate gas kinematics \citep[e.g.][]{Csengeri2011,Serra2014,Dhabal_2018,Punanova2018,Zhou_2022, Li_2022, Xu_2023}. To investigate the kinematics of the observed H$^{13}$CO$^+$ emission we use the Python package \textsc{scousepy} \citep{Henshaw2016, Henshaw19}, which enables the fitting of Gaussian profiles to large numbers of pixels. Kinematically complex regions may be disentangled as many Gaussian components can be fit to a single profile. The fitted line centroids build a detailed position-position-velocity (ppv) space. We summarise our use of \textsc{scousepy} in four distinct steps and refer the reader to the methodology presented in \cite{Henshaw2016} for a complete description of the code. 

The first step is the data input and selection of automated fitting parameters. We use the non-primary beam corrected data so that we include emission across the whole CoCCoA FOV. We restrict our spatial extent to panel (a) of Figure~\ref{fig:continuum_plot}, and our spectral extent to include the complete line profile for each pixel within that spatial area, between $-$3.9 km s$^{-1}$ and 4.6 km s$^{-1}$ around the rest frequency of the H$^{13}$CO$^+$ line. We also apply a mask of 3$\sigma_{\rm nonpb}$, $\sim$16 mJy beam$^{-1}$, where $\sigma_{\rm nonpb}$ is measured in the non primary-beam corrected cube. In our data, the observed H$^{13}$CO$^+$ line profile is not highly kinematically complex and we conservatively set the size of each Spectral Averaging Area (SAA) to 15 pixels ($\sim$1000 au), resulting in 494 SAAs. 

In the second step we fit Gaussian profiles to the mean spectrum of each SAA, the parameters of which will be used as initial guesses for the automated fitting described in step three. \textsc{scousepy} uses derivative spectroscopy to undertake user-assisted Gaussian fitting of each SAA. The user controls a signal to noise ratio (SNR, where the noise is the rms, $\sigma$, measured individually for each spectrum by \textsc{scousepy}) and an $\alpha$ parameter (where $\alpha$ sets the size of a kernel that smooths the spectrum) to achieve the desired fit. By varying the SNR between 3$-$5, $\alpha$ between 1$-$3 and selecting the fit that minimises the reduced chi-squared, $\chi^{2}_{r}$, value, we find satisfactory fits for 488 of the 494 SAAs. For the six remaining SAAs large variation of the parameters was required to achieve satisfactory fits. 

In the third step automated fitting is completed. \textsc{scousepy} uses the component fits of the SAAs to undertake Gaussian decomposition of our 32,997 individual spectra. The SAAs overlap to Nyquist sample the mapped area and each pixel may have multiple possible unique fits. By considering the most significant possible solutions, and minimising the Akaike Information Criterion (AIC), a best fit is found. We find that 31\% of our data is fit with two velocity components, while the rest is fit with 1 component. The fourth step allows for quality control, which we do not make use of as the data are already satisfactorily fit. 

We carry out further analysis of the H$^{13}$CO$^+$ emission, using the Agglomerative Clustering for ORganising Nested Structures (\textsc{acorns}) package \citep{Henshaw19} to find coherent ppv structures. Before inputting the ppv data returned by \textsc{scousepy} we clean the 32,997 fits. We discard pixels for which no model was fit and fits where the uncertainty on the velocity dispersion or amplitude is greater than their fitted values. We also perform a 5$\sigma$ amplitude cut. A total of 2920 fits, $\sim$9\% of the data, are removed by the cleaning process. We input the remaining ppv data, 30,077 fits, to \textsc{acorns}. 

\begin{figure}
	\includegraphics[trim=0 0 1200 0, clip, width=\columnwidth]{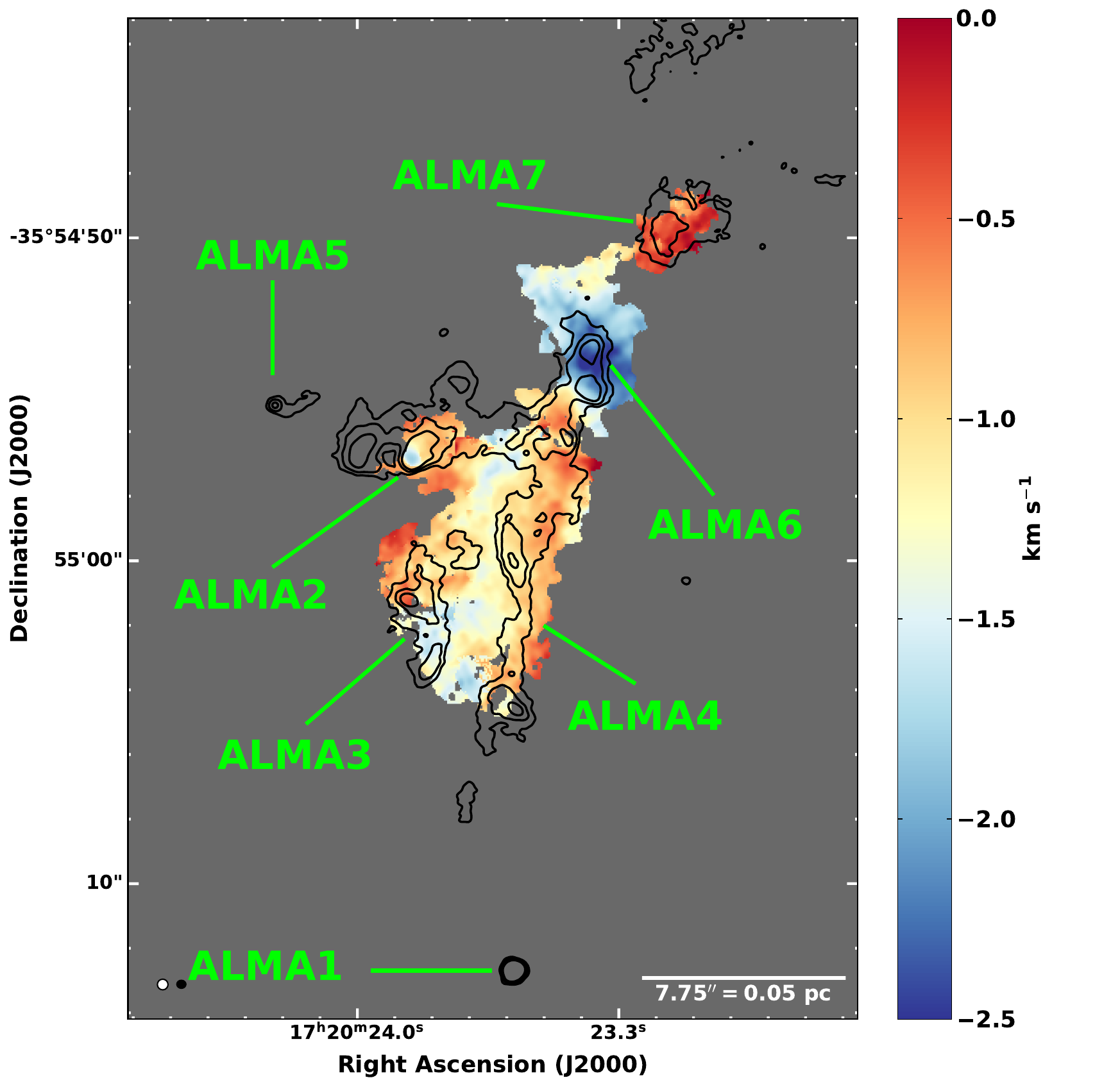}
    \caption{H$^{13}$CO$^+$ centroid velocity map of the largest ppv coherent structure returned by \textsc{Acorns} from the \textsc{scousepy} output. Contours of the CoCCoA ALMA 1.20\,mm continuum image, prior to correction for the primary beam response, are overlaid in black (levels: [5, 10, 20]$\sigma_{\rm centre,nonpb}$, where $\sigma_{\rm centre,nonpb}$ = 0.086 mJy beam$^{-1}$). The colourscale is centred on v$_{\rm LSR}=-$1.25 km s$^{-1}$. The ALMA line and continuum synthesised beams are at bottom left, in white and black respectively.}
    \label{fig:H13CO+scousepy}
\end{figure}

We refer to Appendix B of \cite{Henshaw19} for detail on the clustering carried out by \textsc{acorns} and simply discuss our choice of user parameters. We set the minimum size of a cluster to 49 pixels ($\sim$ the solid angle of the convolved beam), the minimum height above the merge level to 3$\times$ the average rms of the spectra measured by \textsc{scousepy} and the maximum velocity difference between structures to the channel spacing of our dataset, 0.348 km s$^{-1}$. With these parameters, \textsc{acorns} includes $\sim$99\% of the cleaned fits (29,785 of the initial 30,077) and returns 16 distinct structures, only 3 of which have substructure. Notably, a single coherent structure, shown in Figure~\ref{fig:H13CO+scousepy}, contains $\sim$78\% of the clustered data (23,155 components compared to the 6692 assigned to all other structures). We examined the effect that varying the user parameters has on this single structure. Varied combinations of a 50\% relaxation on the minimum radius of structures, a 50\% relaxation on the maximum velocity difference between structures, and the inclusion of a maximum difference in line width parameter (set as 0.348 km s$^{-1}$) find remarkably little change in the number of components included or in the detail of the structure. 

The ppv coherent structure shown in Figure~\ref{fig:H13CO+scousepy} generally traces the large-scale filamentary continuum structure well. Notably, this ppv structure includes the ALMA2a/2b, ALMA3a/3b, ALMA4a/4b, ALMA6a/6b and ALMA7 sources. The average centroid fit of the structure is $-$1.14 km s$^{-1}$. Masking the structure by the 5$\sigma_{\rm centre,nonpb}$ contour of the continuum yields a similar average value, $-$1.05 km s$^{-1}$. These values agree well with the range of estimates for the v$_{\rm LSR}$ of the region in the literature ($-$0.42, $-$0.82 and $-$1.14 km s$^{-1}$, see Section~\ref{introduction}). ALMA2c resides within a different, generally more redshifted (average velocity of 0.07 km s$^{-1}$), \textsc{acorns} structure (not shown). 

We note the presence of a velocity gradient perpendicular to the main north-south filament axis across the southern end of the structure (i.e. below ALMA6); in Figure~\ref{fig:H13CO+scousepy} a central low-velocity ridge of emission generally increases in velocity towards the east (ALMA3a) and west (ALMA4). Velocity gradients perpendicular to the main filament axis have been interpreted as due to accretion onto the filament \cite[e.g.][]{Palmeirim2013, Mundy2020}. We roughly estimate the east-west velocity gradient contained within the 5$\sigma_{\rm centre,nonpb}$ continuum contour shown in Figure~\ref{fig:H13CO+scousepy} at four locations between ALMA4a and ALMA6 using simple position-velocity (pv) cuts, finding between 40$-$90 km s$^{-1}$ pc$^{-1}$. This is somewhat larger than velocity gradients found across filaments of similar width in low-mass star forming regions \citep[e.g.][]{Fern_ndez_L_pez_2014, Dhabal_2018, Chen_2024}. 

In Figure~\ref{fig:H13CO+scousepy} there is also a velocity gradient parallel to the main filament axis: redshifted emission extends to the north and south of the relatively blueshifted emission surrounding ALMA6. Similar velocity gradients have been shown to indicate the presence of an accretion flow \citep[e.g.][]{Zernickel2013,Zhang2015,Chen_2019, Zhou_2022,Li_2022, Wells2024, Rawat2024}, in this case either onto or away from ALMA6. Interestingly, a filamentary accretion flow has recently been observed towards the young high-mass protobinary G11.92$-$0.62 MM2 \citep{Zhang_suinan_2024, sanhueza2025}. 

\subsection{\texorpdfstring{Synthetic CH$_{3}$CN spectra}{Synthetic CH3CN spectra}} \label{CH3CN fitting}

For the derivation of physical gas parameters towards ALMA2a, ALMA2b and ALMA6a, we fit the $k$-ladder of the CH$_{3}$CN J = 13-12 line. To do this, we use the \textsc{weeds} extension of the \textsc{class} software \citep{Maret2011}, which solves the radiative transfer equation assuming a state of local thermodynamic equilibrium (LTE). Specifically, we use the \textsc{weedspymcmc} wrapper\footnote{\hyperlink{https://github.com/gwen-williams/WeedsPy_MCMC}{https://github.com/gwen-williams/WeedsPy\_MCMC}} \citep{Williams2023} to the \textsc{weeds} extension, which through an MCMC algorithm allows the automated exploration of the posterior distributions of the free parameters of the model. The free parameters of the model are the column density (N(CH$_{3}$CN)), the excitation temperature (T$_{\rm ex}$), the centroid velocity (v$_{\rm lsr}$), the velocity width (FWHM), and the source size. We refer the reader to \cite{Maret2011} and \cite{Williams2023} for further details.

Using an image cube produced to cover the full CH$_{3}$CN J = 13-12 ladder (see Section~\ref{sec:Coccoa_obs}) and converted to brightness temperature,\footnote{The image cube and continuum image were converted to brightness temperature using the \textsc{tt.brightnessImage} function of \href{https://safe.nrao.edu/wiki/bin/view/Main/ToddTools}{\textsc{toddtools}} and the Rayleigh–Jeans approximation.} we extract spectra from the pixels corresponding to the fitted centroid positions of ALMA2a, ALMA2b, and ALMA6a (see Table~\ref{tab:fitted_source_properties}). Following \cite{Williams2023}, we estimate the background continuum at these positions using the lower tuning (1.25\,mm) continuum image converted to brightness temperature. For our data, we calculate the projected diameter of the ALMA array, used to calculate the beam dilution/filling factor \citep[see][] {Maret2011,Williams2023}, as 1232 m\footnote{The maximum projected value retrieved using the \texttt{au.getBaselineLengths} function of the \textsc{analysisUtils} package.}.

We perform initial tests to find appropriate parameter ranges for the free parameters. This is especially important for column density, as N(CH$_{3}$CN) can vary by many orders of magnitude towards hot core sources \citep[e.g.][]{Hernandez2014, Gieser2021}. The best-fit solutions for ALMA2a and 2b are found at larger N(CH$_{3}$CN) than for ALMA6a, which is unsurprising as ALMA2a and 2b have stronger CH$_{3}$CN emission (see Figures~\ref{fig:ALMA2ab_WEEDS} and ~\ref{fig:ALMA6a_WEEDS}). We consider a wide range of priors for each of the free parameters: a column density (N(CH$_{3}$CN)) of 10$^{12}$-10$^{19}$ cm$^{-2}$ towards ALMA2a and 2b and 10$^{12}$-10$^{16}$ cm$^{-2}$ for ALMA6a, T$_{\rm ex}$ of 50-800 K, a v$_{\rm lsr}$ of $-$3 to +3 km s$^{-1}$, a FWHM of 2-8 km s$^{-1}$, and a source size between 0$^{\prime\prime}$ and 2$\times$ the convolved beam. We use 50 walkers, 300 burn iterations and 800 production iterations.

\begin{figure}
	\includegraphics[trim =0 0 280 0, clip, width = \columnwidth]{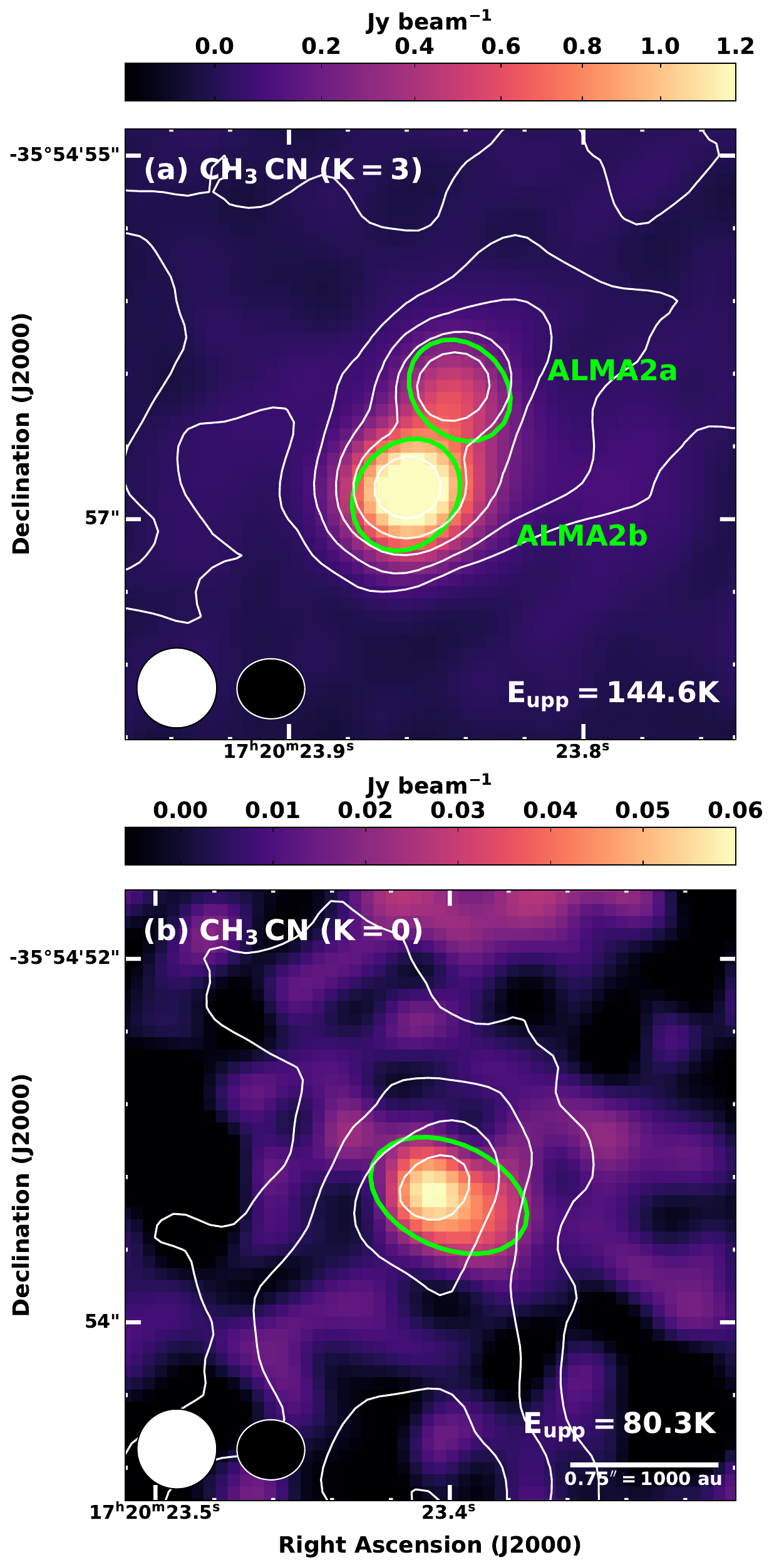}
    \caption{(a): Integrated intensity (moment 0) map of CH$_{3}$CN $k$=3 towards ALMA2a/b (colourscale: power law stretch with exponent = 0.8) overlaid with CoCCoA ALMA 1.20\,mm continuum contours (white, corrected for the primary beam response, levels: [5, 10, 20, 40, 80]$\sigma$, where $\sigma$ = 0.195 mJy beam$^{-1}$). (b): Integrated intensity (moment 0) map of CH$_{3}$CN $k$=0 towards ALMA6a overlaid with CoCCoA ALMA 1.20\,mm continuum contours (white, corrected for the primary beam response, levels: [5, 10, 20]$\sigma$, where $\sigma$ = 0.098 mJy beam$^{-1}$). Both panels: Green ellipses show the FWHM extents of the Gaussian fits described in Section~\ref{CH3CN fitting}. The molecule name and E$_{\rm upper}$ are given at upper left and lower right respectively in each panel. The ALMA line and continuum synthesised beams are at bottom left in each panel, in white and black respectively.}
    \label{fig:mom0_FITS_CH3CN}
\end{figure}

\begin{figure*}
	\includegraphics[trim=0 0 0 0, clip, width = \textwidth]{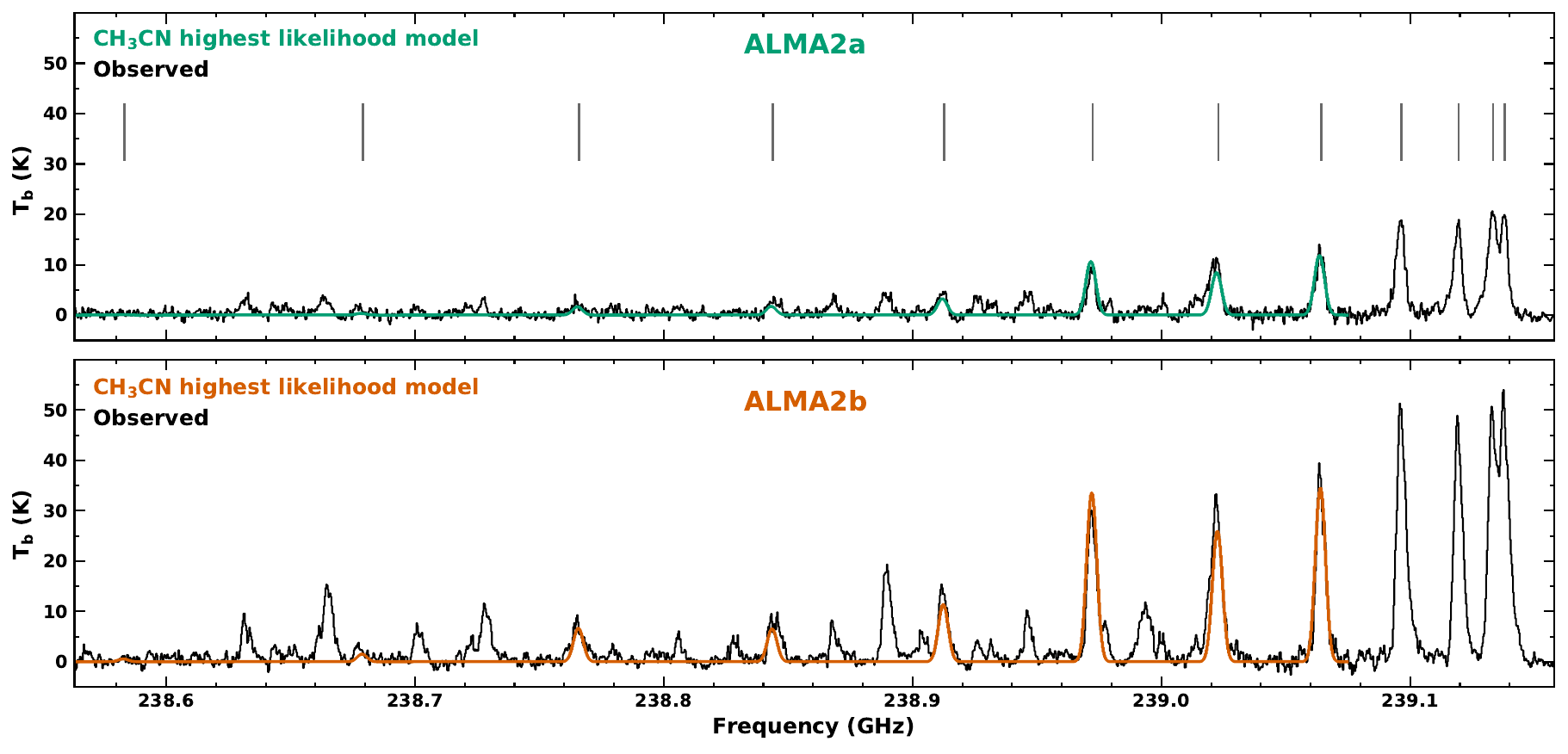}
    \caption{Observed spectrum (black) towards the fitted continuum position of ALMA2a (top) and ALMA2b (bottom), overplotted with the highest likelihood model obtained in Section~\ref{CH3CN fitting} (fitting only k$>$3 components) for each source. The rest frequencies of the $k$ = 0 to $k$ = 11 components of CH$_{3}$CN J = 13-12 are marked with vertical grey lines.}
    \label{fig:ALMA2ab_WEEDS}
\end{figure*}

\begin{figure*}
	\includegraphics[trim=0 0 0 0, clip, width = \textwidth]{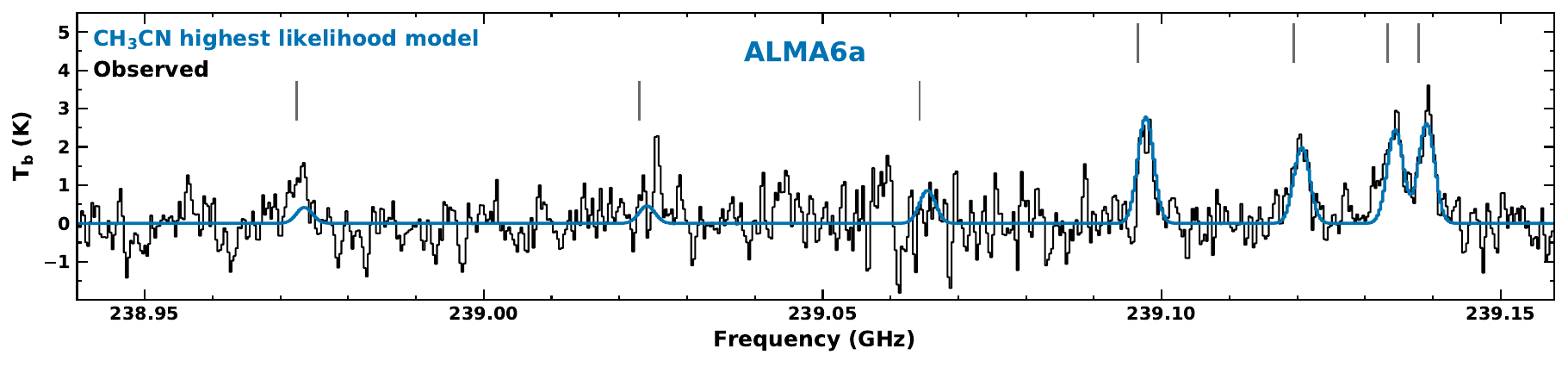}
    \caption{Observed spectrum (black) towards the fitted continuum position of ALMA6a, overplotted with the highest likelihood model obtained in Section~\ref{CH3CN fitting}. The rest frequencies of the $k$ = 0 to $k$ = 6 components of CH$_{3}$CN J = 13-12 are marked with vertical grey lines.
    }
    \label{fig:ALMA6a_WEEDS}
\end{figure*}

For all sources we find that the fitting is poorly constrained when we set the source size as a free parameter. We therefore tested two further cases of setting the source size as a fixed parameter: fixing to the size of the convolved beam or to the geometric mean of the FWHM of a 2D Gaussian fit to the integrated intensity (moment 0) map of a single $k$ transition for each source. We use the $k$=3 transition for ALMA2a and 2b as it is the strongest isolated line and find sizes of (0.24 $\pm$ 0.03)$^{\prime\prime}$ and (0.30 $\pm$ 0.01)$^{\prime\prime}$ respectively. The relevant fits are shown in Figure~\ref{fig:mom0_FITS_CH3CN}. Both values are notably larger than the sizes obtained from the 2D Gaussian fit to the 1.20\,mm continuum emission ($\sim$0\farcs179 and $\sim$0\farcs153 respectively, see Table~\ref{tab:fitted_source_properties}), a trend seen towards discs around high-mass protostars \citep[e.g.][]{Ilee2016} and towards protoplanetary discs \citep[e.g.][]{Ansdell2018, Long2022}. For ALMA6a we use the $k$=0 transition as it is the only transition for which we can deconvolve the fitted size from the beam. This returns (0.40 $\pm$ 0.17)$^{\prime\prime}$, which is consistent with the size of ALMA6a obtained from the 2D Gaussian fit to the 1.20\,mm continuum emission ($\sim$0\farcs42 in Table~\ref{tab:fitted_source_properties}).

Multiple approaches to LTE fitting of CH$_{3}$CN have been taken in the literature. Generally, difficulty is found when a single optical depth value does not appear to apply to all $k$ components. For example, \cite{Feng2015} apply an iterative optical depth correction to satisfactorily fit all $k$ components together, a method that recovers lower best-fit temperatures. \cite{Ahmadi2018} find that restricting their fit to only components above $k$ = 3 and using a fixed source size yields a better fit to the high $k$ components and returns a lower best-fit temperature. \cite{Ahmadi2019} use the same approach but find better results when setting the source size as a free parameter. In order to fit all $k$ components simultaneously, other work has used both an extended cool component and a compact hot component \citep[e.g.][]{Cyganowski_2011, Dedes2011, Hernandez2014}. In the case of ALMA6a, we fit from $k$=0 given that the $k$ > 3 components are weak or undetected above the noise (see Figure~\ref{fig:ALMA6a_WEEDS}). However, for ALMA2a and 2b, where high $k$ components are observed, we tested fitting only the optically thin $k$ components (i.e. those above $k$ = 3) as the low and high $k$ components may arise from multiple gas components with different excitation temperatures, T$_{\rm ex}$ \citep[see][]{Ahmadi2018}.

For ALMA2b there is an independent temperature estimate of between 100 $\pm$ 50 K and 200 $\pm$ 50 K from \cite{Chen2023} modelling of O-bearing COMs. We only obtain a T$_{\rm ex}$ for ALMA2b consistent with the range of values obtained by \cite{Chen2023} when restricting the fit to the $k$ > 3 components. Further, the runs that restrict the fitting to the $k$ > 3 components better reproduce the amplitudes of these components. Fitting all $k$ components towards ALMA2a and 2b results in synthetic spectra that underestimate the amplitudes of the $k$=0,1,2 transitions and overestimate the $k$=3 transition. When fitting only the $k$ > 3 components towards ALMA2a and 2b the only notable difference between the runs fixing the source size to the beamsize and those fixing the source size to the moment 0 size is that N(CH$_{3}$CN) is $\sim$25\% and $\sim$10\% smaller for ALMA2a and 2b respectively in the beamsize runs. We choose to present best-fit parameters for the fit to the $k$ > 3 components using the moment 0 source sizes (see Figure~\ref{fig:mom0_FITS_CH3CN}) as this accounts for the relative difference in the size of the CH$_{3}$CN emission of ALMA2a and 2b. For ALMA6a, fixing the source size to either the beamsize or to the size estimated from the moment 0 map of the $k$=0 transition results in best-fit parameters that agree within the formal uncertainties. Therefore, for consistency with ALMA2a/b, we present the best fit for ALMA6a obtained by fitting all $k$ components using a source size fixed to the size estimated from the moment 0 map (see Figure~\ref{fig:mom0_FITS_CH3CN}).

Highest likelihood model spectra for ALMA2a and 2b are presented in Figure~\ref{fig:ALMA2ab_WEEDS}, and for ALMA6a in Figure~\ref{fig:ALMA6a_WEEDS}. The best-fit model parameters for all three sources are given in Table~\ref{tab:CH3CN_best_fits}. We note that as \textsc{weedspymcmc} assumes LTE and a single body of gas exciting all the observed emission, the uncertainties reported in Table~\ref{tab:CH3CN_best_fits} are likely underestimated. Corner plots of the 1D and 2D posterior distributions of the free parameters are shown in Appendix~\ref{Appendix-CH3CN} (Figures~\ref{fig:ALMA2a_corner_CH3CN}, ~\ref{fig:ALMA2b_corner_CH3CN} and Figure~\ref{fig:ALMA6a_corner_CH3CN}).

We note that the v$_{\rm LSR}$ values for ALMA2a and 2b obtained from our CH$_{3}$CN fitting, 0.75 km s$^{-1}$ and 0.30 km s$^{-1}$ respectively, differ somewhat from estimates obtained from O-bearing COMs (0.0 km s$^{-1}$ for ALMA2a from fitting of CH$_3$OCHO and CH$_3$OCH, Y. Chen priv. comm., and 0.7-0.8 km s$^{-1}$ for ALMA2b, \citealt{Chen2023}). This may suggest that the CH$_{3}$CN emission traces different structure(s) within the scale of the synthesized beam than the O-bearing COMs included in the \citet{Chen2023} study.

\begin{table}
	\centering
	\caption{CH$_{3}$CN best-fit parameters $^{a}$}
	\label{tab:CH3CN_best_fits}
    \begin{tabular*}{\columnwidth}{@{\extracolsep{\fill}}cccc}
        \hline
         Parameter & ALMA2a & ALMA2b & ALMA6a\\
        \hline   
        \rule{0pt}{2.2ex}
        N(CH$_{3}$CN) & \multirow{2}{*}{0.62 (0.02)} & \multirow{2}{*}{1.79 (0.02)} & \multirow{2}{*}{0.012 (0.002)} \\
        ($\times$ 10$^{16}$ cm$^{-2}$) &  &  & \\
        T$_{\rm ex}$ (K) & 215 (8)& 244 (3) & 110 (20) \\
        v$_{\rm lsr}$ (km s$^{-1}$) & $+$0.75 (0.05) & $+$0.30 (0.01)& $-$1.65 (0.08)\\
        FWHM (km s$^{-1}$)& 5.9 (0.1) & 5.75 (0.04) & 3.5 (0.2) \\
        Size$^{b}$ ($^{\prime\prime}$)& 0.24 & 0.30 & 0.40\\
        \hline
    \end{tabular*}
    \begin{tablenotes}
    \small
     \item \bf{Notes} 
     \item{$^{a}$ \textnormal{The larger of the formal uncertainties from the MCMC fitting (see Section~\ref{CH3CN fitting} and Figures~\ref{fig:ALMA2a_corner_CH3CN}-~\ref{fig:ALMA6a_corner_CH3CN}) is given to one significant figure in parentheses.}}
     \item{$^{b}$ \textnormal{Size refers to a diameter. The size is fixed to the geometric mean of the FWHM size obtained from a two-dimensional Gaussian fit to the integrated intensity map of the $k$=3 transition for ALMA2a/b and the $k$=0 transition for ALMA6a (see Section~\ref{CH3CN fitting} and Figure~\ref{fig:mom0_FITS_CH3CN}).}}
     \end{tablenotes}
\end{table}

\subsection{Source properties from dust and line emission} \label{Deriving further source properties}

Assessing the stability of candidate multiple systems (Section~\ref{Stability analysis}) requires mass estimates for the individual member sources. To obtain mass estimates, we first investigate the nature of the mm-wavelength continuum emission by deriving a mm-wavelength spectral index, $\alpha_{\rm 1.25-1.15mm}$, for each CoCCoA source. These values, calculated using the source's peak intensity in the lower and upper tuning continuum images (Section~\ref{sec:Coccoa_obs}), are presented in Table~\ref{tab:derived_source_properties}.

In general, the $\alpha_{\rm 1.25-1.15mm}$ values are consistent with thermal dust emission ($\alpha\gtrsim$2) and with the non-detections of most CoCCoA sources at cm wavelengths (see Table~\ref{tab:cm data table} and Section~\ref{individual_sources}). We note that for ALMA1 the derived mm-wavelength spectral index and cm nondetections are consistent only if the mm spectral index is at the upper limit of our estimated range (i.e. is equal to 2.4) and that the calculated $\alpha_{\rm 1.25-1.15mm}$ for ALMA4b and ALMA7 are unphysical, likely because these sources are extended and weak, respectively, and the estimate of spectral index from peak intensities is less robust. 

For ALMA2b, the only CoCCoA source with a cm-$\lambda$ counterpart, the 1.36\,cm detection from \citet{Yanza2025} is consistent, within uncertainties, with our $\alpha_{\rm 1.25-1.15mm}$. In contrast, the 3.00\,cm detection is not, which is consistent with a change in spectral index across the observed wavelength range due to the presence of both free-free and dust emission. Extrapolating the 3.00\,cm integrated flux density using the spectral index derived by \cite{Yanza2025} ($\alpha$ = $-$0.1, consistent with optically thin free-free emission) corresponds to only 0.18 mJy at 1.20\,mm, $<$1\% of the 1.20\,mm integrated flux density of ALMA2b. Based on the derived $\alpha_{\rm 1.25-1.15mm}$ values and cm non-detections, we conclude that the observed mm continuum emission is attributable to thermal dust emission. 

We estimate the gas mass, M$_{\rm gas}$, of each source assuming isothermal dust emission following \cite{Cyganowski2017}:

\begin{equation} \label{gas_mass_equation}
    M_{gas} (M_{\odot})= \frac{4.79 \times 10^{-14}D^{2}(kpc)S_{\nu}(Jy)C_{\tau_{dust}} R}{B_{\nu,T_{dust}}\kappa_{\nu}},
\end{equation}

\noindent where D is the source distance and $R$ is the gas-to-dust ratio (assumed to be 100). For all sources except ALMA2c and ALMA7 (discussed below) we use the integrated flux density ($S_{\nu}$) from Table~\ref{tab:fitted_source_properties}. The dust optical depth ($\tau_{\rm dust}$) is calculated as ${\tau_{\rm dust}} = -ln(1-\frac{T_{\rm b}}{T_{\rm dust}})$ (see Table~\ref{tab:derived_source_properties} for derivation of brightness temperature T$_{\rm b}$). We include a correction for optical depth as $C_{\tau_{\rm dust}} = \frac{\tau_{\rm dust}}{1-e^{-\tau_{\rm dust}}}$. For ALMA2a/b and ALMA6a we adopt T$_{\rm dust}$=T$_{\rm ex}$(CH$_{3}$CN) (Section~\ref{CH3CN fitting}, Table~\ref{tab:CH3CN_best_fits}). Of the remaining sources, ALMA1 has a high brightness temperature, T$_{\rm b}$ = 84 K, which constitutes a strict lower limit on its physical temperature. As we cannot constrain its temperature further, we adopt a wide range of T$_{\rm dust}=$100$-$200 K for ALMA1. For all other sources (ALMA2c, ALMA3a/b, ALMA4a/b, ALMA5 and ALMA7) we adopt T$_{\rm dust}$ = 20K \citep[as is commonly assumed, see e.g.][]{Motte2022}. 

For all sources we assume dust grains with thin ice mantles and a gas density of 10$^{6}$ cm$^{-3}$ \citep[see e.g.][and references therein]{Coletta2025}, adopting a dust opacity, $\kappa_{\rm 1.20\,mm}$, of 1.04 $\text{cm}^{2} \ \text{g}^{-1}$ \citep[interpolated in wavelength from the values listed in column 5 of Table 1 in][]{Ossenkopf1994}. We note that for all our calculations of M$_{\rm gas}$ the major contributors to the uncertainty are the lack of constraint on T$_{\rm dust}$ and the assumed values of $\kappa_{\nu}$. 

As noted in Sections~\ref{alma2} and ~\ref{alma7}, the flux densities of ALMA2c and ALMA7 are not well characterised by Gaussian fits. To estimate the gas masses of these sources we therefore use the flux densities from Sections~\ref{alma2} and ~\ref{alma7} and assume optically thin emission (and so do not include the $C_{\tau_{dust}}$ term in Equation~\ref{gas_mass_equation}). Using the same assumptions, we also estimate the gas mass of the large-scale filamentary structure discussed in Section~\ref{continuum emission} from its mm continuum emission, finding a gas mass of 15.8 M$_{\odot}$ for T$_{\rm dust}$ = 20 K.

The mass estimates from the thermal dust emission include only circum(proto)stellar material, and do not account for the mass of embedded protostars.  
For the protostellar, outflow-driving sources ALMA2a, ALMA2b and ALMA6a, we thus also estimate virial masses, M$_{\rm vir}$, using the best-fit CH$_{3}$CN parameters \citep[as in e.g.][]{Hernandez2014}. Considering only gravitational forces and a spherically bound cloud with a 1/r$^{2}$ density profile \citep[for protostellar sources, as in][]{Nony2018} we follow \cite{MacLaren1998} and obtain $M_{\rm vir} = 0.305 \ d \ \theta \ \Delta V^{2}$, where $\theta$ and $\Delta V$ are the fitted angular diameter ($^{\prime \prime}$) and FWHM (km s$^{-1}$) respectively from the CH$_{3}$CN fitting and d is the distance in kpc. We correct for a mean inclination angle between the protostellar rotation axis and the line of sight of i = 60$^{\circ}$. We note that if we instead assumed a 1/r density profile the M$_{\rm vir}$ values, which are presented in Table~\ref{tab:derived_source_properties}, would increase by a factor of $\sim$1.5. Our virial mass estimate for ALMA6a has a significant uncertainty (M$_{\rm vir}$ = 2.6 $\pm$ 1.5 M$_{\odot}$), mainly due to the large uncertainty in the source size, (0.40 $\pm$ 0.17)$^{\prime\prime}$ (see Section~\ref{CH3CN fitting}). For comparison, we also calculated the virial mass for each source for a source size set to the convolved beam size (0\farcs33), which yields M$_{\rm vir}$ = 6.2, 5.9 and 2.2 M$_{\odot}$ for ALMA2a, 2b and ALMA6a respectively.

\subsection{Outflow properties} \label{Outflowsubsection}

Outflows are driven by the accretion of material onto protostellar sources \citep[e.g.][]{Arce2007} and studies ranging from the low- to high-mass star formation regimes have found relationships between protostellar and outflow properties \citep[e.g.][]{Ridge2001, Beuther_2004, Wu2004, DuarteCabrat2013, Maud2015}. Information pertaining to the protostar driving an outflow can thus be gained via its outflow; in particular, some constraint on the bolometric luminosity, L$_{\rm bol}$, of the protostar may be obtained from the momentum outflow rate, $\dot{\rm P}$, of the outflow. To this end, we derive a number of properties for the outflows identified in Section~\ref{CO_results}. Due to the large uncertainty in the abundance of SiO with respect to H$_{2}$ \citep[see][and references therein]{towner2023} we derive properties from the $^{12}$CO emission. 

We estimate the total mass of each outflow from its $^{12}$CO emission following the equation given in \cite{Cyganowski_2011}, which assumes optically thin emission:

\begin{equation} \label{equationMout}
    M_{\rm out} = \frac{1.186 \times 10^{-4} \times Q(T_{ex}) e^{\frac{E_{upper}}{T_{ex}}} D^{2} \int S_{\nu} d\nu}{\nu^{3} \mu^{2}S \chi}.
\end{equation}

\noindent M$_{\rm out}$ is the outflow mass in M$_{\odot}$, Q(T$_{\rm ex}$) is the partition function, E$_{\rm upper}$ is the upper energy level of the transition in K, T$_{\rm ex}$ is the excitation temperature in K, D is the source distance in kpc, S$_{\nu}$ is the spectral line flux density in Jy, $\nu$ is the frequency of the transition in GHz, $\mu$ is the dipole moment of the molecule, S is the intrinsic line strength of the transition and $\chi$ is the abundance of the molecule compared to H$_{2}$. We follow previous work \citep[e.g.][]{Qiu2009,Cyganowski_2011, Xu2024} and adopt T$_{\rm ex}$ = 30 K and $\chi$ = 10$^{-4}$ for $^{12}$CO. We use the E$_{\rm upper}$, $\nu$ and $\mu^{2}$S values provided by the Cologne Database for Molecular Spectroscopy \citep[CDMS:][]{Muller2001, MULLER2005} as quoted in Splatalogue\footnote{\hyperlink{https://splatalogue.online/}{https://splatalogue.online/}}. For $^{12}$CO(2-1) these are 16.59608 K, 230.538 GHz and 0.02423 debye$^{2}$ respectively. We find Q(T$_{\rm ex}$) by linearly interpolating between the values provided by CDMS, retrieving Q(30 K) = 11.186. For each outflow lobe we measure S$_{\nu}$ in each channel as the emission within both the 3$\sigma$ contour of the relevant integrated intensity (moment 0) map (produced using the velocity ranges indicated in Table~\ref{tab:outflow_properties}) and a region drawn to encompass the outflow and isolate it from other features. We measure the rms, $\sigma$, of each integrated intensity map locally to each outflow lobe and list the values used in Table~\ref{tab:outflow_properties}.

We estimate the momentum, P$_{\rm out}$, of each outflow lobe following \cite{Qiu2009}:

\begin{equation}
    P_{\rm out} = \sum M_{\rm out}(V) \Delta V,
\end{equation}

\noindent where M$_{\rm out}$($\rm V$) is the mass in each channel and $\rm \Delta V$ is $|V_{\rm channel} - V_{\rm lsr}|$. Calculating the mass and momentum outflow rates, $\dot{\rm M}$ = M$_{\rm out}$/t$_{\rm dyn}$ and $\dot{\rm P}$ = P$_{\rm out}$/t$_{\rm dyn}$, requires an estimate of the dynamical timescale t$_{\rm dyn}$. We estimate t$_{\rm dyn}$ for each outflow lobe as L$_{\rm outflow}$/|V$_{\rm max}$ - v$_{\rm LSR}$| \citep[e.g.][]{Cyganowski_2011}, where L$_{\rm outflow}$, the length of the outflow lobe, and the maximum velocity, V$_{\rm max}$, are reported in Table~\ref{tab:outflow_properties}. We measure the length of each outflow lobe as the projected distance between the driving source's fitted continuum peak and the furthest point of the outflow's moment 0 3$\sigma$ contour produced using the primary beam corrected images. As we do not consider the possible inclination of the outflows to the line of sight the length measurements constitute lower limits. 

For these calculations we use the v$_{\rm LSR}$ values discussed in Section~\ref{CoCCoA_line_emission} and listed in Table~\ref{tab:derived_source_properties}. For ALMA2a/b we assume a v$_{\rm LSR}$ of 0.30 km s$^{-1}$ i.e. the v$_{\rm LSR}$ retrieved in Section~\ref{CH3CN fitting} for the more line rich hot core component ALMA2b. For ALMA1 and ALMA5, for which we could not measure a v$_{\rm LSR}$ from the CoCCoA data, we calculate outflow properties assuming v$_{\rm LSR}$ = 0.0 km s$^{-1}$. We note that, with the exception of $\dot{\rm P}$ (see Section~\ref{outflow_discussion}), we do not correct derived outflow properties for inclination to avoid overcorrection \citep[][]{DownesCabrit2007, Dunham_2014}. As previously noted \citep[e.g.][]{Offner2001, Dunham_2014}, uncorrected derived properties estimated from observations such as ours will be lower limits because interferometric measurements are insensitive to some flux due to spatial filtering, the emission may be optically thick (not optically thin, as assumed in Equation~\ref{equationMout}), and we have discarded channels near the v$_{\rm LSR}$ of the region (see Table~\ref{tab:outflow_properties}). 

\section{Discussion}
\label{sec:Discussion}

\subsection{Sources driving outflows} \label{outflow_discussion}

\begin{figure}
	\includegraphics[trim=0 0 0 0, clip, width=\columnwidth]{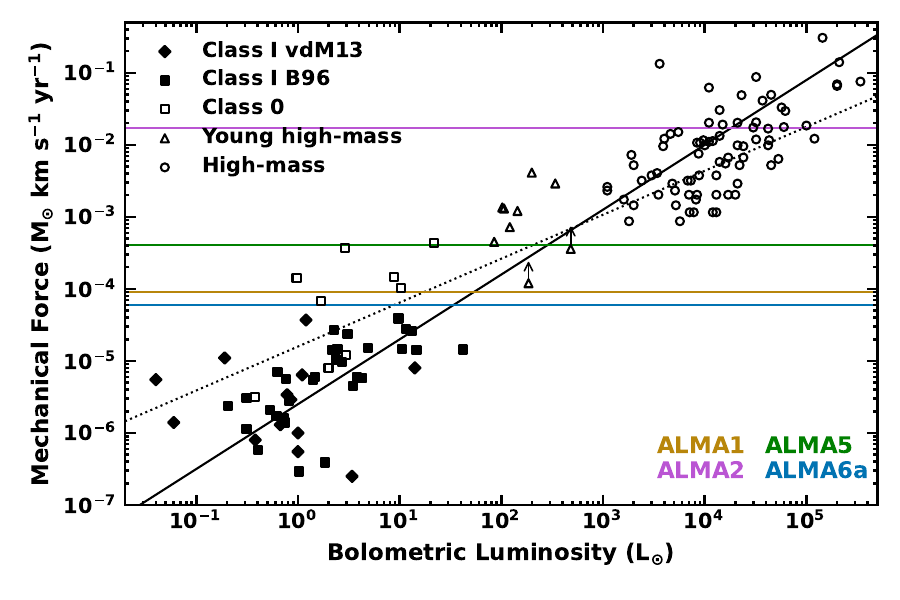}
    \caption{Outflow mechanical force, $\dot{\rm P}$$_{\rm tot}$, plotted against the bolometric luminosity, L$_{\rm bol}$, of the driving source. Symbols show data from \citet{Maud2015} Figure 7: Filled diamonds are the Class I sources of \citet{VanderMarel2013}, filled and open squares are the low-mass Class I and Class 0 sources of \citet{Bontemps1996} respectively, open triangles are the young high-mass sources of \citet{DuarteCabrat2013} (where lower limits are indicated with arrows) and open circles are the high-mass sources of \citet{Maud2015}. The solid and dashed lines indicate the best-fit relations found by \citet{Bontemps1996} and \citet{Maud2015} for their samples. $\dot{\rm P}$$_{\rm tot}$ values from this work are plotted as coloured horizontal lines (see Section~\ref{outflow_discussion}), with colour coded labels on the bottom right.}
    \label{fig:Luke_maud_plot}
\end{figure}

We plot each outflow in $\dot{\rm P}$$_{\rm tot}$$-$L$_{\rm bol}$ space in Figure~\ref{fig:Luke_maud_plot}, overlaid on the data from Figure 7 of \cite{Maud2015}. The plotted quantity $\dot{\rm P}$$_{\rm tot}$ is the sum of the $\dot{\rm P}$$_{\rm out}$ values for all the outflow components driven by each source (see Table~\ref{tab:outflow_properties}). As there is not sufficient high-resolution multi-wavelength data available to estimate the bolometric luminosity of the individual CoCCoA sources, we plot each outflow as a horizontal line in Figure~\ref{fig:Luke_maud_plot}.
For consistency when comparing our results with other work, we follow \cite{Maud2015} and correct for the average inclination of the \cite{Bontemps1996} sample, 57.3$^{\circ}$, resulting in a factor of $\sim$2.9 increase to our $\dot{\rm P}$$_{\rm tot}$ values. 

As shown in Figure~\ref{fig:Luke_maud_plot}, ALMA1 has $\dot{\rm P}$$_{\rm tot}$ = 9 $\times$ 10$^{-5}$ M$_{\odot}$ km s$^{-1}$ yr$^{-1}$ and ALMA6a has $\dot{\rm P}$$_{\rm tot}$ = 6 $\times$ 10$^{-5}$ M$_{\odot}$ km s$^{-1}$ yr$^{-1}$. Both these values sit amongst the $\sim$10$^{0}$$-$10$^{1}$ L$_{\odot}$ Class 0 sources of \cite{Bontemps1996} and above the other low-mass samples. ALMA2a/b has $\dot{\rm P}$$_{\rm tot}$ = 1.7 $\times$ 10$^{-2}$ M$_{\odot}$ km s$^{-1}$ yr$^{-1}$, where we have considered the blueshifted low velocity contribution as only the Collimated+SE component and the blueshifted high velocity contribution as the Collimated component and SE component together. This places ALMA2a/b above the largest $\dot{\rm P}$$_{\rm tot}$ of the \cite{DuarteCabrat2013} sources and amongst the high-mass sources of \cite{Maud2015}, where the corresponding luminosity is $\sim$a few$\times$10$^{3}$-10$^{5}$ L$_{\odot}$. This implied luminosity is notably larger than that derived by \cite{Persi2009} for IR-MM3 (IRS 8E) (see Section~\ref{alma2}). ALMA5 has $\dot{\rm P}$$_{\rm tot}$ = 4.1 $\times$ 10$^{-4}$ M$_{\odot}$ km s$^{-1}$ yr$^{-1}$. This places ALMA5 below the high-mass sources of \cite{Maud2015} and amongst the scatter of both the young high-mass sources of \cite{DuarteCabrat2013} and the low-mass Class 0 sources of \cite{Bontemps1996}. We note, however, that if the redshifted emission to the north of ALMA5 is instead associated with a second bipolar outflow driven by ALMA2a/b (see Sections~\ref{alma2} and ~\ref{alma5}), our outflow properties for ALMA5 will be overestimates. In any case, it is difficult to estimate a corresponding L$_{\rm bol}$ as this area of $\dot{\rm P}$$_{\rm tot}$$-$L$_{\rm bol}$ space is scantly populated and there is a large horizontal scatter. 

We note that ALMA1 has relatively high M$_{\rm gas}$ and T$_{\rm b}$ values (0.75-2.61 M$_{\odot}$ and 84 K in Table~\ref{tab:derived_source_properties}) and a comparatively weak outflow. Further, ALMA1's outflow is detected in both $^{12}$CO and SiO and its $^{12}$CO lobes have rather low t$_{\rm dyn}$ values. We postulate that ALMA1 may be a very young source, where the large majority of its mm continuum emission pertains to an envelope within which a lower mass source has only recently begun to drive its compact outflow.
    
\begin{table*}
	\centering
	\caption{$^{12}$CO(2-1) Outflow properties}
	\label{tab:outflow_properties}
    \begin{tabular*}{\linewidth}{@{\extracolsep{\fill}}cccccccccccccc} 
        \hline
        \rule{0pt}{2.2ex}    
        & & &  V$_{\rm min}$&V$_{\rm max}$&$\sigma$$^{a}$& M$_{\rm out}$  & P$_{\rm out}$& t$_{\rm dyn}$ & Length & $\dot{\rm M}$$_{\rm out}$ & $\dot{\rm P}$$_{\rm out}$ \\
        & & & (km s$^{-1}$)&(km s$^{-1}$)&(Jy beam$^{-1}$ &(M$_{\odot}$) & (M$_{\odot}$ & (years) & (pc) & (M$_{\odot}$ & (M$_{\odot}$km s$^{-1}$\\
        & & & & & km s$^{-1}$) & &km s$^{-1}$) & & & yr$^{-1}$)& yr$^{-1}$)\\
        & & & & & & ($\times$ 10$^{-3}$) & & & & ($\times$ 10$^{-5}$) & ($\times$ 10$^{-4}$)\\
        \hline
        \multirow{2}{*}{\ \ ALMA1$^{b}$} &\multicolumn{2}{c}{Blue} & $-$19.48 & $-$6.04 & 0.273& 2.9& 0.026 &1100& 0.021& 0.27 & 0.24\\
        & \multicolumn{2}{c}{Red}& 7.40 & 20.20 & 0.178& 0.3 & 0.004 & 400& 0.008 & 0.08 & 0.1\\
        \hline \\[-1.8ex] \cline{2-3}
        & \multirow{3}{*}{Blue}&Collimated & $-$50.2 & $-$6.04 & 0.314 & 101.4  & 1.84  & 2620 &  0.135 & 3.87 & 7.02 \\ 
        ALMA2a/b$^{c}$ &&Collimated+SE & ... & ... & ... & 117.4 & 2.022 & 3330 & 0.172 & 3.53 & 6.07 \\
        low velocity&&SW &... & ...&... & 18.4 & 0.215 & 1700 & 0.086 & 1.1 & 1.3\\ \cline{2-3} 
        &\multicolumn{2}{c}{Red}& 6.76 & 50.28 & 0.309 & 82.8 & 1.73 & 3240 & 0.166& 2.56 & 5.95 \\[5pt] \cline{2-3}
        \multirow{2}{*}{ALMA2a/b$^{c}$} &\multirow{2}{*}{Blue} &Collimated & $-$109.72& $-$50.84& 0.188& 22.1 & 1.68 & 850 & 0.095& 2.6 & 20 \\ 
        \multirow{2}{*}{high velocity}&&SE & ...& ...&... & 28.2 & 2.09 & 1510 & 0.170 & 1.87 & 14.2 \\ \cline{2-3} 
        
        &\multicolumn{2}{c}{Red}& 50.92& 139.24 & 0.323 & 18.6 & 1.54 & 1120 & 0.159 & 1.66& 13.7 \\
        \hline

        \ \ ALMA5$^{b}$& \multicolumn{2}{c}{Red}  & 6.76 & 35.56 & 0.249& 11.4& 0.144 & 1000 &\ \ 0.037$^{d}$& 1.1&1.4 \\
        \hline
        ALMA6a$^{e}$&\multicolumn{2}{c}{Blue} & $-$6.04 & $-$27.16 & 0.270 & 0.4  & 0.03& 1400 & 0.036 &  0.03 &0.2 \\
        \hline
    \end{tabular*}
   \begin{tablenotes}
    \small
     \item \bf{Notes} 
    \item{ \textnormal{$^{a}$Measured in moment 0 maps corrected for the primary beam response (see Section~\ref{Outflowsubsection}).}}
    \item{ \textnormal{$^{b}$We assume a v$_{\rm LSR}$ = 0.0 km s$^{-1}$.}}
    \item{ \textnormal{$^{c}$ The velocity ranges for ALMA2a/b's bipolar outflow were chosen such that blue |V$_{\rm max}$ $-$ v$_{\rm LSR}$|$\sim$red |V$_{\rm min}$ $-$ v$_{\rm LSR}$| for v$_{\rm LSR}$ = 0.30 km s$^{-1}$ (see Section~\ref{Outflowsubsection}). Further, we split the red and blue $^{12}$CO and SiO features into low velocity and high velocity components, where a high velocity component corresponds to |V$_{\rm channel}$ $-$ v$_{\rm LSR}$| > 50 km s$^{-1}$. Properties for the collimated low-velocity component are measured within a region drawn to separate the collimated low-velocity emission from the combined collimated+SE emission (see Sections~\ref{alma2} and ~\ref{Outflowsubsection}).}}
    \item{ \textnormal{$^{d}$ Measured towards the north (see Section~\ref{CO_results}).}}
    \item{ \textnormal{$^{e}$ We use a v$_{\rm LSR}$ =$-$1.65 km s$^{-1}$ (see Table~\ref{tab:derived_source_properties}).}}
    \end{tablenotes}
\end{table*}

\subsection{Candidate multiple systems} \label{Stability analysis}

In order to assess whether a group of sources may be considered as a multiple system we follow a simple analysis employed in previous work \citep[e.g.][]{Baumgardt2002, Pineda2015, Li2024, Li2025}. The total gravitational and kinetic energy of each source in a candidate system is estimated in order to determine if together they may be considered bound. The total gravitational energy, $\rm W$, of source $i$ is

\begin{equation}
    W_{i} = - \sum_{i \neq j} \frac{Gm_{i}m_{j}}{r_{ij}}, 
\end{equation}

\noindent where m$_{i}$ is the mass of the source, m$_{j}$ is the mass of another source $j$ and r$_{ij}$ is the projected separation between them. The kinetic energy, $\rm T$, of source $i$ is 

\begin{equation}
    T_{i} = \frac{1}{2} m_{i} (V_{i} - V_{sys})^2,
\end{equation}

\noindent where $\rm V$$_{i}$ is the velocity of the source and V$_{\rm sys}$ is the systemic velocity of the system. A common approach \citep[e.g.][]{Pineda2015, Li2024, Li2025} is to adopt a mass-weighted V$_{\rm sys}$, which for a system with $k$ members is

\begin{equation}
    V_{sys} = \frac{\sum_{k} V_{k} m_{k}}{\sum_{k} m_{k}},
\end{equation}

\noindent where m$_{k}$ and V$_{k}$ are the mass and velocity of each member respectively. A system is considered bound if T$_{i}$ / |W$_{i}| < 1$ for every source within it. This analysis returns a lower limit on T$_{i}$ / |W$_{i}|$ for each source, as |W$_{i}$| decreases for non-zero separations along the line of sight and T$_{i}$ increases for non-zero (V$_{i} - $V$_{\rm sys}$) in the two non-line of sight directions.

The analysis outlined above uses the 1D line of sight velocity and 2D projected separations. We refer to this as a `1D velocity/2D separation' analysis. As in \cite{Li2024}, we compare this to a `3D' case, where simple dimensional corrections for both separation and relative velocity are included. Assuming that the projected separation is similar to the separation along the line of sight, we use a $\sqrt{2}$ correction such that r$_{ij}^{3D}$ = $\sqrt{2}$r$_{ij}^{2D}$. Similarly, if the line of sight velocities of the sources are similar to their two other components of velocity, then (V$_{i} - $V$_{\rm sys}$)$^{3D}$ = $\sqrt{3}$(V$_{i} - $V$_{\rm sys})^{1D}$. 

In our analysis, we consider sources within the 5$\sigma_{\rm centre}$ contour of the large-scale filamentary structure discussed in Section~\ref{continuum emission}, i.e. ALMA2a/b/c, ALMA3a/b, ALMA4a/b and ALMA6a/b. To calculate W and T for each source, we use the v$_{\rm LSR}$ and mass estimates derived above and summarised in Table~\ref{tab:derived_source_properties}, using virial mass estimates where available (for ALMA2a, ALMA2b and ALMA6a).  We derive separations between sources using the centroid positions in Table~\ref{tab:fitted_source_properties} (as outlined in Table~\ref{tab:derived_source_properties}). Due to uncertainty in the derived v$_{\rm LSR}$ values, we detail the range of V$_{\rm sys}$ over which a candidate system is bound and compare it to the mass-weighted value in order to assess whether or not the candidate system may reasonably be considered bound.

\begin{figure*}
	\includegraphics[trim=0 0 0 0, clip, width = \textwidth]{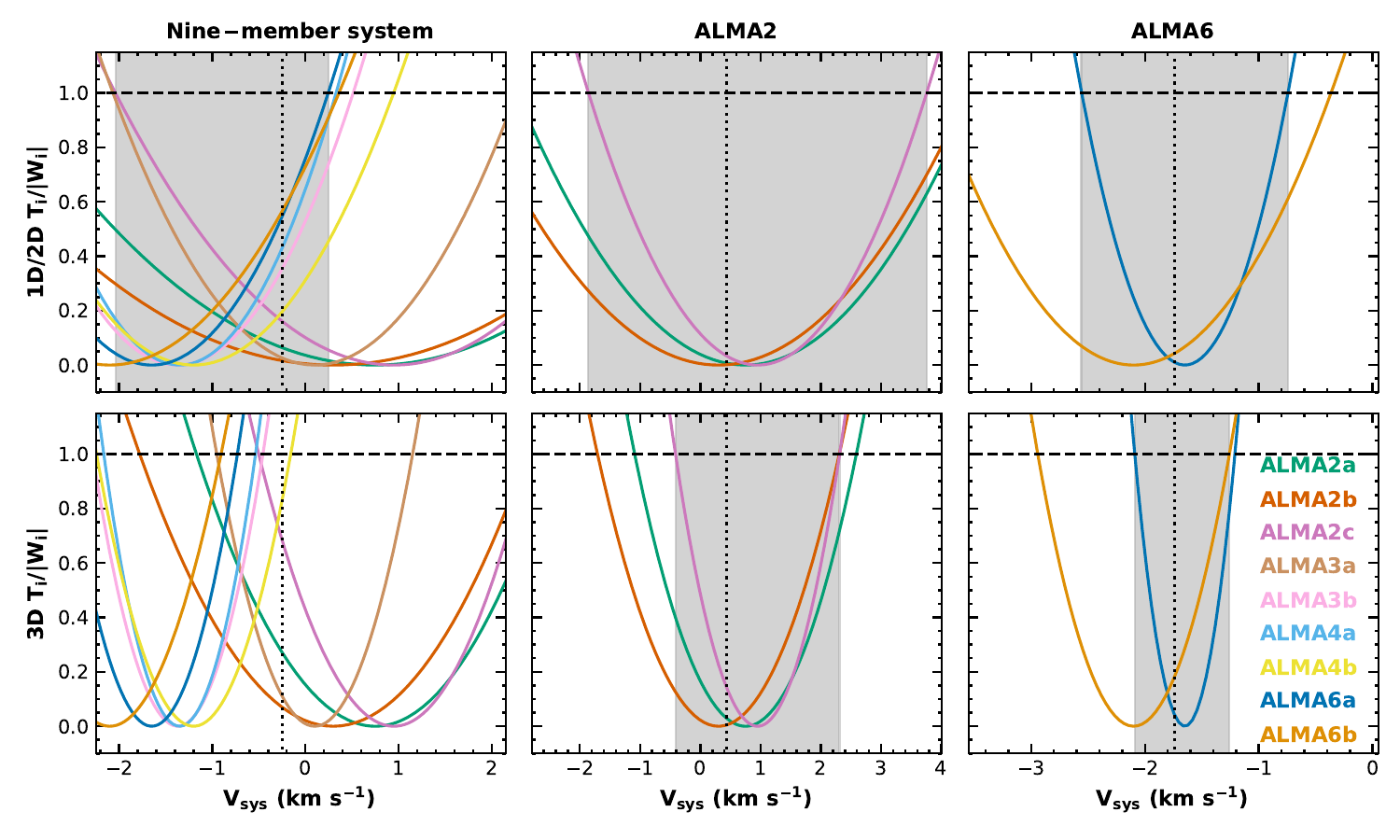}
    \caption{Stability of systems in the 1D velocity/2D separation (top row) and 3D (bottom row) cases. Sources are colour coded as indicated. We plot the ratio of the kinetic energy, T$_{i}$, to the gravitational energy, |W$_{i}$|, for each source in a system as a function of the system's systemic velocity, V$_{\rm sys}$. The mass weighted V$_{\rm sys}$ (from left to right: $-$0.05 km s$^{-1}$, +0.61 km s$^{-1}$ and $-$1.75 km s$^{-1}$) is indicated as a vertical dotted line in each panel. Systems are considered bound when T$_{i}$ / |W$_{i}$| < 1 for every source in the system, indicated by the shaded grey regions in each panel.}
    \label{fig:stability_multiplot}
\end{figure*}

The results of our analysis are summarised in Figure~\ref{fig:stability_multiplot}, where both the 1D velocity/2D separation and 3D cases are considered. Remarkably, we find that the case of a nine-member system comprised of ALMA2a, 2b, 2c, ALMA3a, 3b, ALMA4a, 4b and ALMA6a, 6b is bound in the 1D velocity/2D separation case between V$_{\rm sys}$ $-$2.04 and +0.25 km s$^{-1}$. This range is consistent with the mass weighted V$_{\rm sys}$ of the nine-member system ($-$0.05 km s$^{-1}$), the average centroid fit of the large ppv coherent H$^{13}$CO$^+$ structure ($-$1.14 km s$^{-1}$, see Section~\ref{Large-scale H13CO+}), and literature estimates of the region's v$_{\rm LSR}$ ($-$0.42, $-$0.82 and $-$1.14 km s${-1}$, see Section~\ref{introduction}). The mean separation between pairs of the system is 5\farcs95 (7930 au), while the maximum projected separation is $\sim$11\farcs3 (15,000 au). Whilst the nine-member system is never bound in the 3D case, we have not accounted for the gravitational potential of the large-scale filamentary emission, which may be significant. 

As the observed multiplicity in NGC 6334-43 could be produced via multiple mechanisms acting on various scales (see Section~\ref{introduction}), we also investigate the existence of bound sub-systems within the nine-member multiple. For a robust analysis, we only consider sub-systems where we have a virial mass (see Table~\ref{tab:derived_source_properties}) from a compact velocity tracer (see Section~\ref{CoCCoA_line_emission}) for at least one member. 

As illustrated in Figure~\ref{fig:stability_multiplot}, a triple system comprised of ALMA2a, 2b and 2c is bound in the 1D velocity/2D separation case for V$_{\rm sys}$ between $-$1.86 km s$^{-1}$ and +3.76 km s$^{-1}$, and in the 3D case for V$_{\rm sys}$ between $-$0.41 km s$^{-1}$ and +2.31 km s$^{-1}$. The mass weighted V$_{\rm sys}$ of 0.61 km s$^{-1}$ is consistent with both cases. A bound system is also recovered for all pairs of sources in the system (for both the 1D velocity/2D separation and 3D cases). The mean separation between pairs of the ALMA2 triple is 1\farcs27 (1690 au). We also find that ALMA6a and ALMA6b, which are separated by 1530 au, are a bound binary system in the 1D velocity/2D separation case for V$_{\rm sys}$ between $-$2.56 km s$^{-1}$ and $-$0.74 km s$^{-1}$, and in the 3D case for V$_{\rm sys}$ between $-$2.09 km s$^{-1}$ and $-$1.26 km s$^{-1}$. As shown in Figure~\ref{fig:stability_multiplot}, both ranges are consistent with the mass-weighted V$_{\rm sys}$ of $-$1.75 km s$^{-1}$. 

We note that our analysis pertains to the formation process and that systems that are bound may not remain so, as gravitational interactions in young clusters are likely to result in a reconfiguration of systems and the ejection of some sources \citep[e.g.][]{Reipurth_2012, Reipurth_2014}. Further, as we do not include a gravitational potential associated with the large-scale filamentary structure (which has a mass of $\sim$15.8 M$_{\odot}$, Section~\ref{Deriving further source properties}), this analysis is likely to find a lower limit on the range of V$_{\rm sys}$ over which a system is bound.

\subsection{Formation mechanisms} \label{formation mechanisms}

Here, we discuss the likely formation mechanisms for the bound nine-member system and bound sub-systems identified in Section~\ref{Stability analysis}. Three recognised formation mechanisms of protostellar multiple systems are disk, core, and filament fragmentation \citep[see][]{Offner2022}. The detection of multiple bound sources within a disc-like structure would support the first scenario, while the other scenarios would be favoured by the detection of fragmentation on larger scales, where an aspect ratio criterion, $a$, of $>$3 is typically adopted to distinguish between core and filament structures \citep{Offner2022}.

\subsubsection{The nine-member system} \label{the nine member system}

The estimated width of the large-scale filamentary structure in the field is 1500$-$3000 au (see Section~\ref{continuum emission}) and the nine-member system has a maximum separation between components of $\sim$15,000 au. Its aspect ratio of $a>$3 is consistent with formation via filament fragmentation and its maximum separation of $\sim$15,000 au is comparable to that of an observed quadruple system shown to be formed via filament fragmentation \citep[maximum separation 11,400 au;][]{Pineda2015}.

Sources formed via filament fragmentation are expected to be of a similar age, $\Delta$t $\sim$0.5 Myr \citep{Offner2022}. The nine sources considered here display a range of evolutionary signatures. ALMA2a/b and ALMA6a are clearly protostellar, based on their outflows (Section~\ref{CO_results}) and the high physical temperatures inferred from the CH$_{3}$CN modelling (Table~\ref{tab:CH3CN_best_fits}), which indicate significant heating by embedded protostars. As noted above (Section~\ref{alma2}), ALMA2b is also coincident with cm-$\lambda$ free-free emission, consistent with a more evolved evolutionary state. Conversely, the relative paucity of molecular line emission observed towards ALMA2c, ALMA3a/b, ALMA4a/b and ALMA6b is suggestive of early evolutionary states \citep[e.g.][see also Section~\ref{CoCCoA_line_emission}]{Nony2018,Barnes2023}. However, age differences of $\Delta$t $\sim$0.5 Myr are roughly in line with the total timescale of massive star formation \citep[$\sim$300 kyr:][]{Motte2018}, so a filament-fragmentation origin for the nine-member system remains feasible. 

The nine-member system presented here is notable for its higher order multiplicity and the presence of bound sub-systems.  Interestingly, the more evolved, main-sequence nine-member multiple HD 93206 (QZ Car) also contains bound sub-systems \citep{Mayer2022, Brovz2022}. Of protostellar multiple systems that have been shown to be bound, \cite{Li2024} find remarkable evidence of high-order multiplicity (up to a quintuple system) in the high-mass star forming region G333.23–0.06, where a population of multiples with a mean separation of 731 au is found to be formed via core fragmentation. Their mean separation is rather lower than the mean separation found here (7930 au), although this is likely impacted by their higher angular resolution ($\sim$260 au compared to $\sim$350 au here). Recently, \cite{Li2025} reported a close separation (332 au mean deprojected separation) septuple system formed via disc fragmentation. Our results indicate that filament fragmentation may also produce very high-order multiple systems. 

\subsubsection{The ALMA2 triple system} \label{the alma2 system}

The two hot core components ALMA2a/b, separated by 618\,AU (Table~\ref{tab:derived_source_properties}), appear to be amongst the most evolved mm sources in the NGC 6334-43 region (see Section~\ref{the nine member system}). While there is some indication of possible disc-like signatures towards ALMA2a (Appendix~\ref{Appendix-discs} and Figure~\ref{fig:mom1_ALMA2_CH3CN}), there is no evidence for a shared disc structure encompassing both ALMA2a and ALMA2b in the CoCCoA data. The maximum separation between the components of the ALMA2 system is 2330 au, which is similar to the surrounding structure's width (see Figure~\ref{fig:continuum_plot}c). This is generally consistent with expectations for formation by core fragmentation, particularly if the hot core components have migrated inward \citep[e.g.][]{Tokovinin2019}; simulations in \cite{Kuruwita2023}, for example, find a typical core fragmentation separation that is independent of gas density, and corresponds to a projected 2D separation of $\sim$700$-$2000 au. 

\subsubsection{The ALMA6 binary system} \label{the alma6 system}

The ALMA6 protobinary stands out for its youth and for the presence of a connective bridge and spiral-arm-like feature(s) in the mm continuum (Section~\ref{alma6}). The projected separation between components of the ALMA6 protobinary is 1530 au, similar to the width of its common structure. As with the ALMA2 triple, ALMA6's separation is consistent with the typical scale of core fragmentation found in \cite{Kuruwita2023}. 

Bridge-like features are found in simulations of multiple formation via both disc \citep[e.g.][]{Mignon-Risse2021} and turbulent \citep[e.g.][]{Kuffmeier2019} fragmentation. A distinguishing feature of ALMA6 is the prominent spiral-arm-like structure to its south, which has a potential counterpart to the north (see Figure~\ref{fig:continuum_plot}g). Spiral arms often play an important role in the formation of binaries in simulations of disc fragmentation \citep[e.g.][]{Krumholz2009,Oliva2020,Mignon-Risse2021}, although here the structure (which is $\sim$4400 au long) extends well beyond the expected scale of a disc. 

Comparing ALMA6 to the small existing sample of young high-mass multiple systems (Section~\ref{introduction}), the presence of a linking bridge-like structure is notably similar to the G11.92$-$0.61 MM2 protobinary \citep[formerly a candidate high-mass starless core,][]{Cyganowski14} in the G11.92$-$0.61 protocluster \citep{Cyganowski2022}. This relatively equal-mass, 505 au-separation protobinary system \citep{Cyganowski2022, sanhueza2025} has properties comparable to the super-Alfv\'enic disk fragmentation models of \cite{Mignon-Risse2021}. G11.92$-$0.61 MM2 and ALMA6 also both show evidence of ongoing accretion, with MM2 fed significant material from a large-scale ($\sim$35,000 au) filamentary structure ($\sim$ 1.8 $\times$ 10$^{-4}$-1.2 $\times$ 10$^{-3}$ M$_{\odot}$ yr$^{-1}$, \citealt{Zhang_suinan_2024}; see also \citealt{sanhueza2025}).

\cite{Kong2023} find that the C1-Sa system, a $\sim$1400 au binary residing in the dragon cloud (G028.37+00.07), has likely formed via turbulent fragmentation. The authors note that only one component of C1-Sa exhibits any mm line emission and that N$_{2}$D$^+$ and ortho-H$_{2}$D$^+$ in the vicinity of the binary indicates an early stage of star formation. The separation of C1-Sa is very similar to the 1530 au ALMA6 system and both systems show evidence of differences in evolutionary state amongst their members. Based on the detection of outflows, \citet{Kong2023} find that C1-Sa1 is protostellar and that C1-Sa2 is likely starless, an evolutionary discrepancy also seen in ALMA6 (see Section~\ref{the nine member system}). 

\section{Conclusions}\label{conclusions}

We have used high resolution ($\sim$350 au) ALMA data from the CoCCoA survey to characterise the protostellar multiplicity within a $\sim$0.05 pc$^{2}$ FOV near the hot core NGC 6334-43. The high-resolution CoCCoA observations reveal 12 compact mm sources, three of which, ALMA1, ALMA5 and ALMA7 appear to be isolated. Our main results are summarised below.

\begin{enumerate}
    \item [(i)] We find that a nine-member protostellar system appears to have formed via the fragmentation of a single large-scale filamentary structure traced in 1.20\,mm continuum and H$^{13}$CO$^+$ emission. A simple stability analysis, comparing gravitational and kinetic energies, shows that the system, which has a mean separation between pairs of 7930 au, is bound.

    \item [(ii)] The sources in the nine-member system display a variety of current masses and evolutionary stages. By modelling CH$_{3}$CN emission observed towards ALMA6a and towards the line-rich hot cores ALMA2a and ALMA2b, we derive M$_{\rm vir}$ = 2.6, 4.5 and 5.4 M$_{\odot}$ respectively. Best-fit model parameters also indicate significant heating by embedded protostar(s) in these sources. The other compact mm sources in the nine-member system have gas masses, derived from their mm continuum emission, of M$_{\rm gas}$ = 0.50-1.87 M$_{\odot}$ and display weak and sparse mm line emission, indicating earlier evolutionary stages. 
    
    \item [(iii)] Using archival ALMA data we detect outflows in $^{12}$CO driven by ALMA1, ALMA2a/b, ALMA5 and ALMA6a. The derived outflow properties of  ALMA2a/b are consistent with those of other high-mass protostars, consistent with these sources' rich hot-core line emission. The outflow properties of ALMA6a and ALMA1 are similar to those of low-mass Class 0 sources, while those of ALMA5 resemble low-mass Class 0 and young high-mass sources. The apparent discrepancy between ALMA1's outflow properties and its high mm continuum brightness temperature (84 K) suggests that this may be a very young source which has only recently begun driving its outflow. The outflow lobes driven by ALMA1 and ALMA2a/b are also strongly detected in SiO, indicating that these are active outflows and these sources are actively accreting. 
     
    \item [(iv)] The ALMA2 sub-system, comprised of the 618-au separation hot cores ALMA2a/b and the younger source ALMA2c, has a mean separation of 1690 au, which is consistent with formation via core fragmentation. ALMA2b is the only mm source in the NGC 6334-43 field detected in recent cm observations. 

    \item [(v)] The ALMA6 sub-system is a young 1530 au-separation protobinary that displays an interesting $\sim$4400 au long spiral-arm-like structure and appears to have formed via core fragmentation. The protostellar source ALMA6a is apparently more evolved than the candidate prestellar source ALMA6b. 
    
\end{enumerate}

\section*{Acknowledgements}

We thank Yuan Chen for their help and discussion concerning line emission fitting and Mairi Nonhebel for their input on using \textsc{weedspymcmc}. We also thank Luke Maud for sharing data pertaining to Figure~\ref{fig:Luke_maud_plot} and Vanessa Yanza for sharing relevant cm images. This paper makes use of the following ALMA data: ADS/JAO.ALMA\#2017.1.00180.S and  ADS/JAO.ALMA\#2019.1.00246.S. ALMA is a partnership of ESO (representing its member states), NSF (USA) and NINS (Japan), together with NRC (Canada), NSTC and ASIAA (Taiwan), and KASI (Republic of Korea), in cooperation with the Republic of Chile. The Joint ALMA Observatory is operated by ESO, AUI/NRAO and NAOJ. The National Radio Astronomy Observatory is a facility of the National Science Foundation operated under agreement by the Associated Universities, Inc. D.J.T acknowledges financial support from STFC studentship ST/V507088/1 and C.J.C acknowledges support from the STFC (grant ST/Y002229/1). For the purpose of open access, the author has applied a Creative Commons Attribution (CC BY) licence to any Author Accepted Manuscript version arising. This work made use of NASA Astrophysics Data System, funded by NASA under Cooperative Agreement 80NSSC25M7105, and Python packages \textsc{analysisUtils} \citep{Hunter_2023}, APLpy \citep{Robitaille2012}, Astropy \citep{Astropy2022}, matplotlib \citep{Hunter2007} and NumPy \citep{Harris2020}.

\section*{Data availability}

The CoCCoA ALMA raw data and pipeline-processed image cubes are publicly available from the ALMA archive (project number 2019.1.00246.S). The combined CoCCoA 1.20\,mm continuum image used in this work and the combined 7m+12m array $^{12}$CO(2-1) and SiO(5-4) image cubes image cubes are available at doi:10.5281/zenodo.20202233.



\bibliographystyle{mnras}
\bibliography{example} 




\appendix

\section{Search for disc signatures} \label{Appendix-discs}

\begin{figure*}
	\includegraphics[trim =0 0 0 670, clip, width = \textwidth]{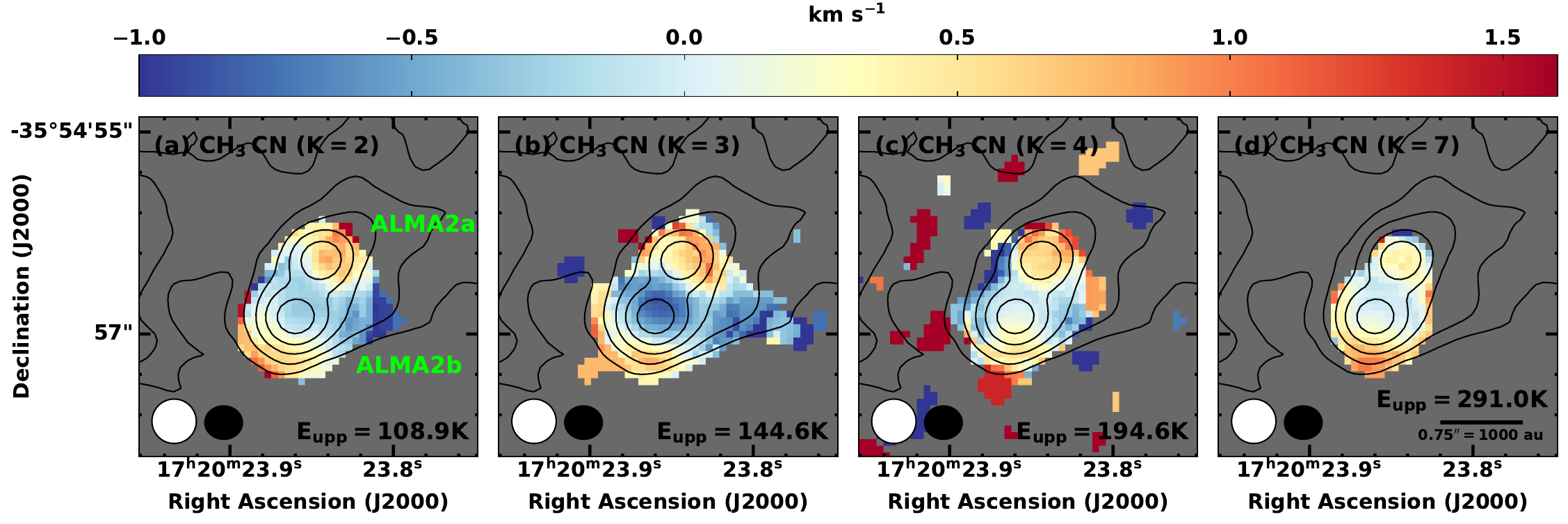}
    \caption{Primary beam corrected intensity weighted velocity (moment 1) maps for selected unblended CH$_{3}$CN transitions listed in Table~\ref{tab:line_emission_properties} towards ALMA2a/b, overlaid with CoCCoA ALMA 1.20\,mm continuum contours (in black, corrected for the primary beam response, levels: [5, 10, 20, 40, 80]$\sigma$, where $\sigma$ = 0.195 mJy beam$^{-1}$). A common mask of 4$\sigma$ is used, where $\sigma$ is measured locally as 4.05 mJy beam$^{-1}$ (see Section~\ref{sec:Coccoa_obs}). The colourscale is centred on a v$_{\rm LSR}$ = 0.30 km s$^{-1}$. The molecule name is given at upper left in each panel and the panels are ordered by increasing E$_{\rm upper}$, displayed at lower right in each panel. The ALMA line and continuum synthesised beams are at bottom left in each panel, in white and black respectively.}
    \label{fig:mom1_ALMA2_CH3CN}
\end{figure*}

\begin{figure*}
	\includegraphics[trim =0 0 0 670, clip, width = \textwidth]{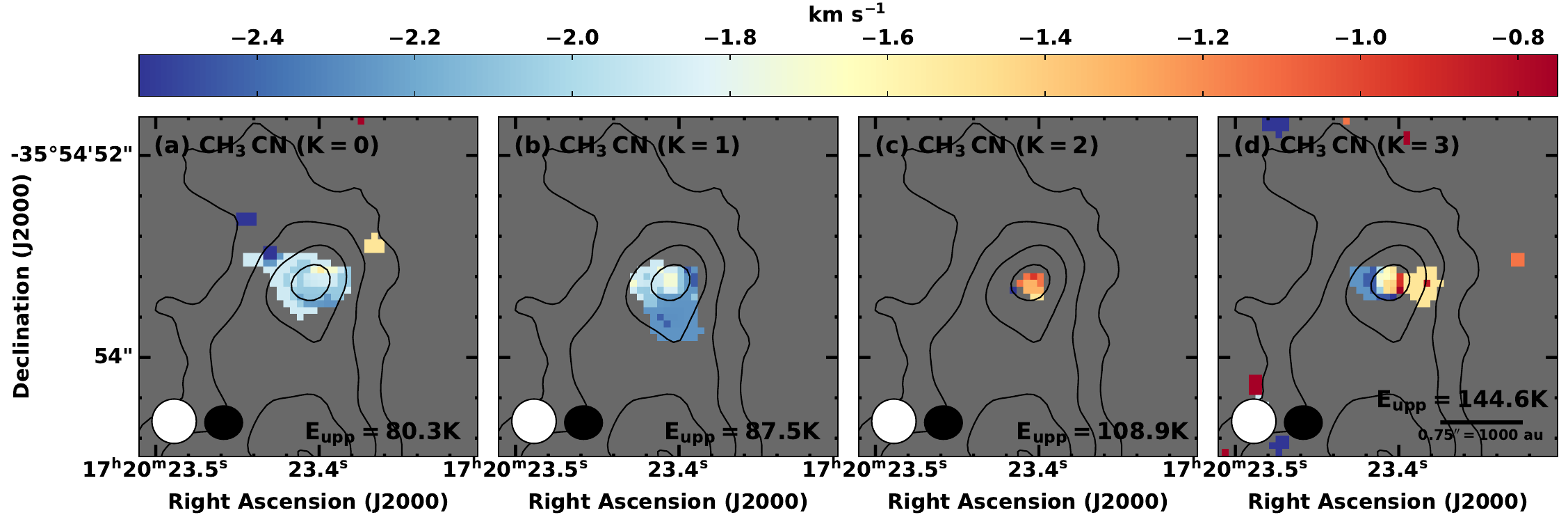}
    \caption{Primary beam corrected intensity weighted velocity (moment 1) maps for CH$_{3}$CN transitions listed in Table~\ref{tab:line_emission_properties} towards ALMA6a, overlaid with CoCCoA ALMA 1.20\,mm continuum contours (in black, corrected for the primary beam response, levels: [5, 10, 20, 40]$\sigma$, where $\sigma$ = 0.098 mJy beam$^{-1}$). A common mask of 4$\sigma$ is used, where $\sigma$ is measured locally as 2.57 mJy beam$^{-1}$ (see Section~\ref{sec:Coccoa_obs}). The colourscale is centred on a v$_{\rm LSR}$ = $-$1.65 km s$^{-1}$. The molecule name is given at upper left in each panel and the panels are ordered by increasing E$_{\rm upper}$, displayed at lower right in each panel. The ALMA line and continuum synthesised beams are at bottom left in each panel, in white and black respectively.}
    \label{fig:mom1_ALMA6_CH3CN}
\end{figure*}

Sources ALMA1, ALMA2a/b, ALMA5 and ALMA6a launch outflows (see Section~\ref{CO_results}), motivating investigation into the presence of a disc. We look for signatures of Keplerian rotation using strong unblended lines among ALMA2a/b's copious hot core line emission and in the CH$_{3}$CN emission towards ALMA6a in both the unconvolved and convolved line cubes (Section~\ref{sec:Coccoa_obs}). As no compact line emission is observed towards ALMA1 or ALMA5 we do not search for disc signatures. We find that moment 1 maps made from the unconvolved and convolved cubes show similar velocity gradients, and present maps made from the convolved cubes in Figures~\ref{fig:mom1_ALMA2_CH3CN} and ~\ref{fig:mom1_ALMA6_CH3CN}. 
Towards ALMA2a, there is tentative evidence of disc-like velocity structure in CH$_{3}$CN emission, with a velocity gradient roughly perpendicular to the low-velocity blueshifted SW component of the outflow (see Figures~\ref{fig:mom1_ALMA2_CH3CN} and ~\ref{fig:Louvet_outflow_plot} and Section~\ref{alma2}). We also find some evidence of an E-W velocity gradient across ALMA6a in CH$_{3}$CN $k$ = 3 (see Figure~\ref{fig:mom1_ALMA6_CH3CN}d). However, the direction of the gradient is not obviously perpendicular to the outflow lobe (see Figure~\ref{fig:Louvet_outflow_plot}). In each case higher resolution line data are required to confirm the possible presence of a disc. 

\section{\textsc{weedspymcmc} posterior distributions} \label{Appendix-CH3CN}

\begin{figure*}
	\includegraphics[trim=0 0 0 0, clip, width = \textwidth]{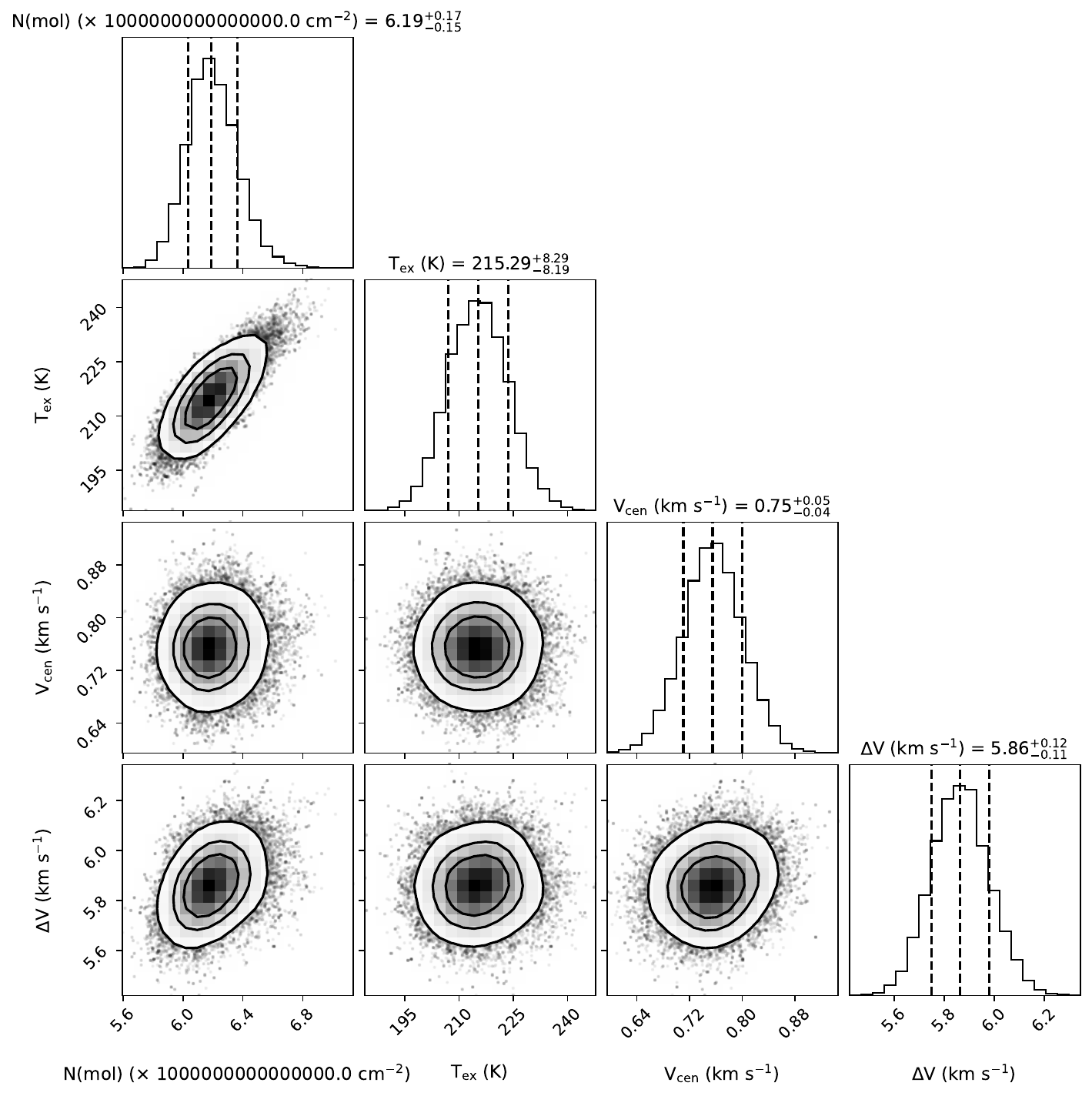}
    \caption{One and two dimensional histograms show the posterior distributions for the free parameters of the modelling of the ALMA2a spectrum, presented in Section~\ref{CH3CN fitting}. Contours on the 2D histograms are placed at [2, 1.5, 1]$\sigma$ and the dashed vertical lines on the 1D histograms are placed at the 0.16, 0.5, and 0.84 quantiles.}
    \label{fig:ALMA2a_corner_CH3CN}
\end{figure*}

\begin{figure*}
	\includegraphics[trim=0 0 0 0, clip, width = \textwidth]{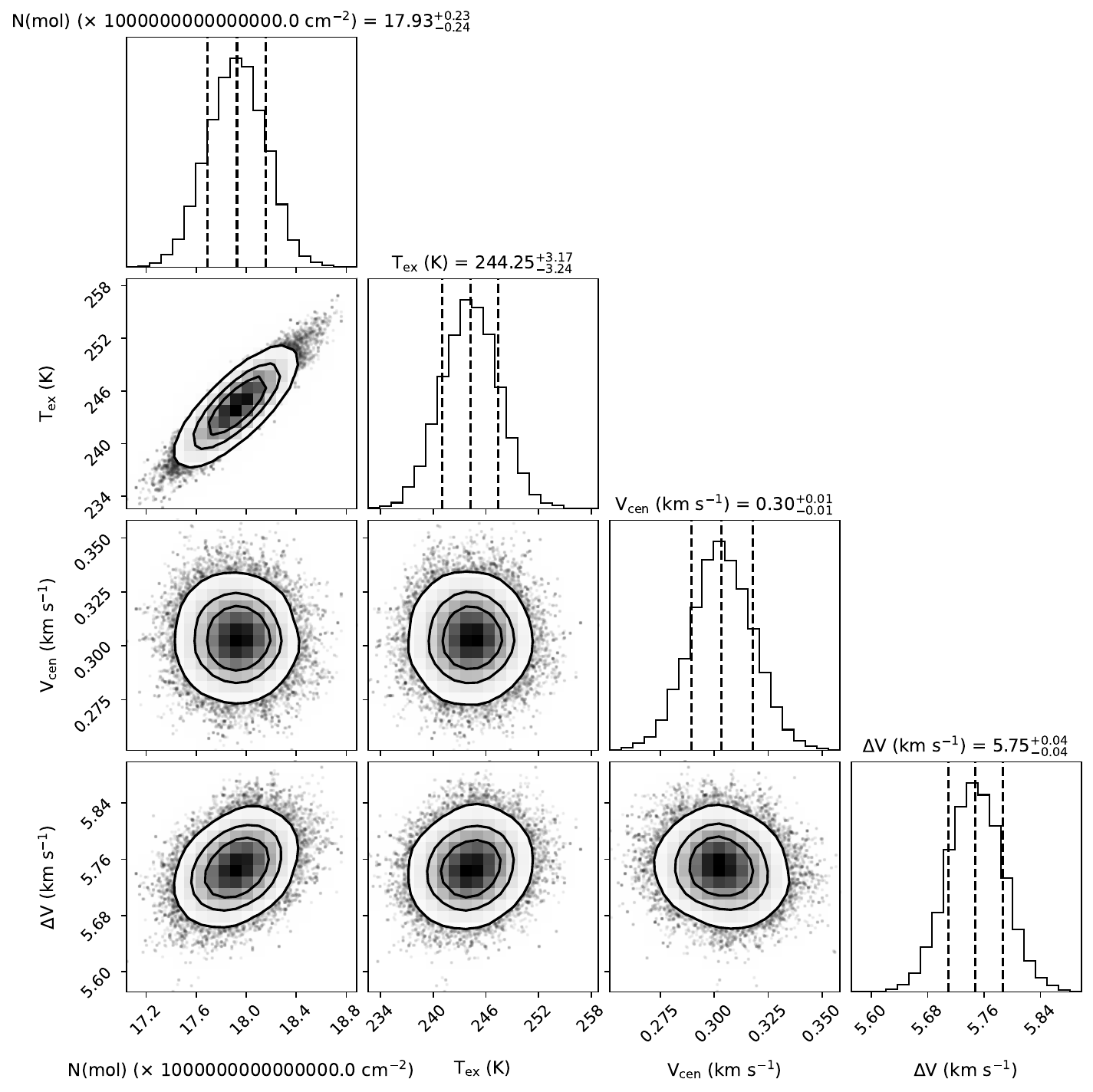}
    \caption{One and two dimensional histograms show the posterior distributions for the free parameters of the modelling of the ALMA2b spectrum, presented in Section~\ref{CH3CN fitting}. Contours on the 2D histograms are placed at [2, 1.5, 1]$\sigma$ and the dashed vertical lines on the 1D histograms are placed at the 0.16, 0.5, and 0.84 quantiles.}
    \label{fig:ALMA2b_corner_CH3CN}
\end{figure*}

\begin{figure*}
	\includegraphics[trim=0 0 0 0, clip, width = \textwidth]{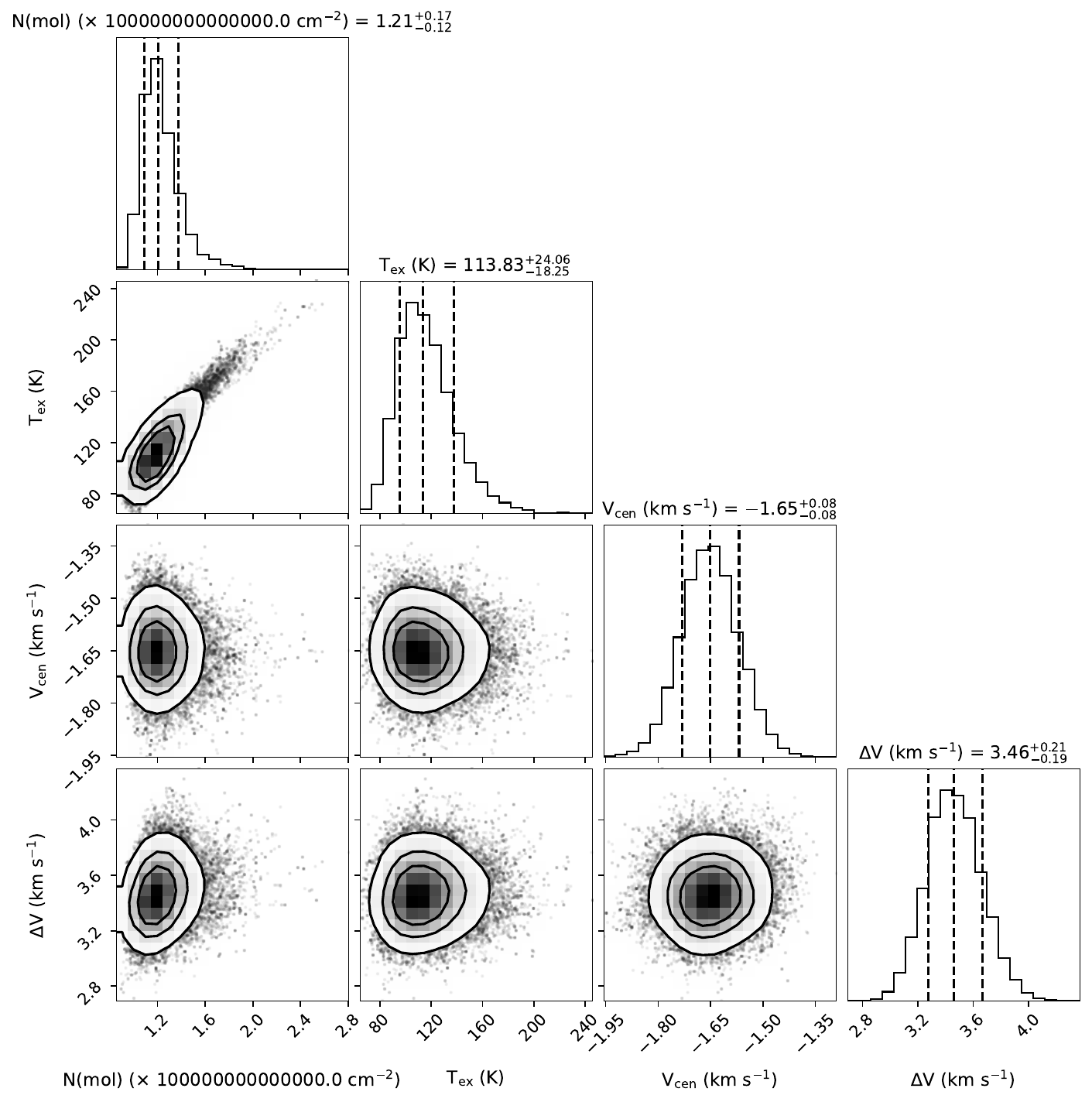}
    \caption{One and two dimensional histograms show the posterior distributions for the free parameters of the modelling of the ALMA6a spectrum, presented in Section~\ref{CH3CN fitting}. Contours on the 2D histograms are placed at [2, 1.5, 1]$\sigma$ and the dashed vertical lines on the 1D histograms are placed at the 0.16, 0.5, and 0.84 quantiles.}
    \label{fig:ALMA6a_corner_CH3CN}
\end{figure*}

We show corner plots of the posterior distributions produced from the \textsc{weedspymcmc} fits to the CH$_{3}$CN emission of ALMA2a, 2b and ALMA6a presented in Section~\ref{CH3CN fitting} in Figures~\ref{fig:ALMA2a_corner_CH3CN}, ~\ref{fig:ALMA2b_corner_CH3CN} and ~\ref{fig:ALMA6a_corner_CH3CN} respectively. 

\bsp	
\label{lastpage}
\end{document}